\documentclass[12pt]{article}
\usepackage{latexsym,epsfig,graphicx,amsmath,amssymb,amscd,undertilde,multirow}
\usepackage{chicago,psfrag,paralist,dsfont,url,comment}
\usepackage{subfig}
\usepackage{graphicx}
\usepackage[titletoc]{appendix}

\usepackage[american]{babel}

\newcommand{\MVN}{\textrm{MVN}}

\newcommand{\IG}{\textrm{IG}}

\newcommand{\tlog}{\textrm{log}}

\newcommand{\bx}{\boldsymbol{x}}
\newcommand{\bz}{\boldsymbol{z}}
\newcommand{\bw}{\boldsymbol{w}}
\newcommand{\bu}{\boldsymbol{u}}
\newcommand{\muvec}{\boldsymbol{\mu}}
\newcommand{\deltavec}{\boldsymbol{\delta}}
\newcommand{\tmin}{\textrm{min}}
\newcommand{\tmax}{\textrm{max}}
\newcommand{\trace}{\textrm{tr}}
\newcommand{\loglik}{\mathcal{L}}
\newcommand{\E}{\mathds{E}}
	
\newcommand{\btheta}{\boldsymbol{\theta}}
\newcommand{\bt}{\boldsymbol{t}}
\newcommand{\bbeta}{\boldsymbol{\beta}}
\newcommand{\bPsi}{\boldsymbol{\Psi}}

\newtheorem{preprop}{Proposition}
\newenvironment{proposition}%
{\begin{preprop}\upshape}{\end{preprop}}

\begin{document}

%%%%%%%%%%%%TITLE%%%%%%%%%%%%%%%%%%%%%%%%%%%%%%%%%%%

\title{A Spatially Correlated Competing Risks Time-to-Event Model for Supercomputer GPU Failure Data}

%%%%%%%%%%%%%%%%%%%%%%%%%%%%%%%%%%%%%%%%%%%%%%%%%%%%%%%%%%%%%%%%%%%%%%%%%%%%%%%%%%%%%%%%%%%%%%%%%%%%%%%%%%%%%%%%%

\author{
Jie Min$^1$, Yili Hong$^1$, William Q. Meeker$^2$, and George Ostrouchov$^3$ \\
{\small $^1$Department of Statistics, Virginia Tech, Blacksburg, VA 24061}\\
{\small $^2$Department of Statistics, Iowa State University, Ames, IA 50011}\\
{\small $^3$Computer Science and Mathematics Division, Oak Ridge National Laboratory,}\\
{\small Oak Ridge, TN 37831}
}

%\date{\today}
\date{}
\maketitle
	%%%%%%%%%%%%%%%%%%%%%%%%%%%%%%%%%%%%%%%%%%%%%%%%%%%%%%%%%%%%%%%%%%%%%%%%%%%%%%%%%%%%%%%%%%%%%%%%%%%%%%%%%%%%%%%%
\begin{abstract}
Graphics processing units (GPUs) are widely used in many high-performance computing (HPC) applications such as imaging/video processing and training deep-learning models in artificial intelligence. GPUs installed in HPC systems are often heavily used, and GPU failures occur during HPC system operations. Thus, the reliability of GPUs is of interest for the overall reliability of HPC systems. The Cray XK7 Titan supercomputer was one of the top ten supercomputers in the world. The failure event times of more than 30{,}000 GPUs in Titan were recorded and previous data analysis suggested that the failure time of a GPU may be affected by the GPU's connectivity location inside the supercomputer among other factors. In this paper, we conduct in-depth statistical modeling of GPU failure times to study the effect of location on GPU failures under competing risks with covariates and spatially correlated random effects. In particular, two major failure types of GPUs in Titan are considered. The connectivity locations of cabinets are modeled as spatially correlated random effects, and the positions of GPUs inside each cabinet are treated as covariates. A Bayesian framework is used for statistical inference. We also compare different methods of estimation such as the maximum likelihood, which is implemented via an expectation-maximization algorithm. Our results provide interesting insights into GPU failures in HPC systems.

\end{abstract}

\newpage
	
	%%%%%%%%%%%%%%%%%%%%%%%%%%%%%%%%%%%%%%%%%%%%%%%%%%%%%%%%%%%%%%%%%%%%%%%%%%%%%%%%%%%%%%%%%%%%%%%%%%%%%%%%%%%%%%%%%%%%%
\section{Introduction}\label{sec:introduction}
%%%%%%%%%%%%%%%%%%%%%%%%%%%%%%%%%%%%%%%%%%%%%%%%%%%%%%%%%%%%%%%%%%%%%%%% %%%%%%%%%%%%%%%%%%%%%%%%%%%%%%%%%%%%%%%%%%%%%%%%%%%%%%%%%%%%%%%%%%%%%%%%%%%%%%%%%%%%%%%%%%%%%%%%%%%%%%%%%%%%%%%%%%%%%
%\subsection{The Problem} %%%%%%%%%%%%%%%%%%%%%%%%%%%%%%%%%%%%%%%%%%%%%%%%%%%%%%%%%%%%%%%%%%%%%%%%%%%%%%%%%%%%%%%%%%%%%%%%%%%%%%%%%%%%%%%%%%%%%
	
Graphics processing units (GPUs) are widely used in high-performance computing (HPC). In many applications such as imaging/video processing and training deep-learning models, GPUs are important hardware components in computing systems. Supercomputers with GPUs provide a capability for doing massive computing, dealing with large-scale data, and training complicated models. These capabilities are important computational infrastructure for big data, machine learning, and artificial intelligence technologies. GPUs installed in HPC systems are often heavily used (i.e., in a continuously running mode), and failures occur during the service period of the HPC systems. Thus, the reliability of GPUs is of interest for the overall reliability of HPC systems.

The Cray XK7 Titan supercomputer was one of the top ten supercomputers in the world from November 2012 to November 2018 (\shortciteNP{top500}). The Titan supercomputer system had been operating since 2012 and was decommissioned in the summer of 2019.
Operating data collected over the nearly seven-year period provides rich information in
analyzing the failure event times of GPUs in the Titan supercomputer. The failure event times of more than 30{,}000 GPUs (as well as running times for the GPUs that had not failed by the end of the study) in Titan were recorded and made available in~\shortciteN{ostrouchov2020gpu}, providing an opportunity to study the reliability of GPUs in the Titan supercomputer.

During the operating period of the Titan supercomputer, two major GPU failure event types were observed. These were double-bit errors (DBE) and off-the-bus (OTB) failures (\shortciteNP{ostrouchov2020gpu}). A failure caused by either failure mode excludes the possibility of observing the failure caused by another failure mode; this suggests a competing risks model. In particular, DBE is related to the correction of a single-bit flip, or the detection of a double-bit flip, and OTB is related to the loss of host CPU connection to the GPU. Both DBE and OTB were found to be signature events of GPU board failing resistors that often lead to GPU replacement (\shortciteNP{ostrouchov2020gpu}). Thus, a good understanding of the occurrence of both event types can lead to a better understanding of GPU reliability. In addition to the event types (i.e., failure modes) and spatial location, there are also other covariates available such as the cage and slot information of the GPU within the cabinet, which will be detailed in Section~\ref{sec:GPU.study}.

	%%%%%%%%%%%%%%%%%%%%%%%%%%%%%%%%%%%%%%%%%%%%%%%%%%%%%%%%%%%%%%%%%%%%%%%%%%%%%%%%%%%%%%%%%%

The Titan GPU dataset has been analyzed by several authors in engineering literature, and descriptive statistics and elementary statistical tools were used. \shortciteN{gupta2015understanding} and \shortciteN{tiwari2015reliability} used descriptive plots such as histograms and heatmaps to show different types of failures have different failure proportions across GPU positions. \shortciteN{nie2018machine} used methods such as logistic regression and support vector machine to show GPU position influences single-bit error rates. More recently, \shortciteN{ostrouchov2020gpu} used Kaplan-Meier estimates and the Cox proportional hazards (PH) model to analyze the influence of GPU position on DBE and OTB failure times. For other HPC applications, \shortciteN{wang2017can} analyzed hardware failures of supercomputers from multiple data centers and concluded that the failure-time distribution of a server is related to the server's rack position. \shortciteN{di2017logaider} developed a novel $K$-meaning clustering algorithm based on the spatial correlation of failure events for the Mira supercomputer.

In the previous research, there has been no in-depth statistical modeling for GPU failure events. In this paper, we propose to use state-of-art statistical modeling for the Titan GPU failure data to investigate the relationship between locations and GPU failures, considering both spatial random effects and competing risks. Including spatial random effects introduces difficulty in estimation, and in the existing literature, there have been no spatial survival models used to model HPC systems.

	Although not used in modeling HPC systems, spatial survival models have become popular in biostatistics applications, such as modeling cancer data (\shortciteNP{onicescu2018spatially}, \shortciteNP{carroll2019temporally}, and \shortciteNP{wang2016bayesian}), stroke incidents (\shortciteNP{mlynarczyk2021bayesian}), and AIDS data (\shortciteNP{momenyan2020modeling}).  In geostatistics, spatial random effects are usually assumed to have a normal distribution and are added to the linear predictors in the hazard function (HF) of the Cox PH model.  \citeN{li2002modeling} added spatial random effects to the HF of the Cox PH model with a nonparametric baseline HF. \shortciteN{hennerfeind2006geoadditive} proposed a similar model using Bayesian analysis. Markov random field priors, P-spline priors, and Gaussian random field priors were used for the coefficients in the spatial random effects model. \shortciteN{li2015survival} also added spatial random effects to the semiparametric PH model and considered various correlation functions. \shortciteN{pan2014bayesian} considered spatial interval-censored data and used a conditional autoregressive (CAR) model to describe the spatial structure.  \shortciteN{motarjem2019bayesian} used spatial random effects with non-Gaussian distributions.

In some papers, spatial correlation terms have been added to the baseline HF of the Cox PH model. \citeN{geng2021bayesian} proposed a model with different baseline HFs
for different locations, and used a geographically weighted Chinese restaurant process prior to capture the spatial structure. \shortciteN{chang2013spatial} linked a discrete event baseline HF with spatial random effects using a probit link.  \citeN{li2006semiparametric} used a probit link to describe the marginal cumulative HF for each observation and jointly assumed them to have a multivariate normal distribution with spatially correlated covariance matrices.  \shortciteN{henderson2002modeling} proposed a model allowing the individual frailty terms in the Cox PH model to have a marginal Gamma distribution and jointly have a covariance matrix with a spatial correlation structure. The accelerated failure time (AFT) model with spatial random effects as linear predictors has also been used in the literature. For example, \citeN{zhou2018unified} proposed a framework to model arbitrarily censored spatial survival data considering both areal and georeferenced spatially correlated data using an AFT model with spatial random effects.
Also, \shortciteN{wang2016bayesian} proposed a normal mixture AFT model, using a Dirichlet prior for mixture weights and a CAR model for the random effects.

    Although there are many papers focusing on combining time-to-event data and spatial random effects together, there are few of them that consider competing risks at the same time. \shortciteN{hesam2018cause} and \shortciteN{momenyan2020modeling} are two papers considering this combination, and CAR or multivariate conditional autoregressive (MCAR) model priors are used for spatial random effects. This approach is not suitable for the Titan GPU data, because the distance in Titan is computed based on a point-reference structure.

In summary, existing methods in the statistical literature are not directly applicable to the Titan GPU data. The special features of the Titan GPU data motivate us to develop a new time-to-event model using spatial random effects and a competing risk model. We consider several commonly used spatial correlation functions to model the spatial random effects for the two failure modes, taking the correlation between the two failure modes into consideration. Because of the large number of observations and spatial locations, we use Bayesian methods with noninformative or weakly informative priors for model fitting and inference. We also compare different statistical methods to ensure that the best available statistical methods are used for the data analysis. Our contribution can be highlighted as follows. This paper is the first work that conducts an in-depth statistical analysis of the large-scale Titan GPU data. Our proposed time-to-event model with spatial random effects under competing risks enables us to answer the two important questions: 1) how do different spatial locations affect the GPU failure time, and 2) how do spatial effects of the different failure modes interact with each other. Answering these questions would allow the designers of HPC systems to modify system design (e.g, physical layout and thermal management) to improve the HPC system reliability. In addition to this scientific problem contribution, our proposed model is also new to the statistical literature. That is, the developed spatially correlated time-to-event model under competing risks is not limited to GPU failure event analysis, but it should be applicable to a wide range of other applications in reliability and survival analysis.

	%%%%%%%%%%%%%%%%%%%%%%%%%%%%%%%%%%%%%%%%%%%%%%%%%%%%%%%%%%%%%%%%%%%%%%%%%%%%%%%%%%%%%%%%%%%%%%%%%%%%%%%%%%%%%%%%%%%%%
	The rest of the paper is organized as follows. Section~\ref{sec:GPU.study} introduces the Titan GPU dataset and the data notation. Section~\ref{sec:stat.model} describes the proposed time-to-event model with spatial random effects under competing risks, parameter estimation, and inference procedures. Section~\ref{sec:simulation.study} presents a simulation study in model estimation and inference. Section~\ref{sec:data.analysis} describes the data analysis, comparisons with the AFT model and maximum likelihood estimation via the expectation-maximization (EM) algorithm, and the interpretation of the results. Section~\ref{sec:conclusoin} gives some concluding remarks.

	%%%%%%%%%%%%%%%%%%%%%%%%%%%%%%%%%%%%%%%%%%%%%%%%%%%%%%%%%%%%%%%%%%%%%%%%%%%%%%%%%%%%%%%%%%%%%%%%%%%%%%%%%%%%%%%%%%%%%
	\section{The GPU Data for the Titan Supercomputer}\label{sec:GPU.study}
	%%%%%%%%%%%%%%%%%%%%%%%%%%%%%%%%%%%%%%%%%%%%%%%%%%%%%%%%%%%%%%%%%%%%%%%%%%%%%%%%%%%%%%%%%%%%%%%%%%%%%%%%%%%%%%%%%%%%%
	\subsection{Data Summary and Visualization}
	%%%%%%%%%%%%%%%%%%%%%%%%%%%%%%%%%%%%%%%%%%%%%%%%%%%%%%%%%%%%%%%%%%%%%%%%%%%%%%%%%%%%%%%%%%%%%%%%%%%%%%%%%%%%%%%%%%%%%
The data used in this paper is described in \shortciteN{ostrouchov2020gpu}, which provide failure-time data for more than 30{,}000 GPUs that were in use during the service period of the Titan supercomputer, ranging from 2012 to 2019. We consider the OTB and DBE failure modes to obtain a more complete understanding of GPU reliability.

\begin{figure}
\centering
\begin{tabular}{ccc}
\includegraphics[width=.3\textwidth]{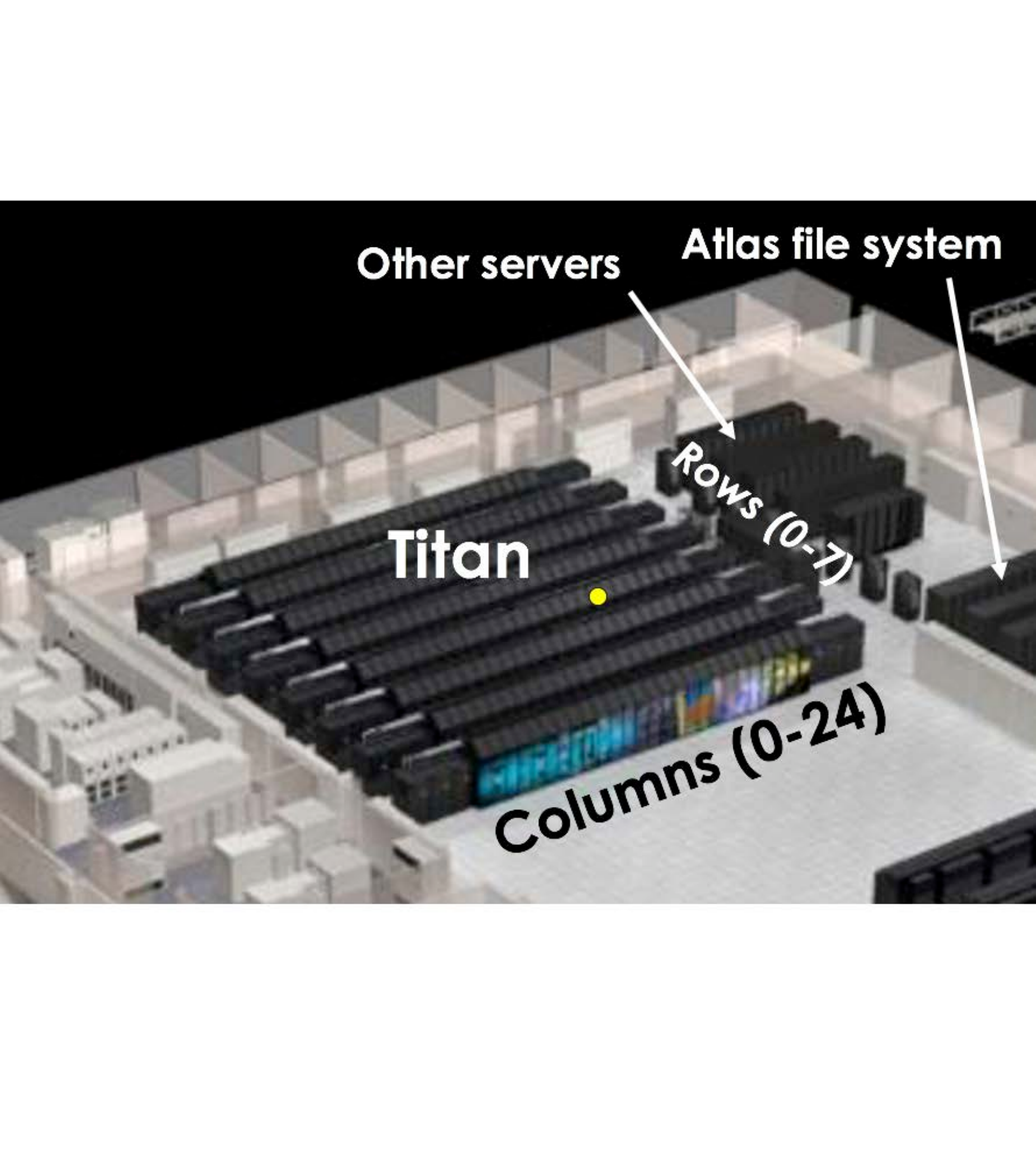} &
\includegraphics[width=.3\textwidth]{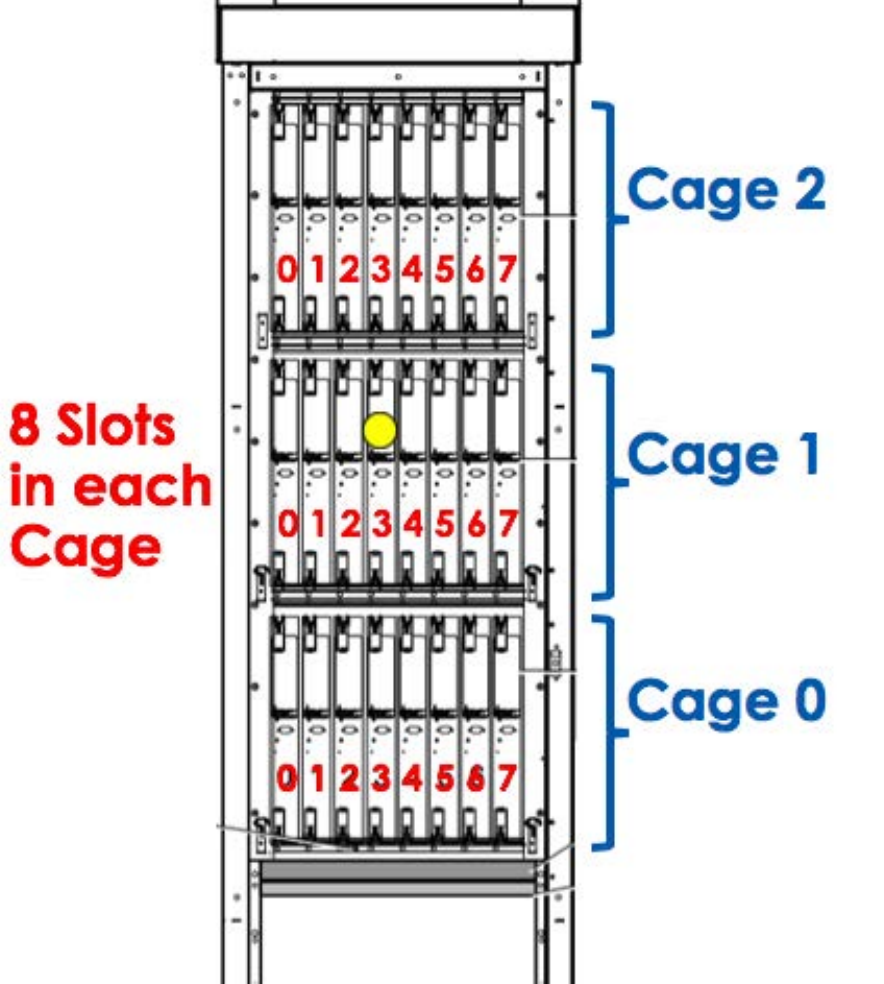}&
\includegraphics[height=.3\textwidth]{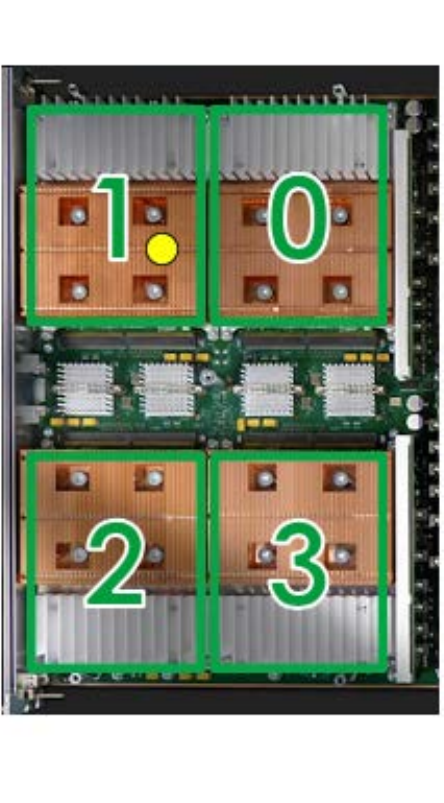} \\
(a) Cabinet Array &
(b) Cabinet Structure &
(c) Slot Structure \\
\end{tabular}
\caption{The structure of the Titan supercomputer (images reused with permission from Ostrouchov et al. (2020)).}\label{fig:tts}
\end{figure}

Inside the Titan supercomputer, there were 25 columns and 8 rows of cabinets, with 3 cages inside each cabinet, 8 slots in each cage, 4 nodes in each slot, and one GPU was installed on each node. Figure~\ref{fig:tts} shows the structure of the Titan supercomputer from \shortciteN{ostrouchov2020gpu}. The yellow dots on the three sub-figures represent the GPU at column 17, row 4, cage 1, slot 3 and node 1. The GPU dataset contains the following information: the serial number of GPUs in Titan, the positions of GPUs inside cabinets (i.e., the cage, slot, and node information), the row and column locations of cabinets, the failure/service time of GPUs, and the GPU failure mode (i.e., DBE, OTB, or censored for GPUs that did not fail).

According to \shortciteN{ostrouchov2020gpu}, GPUs that were first installed before 01 January 2016 were labeled as the old batch, and those that were installed after 01 January 2016 were labeled as the new batch. \shortciteN{ostrouchov2020gpu} indicated that the failure-time distributions of the old and new batches are different, because of different failure mechanisms. Thus, it would be misleading to pool the data for analysis. In addition, there were only a few new-batch GPUs that failed, and most locations had not observed any failures yet. Thus, there is not enough information in the data for spatial modeling of new-batch GPUs. Based on those considerations, we use the GPU data from the old batch for the modeling and analysis in this paper. As a summary, there are  1{,}127 OTB events, 3{,}093 DBE events, and 15{,}099 censored observations, and in total, there are 19{,}319 GPU in the old batch that are used in analysis.

\begin{figure}
\begin{center}
\begin{tabular}{cc}
\includegraphics[width = 0.46\textwidth]{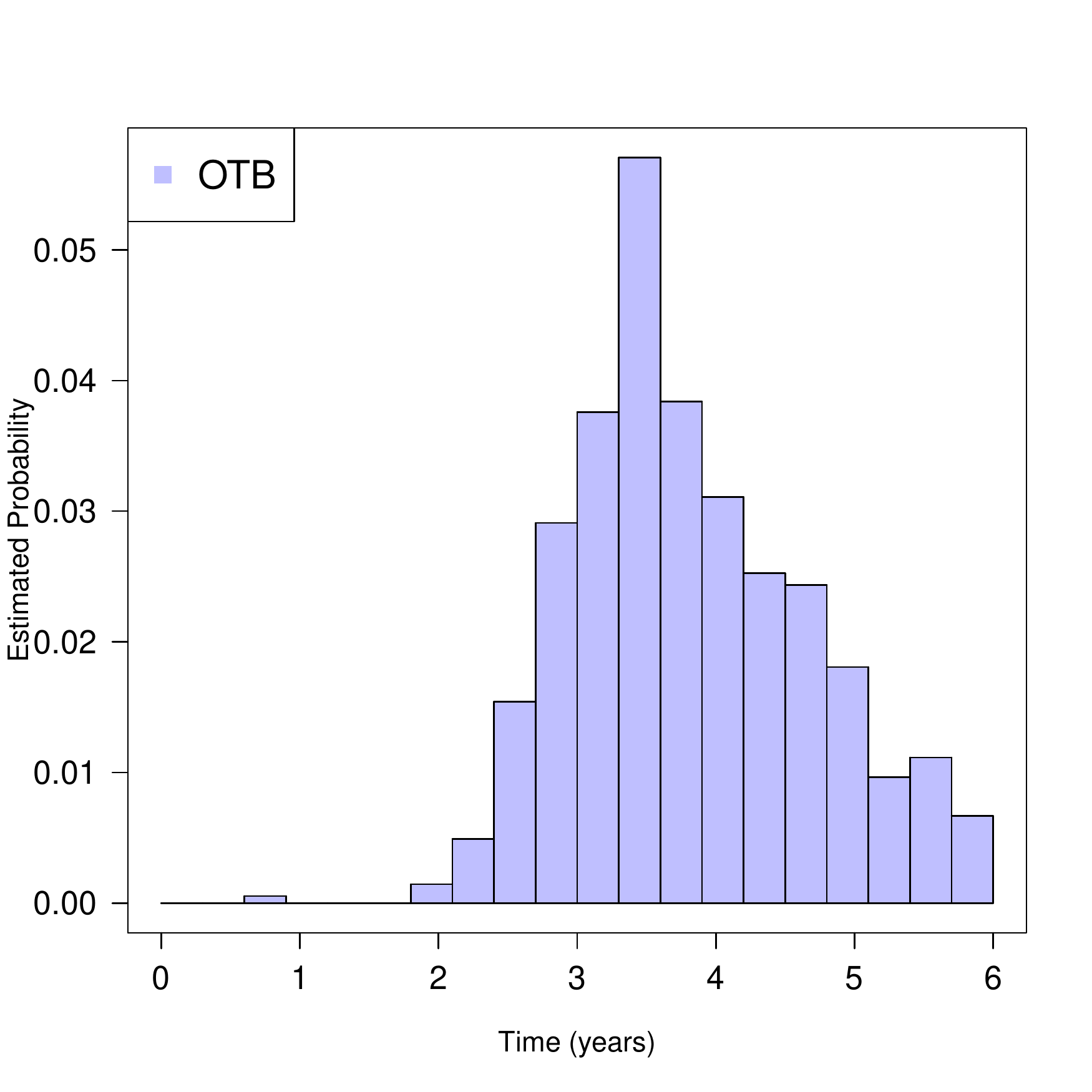} &
\includegraphics[width = 0.46\textwidth]{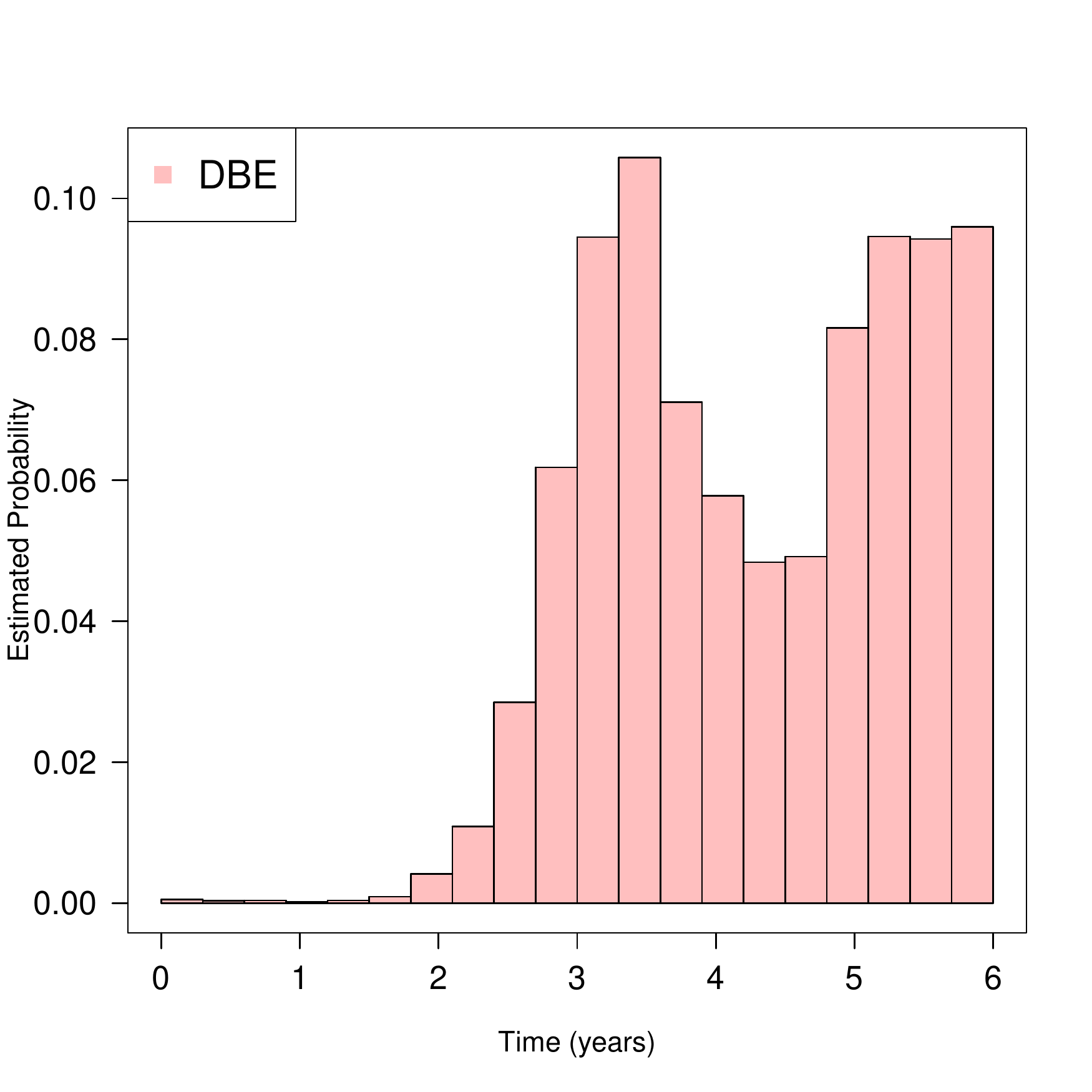}
\end{tabular}
\caption{Barplots of discretized estimated probability mass functions of OTB and DBE failures based on Kaplan-Meier estimates.}\label{fig:visual1}
\end{center}
\end{figure}

Figure~\ref{fig:visual1} shows barplots of discretized estimated probability mass function of OTB and DBE failures using Kaplan-Meier estimator. The barplots are built according to the method introduced in \shortciteN{huzurbazar2005censored}. From the barplot of DBE, we can clearly see a bimodal behavior of failure times, and the mode for the first DBE failure component nears the mode of OTB failure, which is around $3.5$ years. It is interesting to point out that the bimodal behavior of DBE failure times could indicate that there are two causes for DBE failures (i.e., one cause for early failures and another cause for later failures). Because DBE failures are memory errors and OTB failures are disconnects, it is possible that one of the causes for DBE failures is similar to the cause of OTB failures. We, however, do not have any explicit further information about the causes. Thus, we model the bimodal behavior using a mixture distribution with two components.

\begin{figure}[ht]
\begin{center}
\begin{tabular}{c}
\includegraphics[width = 0.75\textwidth]{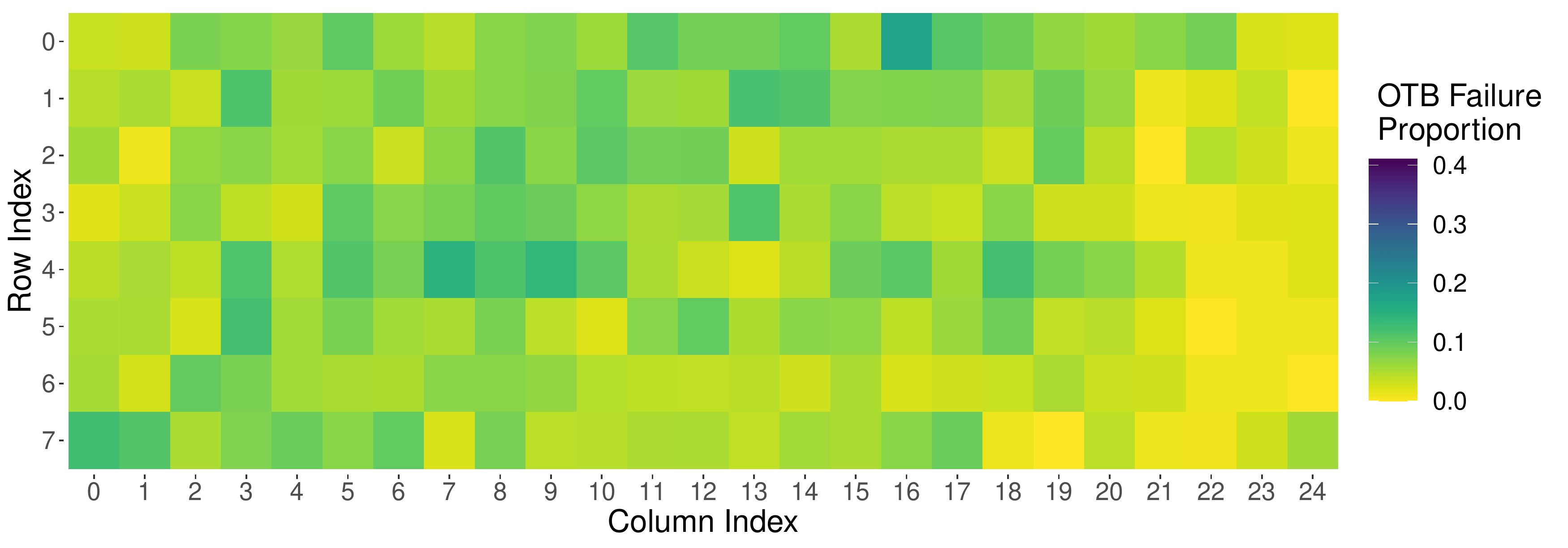}\\
\includegraphics[width = 0.75\textwidth]{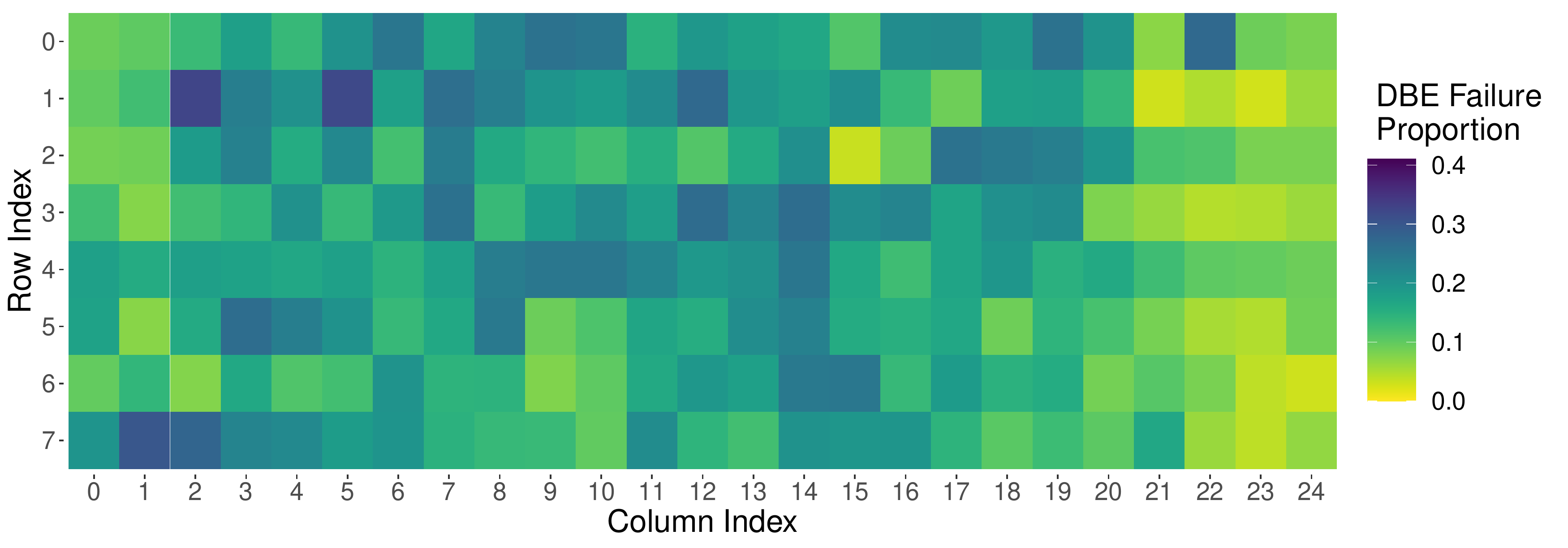}
\end{tabular}
\caption{Visualization of the OTB and DBE failure proportions on 200 different cabinet locations.}\label{fig:visual.heat.map}
\end{center}
\end{figure}

To visualize the spatial effects, Figure~\ref{fig:visual.heat.map} shows the marginal OTB and DBE failure proportions at $200$ different cabinet locations ($25$ columns and $8$ rows). The marginal failure proportion on one location is the number of OTB or DBE failed GPUs on the location divided by the total number of GPUs on the location. It is clear that the marginal failure proportions at different locations are different, indicating the existence of spatial effects.

\subsection{Notation for Data}
%%%%%%%%%%%%%%%%%%%%%%%%%%%%%%%%%%%%%%%%%%%%%%%%%%%%%%%%%%%%%%%%%%%%%%%%%%%%%%%%%%%%%%%%%%
We group GPUs by different cabinet locations. Let $i$ be the location of a GPU cabinet, and $j$ be the index for a GPU inside one cabinet. There are $25$ columns and $8$ rows of cabinets for the overall system, so we have $i={1,2,\ldots,n}$, and there are $n=200$ locations. We have $j={1,2,\ldots,n_i}$, and $n_i$ is the total number of GPUs at location $i$. Note that $n_i$ can vary from location to location, and the sum of the $n_i$ values is equal to the total number of GPUs in the dataset. Let $T_{ijk}$ be the random variable of failure time for unit with cabinet location index $i$ and inside cabinet index $j$ having failure mode $k$. Let $t_{ij}$ be the observed failure time for failed (censored) unit with cabinet location index $i$ and inside cabinet index $j$. Let $\delta_{ijk}$ be the event-type indicator. That is, $\delta_{ijk}=1$ when unit $(i,j)$ failed because of failure mode $k$, and $k={1,2}$ for the two failure modes. Note that for a censored observation $\delta_{ij1}=\delta_{ij2}=0.$

Let $\bx_j=(x_{j1},\ldots,x_{jp})'$ be the covariate vector with inside cabinet index $j$. Because the cage, slot, and node positions are categorical with $3$ cage levels, $8$ slot levels, and $4$ node levels, respectively, we use dummy variable coding. In total, we have $p=12$ where $p$ is the total number of covariates.
Let $\bz_i=(r_i,c_i)'$ be the row and column location index for units with location index $i$, where $r_i$ is the row index and $c_i$ is the column index. Here, $r_i$ can take values ${0,1,\ldots,7}$, and $c_i$ can take values ${0,1,\ldots,24}$. During the service period of the Titan supercomputer, some GPUs were swapped with GPUs from another location. For those GPUs, we follow the convention in \shortciteN{ostrouchov2020gpu} and use the location where the GPU spent the majority of its service time to determine the value of $c_i$ and $r_i$ for that unit. In summary, the time-to-event data are denoted by $\bt=\{t_{ij},\delta_{ijk},\bz_i,\bx_j\}$, $i= 1,\ldots,n$, $j=1,\ldots,n_i$, and $k=1,2$.

%%%%%%%%%%%%%%%%%%%%%%%%%%%%%%%%%%%%%%%%%%%%%%%%%%%%%%%%%%%%%%%%%%%%%%%%%%%%%%%%%%%%%%%%%%
\section{Statistical Models and Inference}\label{sec:stat.model}
%%%%%%%%%%%%%%%%%%%%%%%%%%%%%%%%%%%%%%%%%%%%%%%%%%%%%%%%%%%%%%%%%%%%%%%%%%%%%%%%%%%%%%%%%%
\subsection{Statistical Models}	\label{sec:sm}

We model the time to event $T_{ijk}$ using distributions from the log-location-scale family. In particular, we use a single log-location-scale distribution to describe the distribution of $T_{ij1}$ for OTB (failure mode 1). Because of the multimodal pattern of DBE failure (failure mode 2) seen in Figure~\ref{fig:visual1}, we model failure mode 2 with a mixture of two different log-location-scale distributions. Specifically, for OTB failures  (failure mode $1$, $k = 1$), the probability density function (pdf)  and cumulative distribution function (cdf) can be represented as,
\begin{align}
f(t_{ij}|w_{ik})= \frac{1}{\xi_{k} t_{ij}}\phi\left [\frac{\log (t_{ij})-\mu_{ijk}}{\xi_{k}}\right],
\text{ and }
F(t_{ij}|w_{ik})= \Phi\left[\frac{\log(t_{ij})- \mu_{ijk}}{\xi_{k}} \right],\label{eqpdf1}
\end{align}
where $\mu_{ijk}$ and $\xi_{k}$ are the location and scale parameters for the location-scale distribution corresponding to observation $i$ with inside cabinet index $j$ and failure mode $k$, respectively. We consider two widely used distributions from the log-location-scale family, namely the Weibull distribution and lognormal distributions (e.g., Chapter 4 in \citeNP{meeker2022statistical}). In \eqref{eqpdf1}, $\phi(z)$ and $\Phi(z)$ are the standard pdf and cdf from the location-scale distribution, which can be the smallest extreme value (SEV) distribution or the normal distribution in the paper. Also, $w_{ik}$ is a spatial random effect that will be discussed later.

We model the distribution of DBE failures (failure mode $2$, $k=2$) with a mixture distribution and the pdf is
\begin{align}
	f(t_{ij}|w_{ik}) = \lambda\left[\frac{1}{\xi_{k1} t_{ij}}\phi\left (\frac{\log (t_{ij})-\mu_{ijk1}}{\xi_{k1}}\right)\right] + (1- \lambda) \left[\frac{1}{\xi_{k2} t_{ij}}\phi\left (\frac{\log (t_{ij})-\mu_{ijk2}}{\xi_{k2}}\right) \right],\label{reflab}
\end{align} where  $\mu_{ijk1}$ and $\mu_{ijk2}$ are the two location parameters, and $\xi_{k1}$  and $\xi_{k2}$ are the two scale parameters of the corresponding location-scale distributions for the mixture components. We use $\lambda$ to represent the mixture proportion.
One commonly known issue of mixture modeling is the label switching problem, which makes the model not identifiable. A simple way to solve this problem is to add constraints on the parameters to make the likelihood  asymmetric with respect to the two mixed distributions \shortcite{jasra2005markov}. In particular, for the mixture distribution in \eqref{reflab}, we use the restriction $\mu_{ijk1} < \mu_{ijk2}$ to avoid identifiability problems in estimation.

In summary, the location parameters are modeled as follows,
\begin{align}
\mu_{ijk}  = &\mu_k+\bx_j' \bbeta_k+ w_{ik}, k=1\label{eqrfff0}\\
\mu_{ijk1} = &\mu_k+\bx_j' \bbeta_k+ w_{ik}, k=2\label{eqrfff1}\\
\mu_{ijk2} = &\mu_k+ \eta + \bx_j' \bbeta_k+ w_{ik}, k=2.\label{eqrff2}
\end{align}
Under failure mode $k$, $\bbeta_k$ contains the corresponding coefficients for $\bx_j$, $w_{ik}$ corresponds to the location random effect at cabinet location $\boldsymbol{z}_i$, and $\eta$ represents the difference between two components of the mixture distribution. In \eqref{eqrfff0}, $\mu_1$ is the baseline of the $\mu_{ij1}$ parameter for event time when $w_{i1}=0$. In \eqref{eqrfff1} and \eqref{eqrff2}, $\mu_2$ is the baseline for the first mixture component when $w_{i2}=0$ and $\mu_2 + \eta$ is the baseline for the second mixture component when $w_{i2}=0$. By setting $\eta>0$, we can avoid the identifiability problem. Note that the pdf of the two failure modes can be separated by conditioning on the spatial random effects.

We now discuss the modeling of spatial random effects. Let $\bw_k=(w_{1k},\ldots,w_{nk})', k=1, 2$ and $\bw=(\bw_1', \bw_2')'$. We model $\bw$ by using a multivariate normal distribution $\MVN(\boldsymbol{0},\Sigma_{\bw})$, and the covariance matrix $\Sigma_{\bw}$ is
$
\Sigma_{\bw} =  \Sigma_f \otimes \Omega,
$	
where $\otimes$ is the Kronecker product. The covariance matrix of $(w_{i1}, w_{i2})', i = 1,\ldots, n$, is
	\begin{equation*}
		\Sigma_{f} =  \begin{pmatrix}
			\sigma_{1}^2 & \rho_{12}\sigma_{1} \sigma_{2}\\
			\rho_{12}\sigma_{1} \sigma_{2} & \sigma_{2}^2
		\end{pmatrix},
	\end{equation*}
where $\sigma_{1}^2$ and $\sigma_{2}^2$ are variances for $w_{i1}$ and $w_{i2}$, respectively, and $\rho_{12}$ is the correlation between $w_{i1}$ and $w_{i2}$, which is used to model the correlation between the two failure modes. The spatial correlation matrix is,
\begin{align}\label{eqref2}
\Omega = (\omega_{sl}) = \exp\left[ - \left(\frac{d_{sl}}{\nu}\right)^\kappa \right], s=1,\ldots, n, l=1,\ldots, n,
\end{align}
with parameters $0<\kappa\leq 2$ and $\nu >0$, and $d_{sl}$ is the distance between cabinet locations. We consider three commonly used spatial correlation functions here, which are the Gaussian, exponential, and power exponential correlation functions \shortcite{sherman2011spatial}. These three different correlation functions have different $\kappa$ values in \eqref{eqref2}. The Gaussian correlation function has fixed $\kappa = 2$, the exponential correlation function has fixed $\kappa = 1$, and the power exponential correlation function has a power $0<\kappa \leq 2$ that is not fixed.

%%%%%%%%%%%%%%%%%%%%%%%%%%%%%%%%%%%%%%%%%%%%%%%%%%%%%%%%%%%%%%%%%%%%%%%%%%%%%
\subsection{Quantifying Distance}
\begin{figure}
\begin{center}
	\includegraphics[width = 0.5\textwidth]{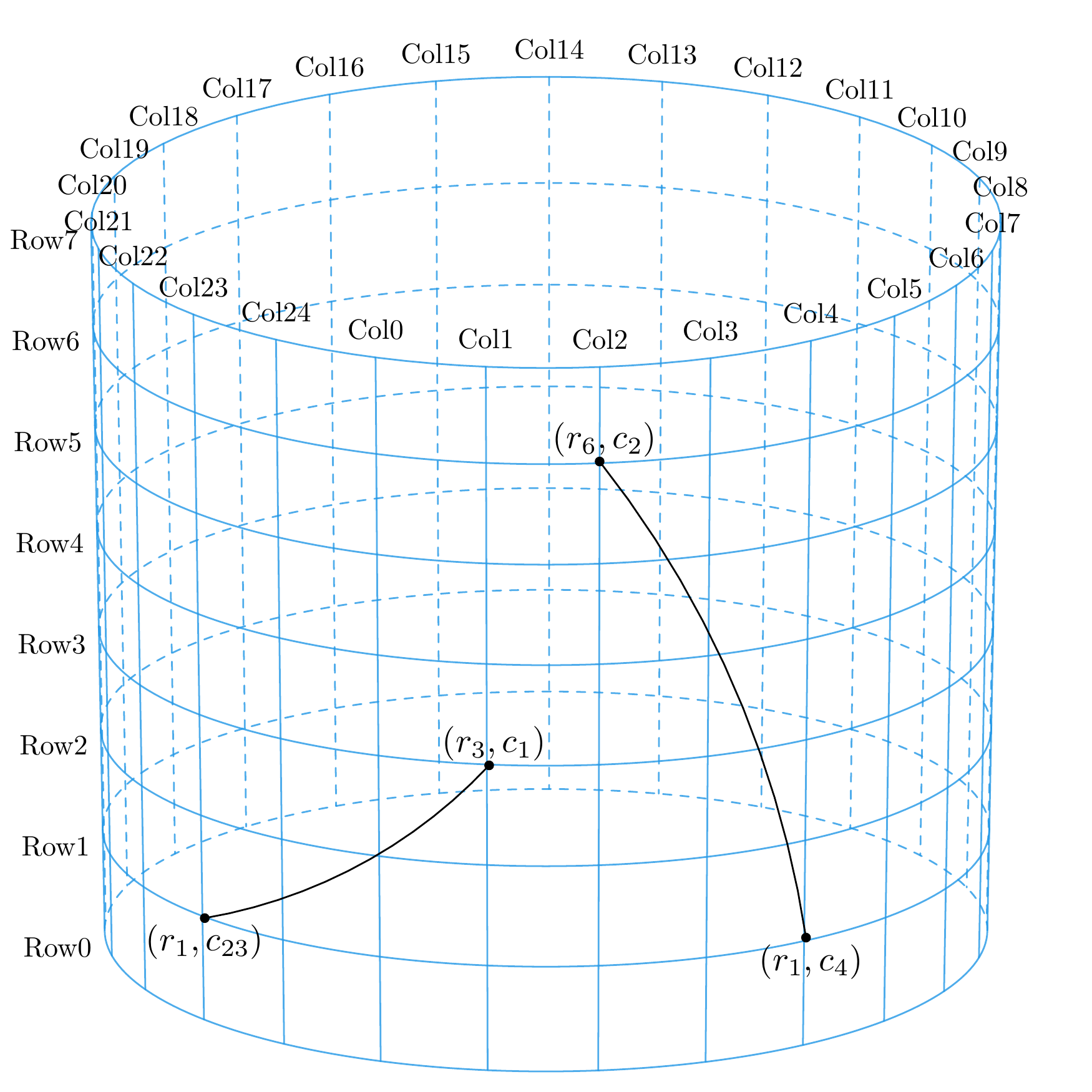}
	\caption{Visualization of the distance calculation. The black lines represent the distance between position $(r_1,c_{23})$ to $(r_3,c_1)$, and $(r_6, c_2)$ to $(r_1, c_4)$.}\label{fig:dist}
\end{center}
\end{figure}

%%%%%%%%%%%%%%%%%%%%%%%%%%%%%%%%%%%%%%%%%%%%%%%%%%%%%%%%%%%%%%%%%%%%%%%%%%%%%
Our method of quantifying the distances between GPU cabinets requires some discussion.  We want to point out that here the column index of the cabinet is labeled according to the connectivity of the cabinets but not the physical locations of the cabinets. In Titan, column~0 (under the current labeling) is connected to column~1 (under the current labeling), column~1 is connected to column 2, and so on. Then column~24 is connected back to column 0. Figure~\ref{fig:dist} visualizes the spatial connectivity, as indexed by row and column. Each intersecting point represents a cabinet. Our preliminary analysis showed that relabeling the column index using this connectivity is necessary. Failure proportions of different rows and columns without relabeled column indices (i.e., using the original physical location labels) are shown in Appendix~\ref{apd1}. The figures show that the pattern of failure proportions is more clearly related to the column index based on connectivity. For the row index, we use the original row value to indicate the physical location.

To compute the distance between cabinet locations $d_{sl}$, we take into account the spatial connectivity of the columns of cabinets and the physical location of the rows of cabinets. The distance is defined as
\begin{align}\label{eqref3}
d_{sl}=
\sqrt{((r_s-r_l)/7)^2+\left(\min\{|c_s-c_l|, 25-|c_s-c_l|\}/12\right)^2},
\end{align}
because column~24 is connected back to column~0. Figure~\ref{fig:dist} provides a visualization of calculating distance between location $(r_1, c_{23})$ to $(r_3, c_1)$, and $(r_6, c_2)$ to $(r_1, c_4)$. Using \eqref{eqref3}, the distance between column~0 and column~24 is 1. Thus, the column labels form a circle and the rows and columns together form a cylinder. Note that we also normalize $(r_s - r_l)$ and $\min\{|c_s-c_l|, 25-|c_s-c_l|\}$ to range $[0, 1]$ in calculating $d_{sl}$ to make the connectivity and physical distance have an equal contribution in the distance calculation. The largest distance is $d_{sl} = \sqrt{2}$.
As noted by \shortciteN{gneiting2013strictly}, distances defined on a sphere can result in a non-positive definite correlation matrix. Our distance is different from that used in existing work because it is like a cylinder. Thus, some special attention is needed to assure positive definiteness. We describe the accommodations in Section~\ref{sec:prior} as part of the modeling process.

%%%%%%%%%%%%%%%%%%%%%%%%%%%%%%%%%%%%%%%%%%%%%%%%%%%%%%%%%%%%%%%%%%%%%%%%%%%%%%%%%%%%%%%%%%%
\subsection{Likelihood and Prior Specification} \label{sec:prior}	%%%%%%%%%%%%%%%%%%%%%%%%%%%%%%%%%%%%%%%%%%%%%%%%%%%%%%%%%%%%%%%%%%%%%%%%%%%%%%%%%%%%%%%%%%%

Let $\btheta_{\bt}$ be the unknown parameters corresponding to the time-to-event distribution of $T_{ijk}$ and let $\btheta_{\bw}$ be the unknown parameters for the distribution of random effects. Specifically, 	
	$\btheta_{\bt}=(\mu_1,\mu_2, \bbeta_1',\bbeta_2', \xi_1, \xi_{21},\xi_{22}, \lambda,\eta)'$ and $\btheta_{\bw}=(\sigma_{1},\sigma_{2},\rho_{12},\nu,\kappa)'$. The likelihood, conditional on fixed values of the spatial random effects $\bw$ is,
	\begin{align}\label{eqn:lik.cond.w}
		L(\btheta_{\bt}|\bt,\bw)&= \prod_{i,j}\prod_{k=1}^{2}f(t_{ij}|w_{ik})^{\delta_{ijk}}[1-F(t_{ij}|w_{ik})]^{1-\delta_{ijk}}.
	\end{align}	
The joint density for  $\bw$ is,
	\begin{align*}
		f_{\bw}(\bw|\btheta_{\bw}) \propto |\Sigma_{\bw}|^{-\frac{1}{2}}\exp\left(-\frac{1}{2} \bw' \Sigma_{\bw}^{-1}\bw\right).
	\end{align*}
	Then, the likelihood for $\btheta=(\btheta_{\bt}',\btheta_{\bw}')^ \prime$ is,
	\begin{align*}
		L(\btheta|\bt,\bw)& \propto \prod_{i,j}\prod_{k=1}^{2}f(t_{ij}|w_{ik})^{\delta_{ijk}}[1-F(t_{ij}|w_{ik})]^{1-\delta_{ijk}}f_{\bw}(\bw|\btheta_{\bw}).
	\end{align*}

For $\btheta_{\bt}$, we choose flat priors for the unrestricted parameters $\mu_k$ and $\bbeta_k$, and uniform priors for the restricted parameters $\lambda$ and $\eta$. Because the domain of the $\xi_k$'s is $(0,\infty)$, we use the noninformative prior for $\xi_k$'s. In particular, the priors are specified as,
\begin{align*}
& \pi (\mu_k) \propto 1, k=1,2, \quad \pi(\bbeta_k) \propto \boldsymbol{1}, k=1,2, \quad \pi (\xi_{1}) \propto 1/\xi_1, \, \xi_1 >0,\\
& \pi (\xi_{21}) \propto 1/\xi_{21}, \, \xi_{21} > 0,\quad \pi (\xi_{22}) \propto 1/\xi_{22}, \, \xi_{22} >0,\\
& \pi(\lambda) \propto 1, \, 0 \leq \lambda \leq 1, \quad \pi(\eta) \propto 1, \, \eta >0.
\end{align*}
For $\btheta_{\bw}$, we use noninformative priors for most parameters, and use an inverse Gamma (IG) prior for $\nu$, and a Beta prior for $\kappa/2$. Specifically,
\begin{align*}
&\pi(\sigma_{1}) \propto 1/\sigma_{1}, \, \sigma_{1} >0, \quad \pi (\sigma_{2}) \propto 1/\sigma_{2}, \, \sigma_{2} >0, \\
&\pi (\rho_{12}) \propto 1, \, -1 \leq \rho_{12} \leq 1, \quad \nu \sim \IG(a,b), \, \nu > 0, \quad \pi(\kappa/2) \sim \textrm{Beta}(a,b), \,0 < \kappa \leq 2.
 \end{align*}

We can choose the values of $a,b,c$, and $d$ for specific priors of $\nu$ and $\kappa$. In the simulation study and the GPU data application, we set $c=d=1$ so that $\kappa$ has a proper flat prior. It is common to use the IG distribution as a prior distribution for $\nu$, and we set $a=5$ and $b=1$ to penalize large values of $\nu$ to guarantee the positive definiteness of the $\Omega$ matrix.

\begin{figure}[ht]
    \begin{center}
    	\includegraphics[width = 0.5\textwidth]{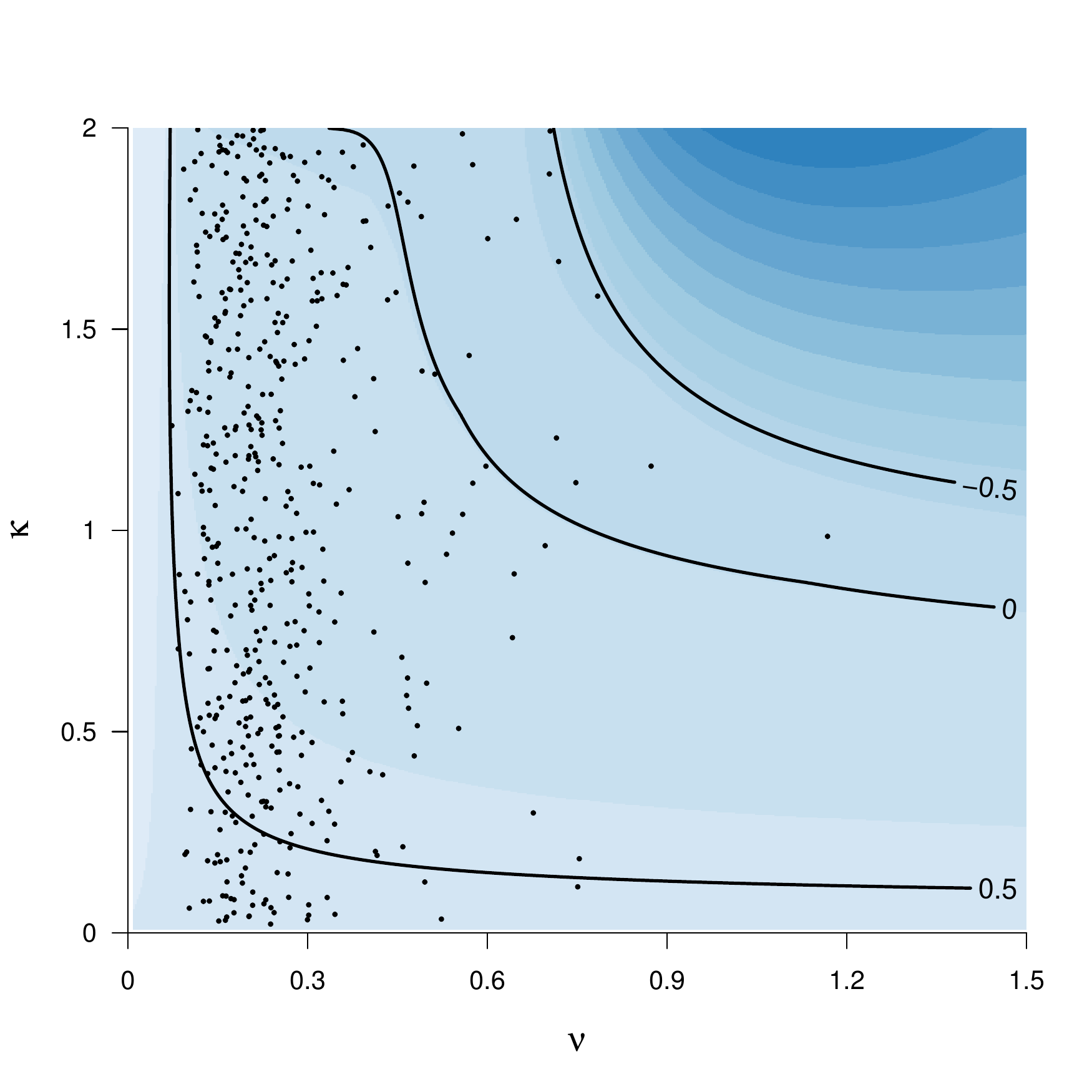}
    \end{center}
		\caption{Heatmap and contour lines for the smallest eigenvalue of $\Omega$ as a function of $\nu$ and $\kappa$. Darker colors represent smaller values. The black dots are 500 draws from the joint prior distribution of $\nu$ and $\kappa$.}\label{fig:eiegen}
    \end{figure}

 Essentially the correlation matrix $\Omega$ is a function of $\nu$ and $\kappa$, given the distance matrix $(d_{sl})$. Thus, we can compute the smallest eigenvalue of $\Omega$ given $\nu$ and $\kappa$. Figure~\ref{fig:eiegen} shows the heatmap and contour lines of the smallest eigenvalue for $\Omega$ as a function of $\nu$ and $\kappa$. The darker the color, the smaller the smallest eigenvalue. The figure indicates that when $\nu$ is large, it is possible that the correlation matrix is not positive definite. Therefore we give low prior density to large $\nu$ values. The probability that $\nu$ is less than or equal to $0.5$ is about $0.95$ under its prior distribution. The black dots represent $500$ draws of $\kappa$ and $\nu$ based on the joint prior distribution. Although the correlation function of the spatial random effects decreases rapidly for values of $\nu$ less than $0.5$ and values of $\kappa$ close to $2$, the power exponential correlation function is still flexible, because both $\nu$ and $\kappa$ are not fixed. That is, different combinations of $\nu$ and $\kappa$ values can achieve correlation functions that are similar to each other.

%%%%%%%%%%%%%%%%%%%%%%%%%%%%%%%%%%%%%%%%%%%%%%%%%%%%%%%%%%%%%%%%%%%%%%%%%%%%%%%%%%%%%%%%%%%%%%%%%%%%%%%%%%%%%%%%%%%%%
\subsection{Posterior and Model Inference}\label{sec:model.est.inf}		%%%%%%%%%%%%%%%%%%%%%%%%%%%%%%%%%%%%%%%%%%%%%%%%%%%%%%%%%%%%%%%%%%%%%%%%%%%%%%%%%%%%%%%%%%%%%%%%%%%%%%%%%%%%%%%%%%%%%	
Based on the likelihood and the prior specification, the joint posterior distribution is  $$f(\btheta|\bt,\bw) \propto  L(\btheta_{\bt}| \bt,\bw)f_{\bw}(\bw|\btheta_{\bw}) f(\btheta_{\bt})f(\btheta_{\bw}),$$
where $f(\btheta_{\bt})$ and $f(\btheta_{\bw})$ are the joint prior distribution for $\btheta_{\bt}$ and $\btheta_{\bw}$, respectively. Although we used improper priors for $\mu_k$, $\bbeta_k$, $\xi_k$, $\sigma_{1}$ and $\sigma_{2}$, the posterior is proper. That is,
\begin{align*}
	\int f(\btheta| \bt, \bw) d \btheta \propto \int L(\btheta_{\bt}|\bt,\bw)f(\btheta_{\bt}) d\btheta_{\bt} \int f(\bw | \btheta_{\bw})f(\btheta_{\bw}) d\btheta_{\bw} < \infty.
\end{align*}
For the $\int f(\bw| \btheta_{\bw})f(\btheta_{\bw}) d\btheta_{\bw}$ part, it is commonly known that the improper priors for $\sigma_{1}$ and $\sigma_{2}$ lead to proper posterior with more than two random effects. For the integral $\int L(\btheta_{\bt}| \bt,\bw)f(\btheta_{\bt}) d\btheta_{\bt}$, with a sufficient amount of data, the improper priors lead to a proper posterior for a Weibull or a lognormal time-to-event distribution. For the Weibull distribution, \shortciteN{ramos2020posterior} proved that, for a single distribution, the posterior is proper when there are at least two failures in the data. We extend the result for a regression model by using a different parametrization and using an improper noninformative prior for the regression coefficient $\bbeta_k$. The proof is in Appendix~\ref{apd0}. For the lognormal distribution, the posterior is proper when the number of failures for each failure mode is larger than $(p+3)/2$, and $p$ is the dimension of the covariates. The proof is similar to that the normal regression model with flat priors on covariates and variances lead to proper posteriors.

Because the likelihood $L(\btheta_{\bt}| \bt,\bw)$ for the two failure modes in \eqref{eqn:lik.cond.w} can be factored into likelihoods for the two failure modes, the posterior $L(\btheta_{\bt}| \bt,\bw)f(\btheta_{\bt})$, for the two failure modes can be separated, and the integral involved in showing the properness of the posterior can be done in parts. Thus, it is sufficient to prove that the posterior for one failure mode is proper. Also, because we use a proper prior on the mixture probability $\lambda$, we focus on the situation where there is only one failure mode in the failure-time distribution.

Based on the above discussion, for notational simplicity in the statement of the result and its proof, we drop the index $k$ and resort the data by one index $i$. That is,  we use $t_i$ to denote the event time, $\delta_i$ for the event-type indicator, and $w_{i}$ for the corresponding random effect, $\mu$ for the log-location parameter, $\xi$ for the log-scale parameter, and $\bbeta$ for the regression coefficients. Let $\bPsi = (\mu, \bbeta',\xi)^{'}$. Suppose that  $\sum_{i}\delta_{i} = m$ is the number of observed failures, and $p$ is the length of $\bbeta$. We summarize the result in the following proposition.

\begin{proposition}\label{prop1}
Under noninformative priors $\pi(\mu) \propto 1$, $\pi(\bbeta) \propto \boldsymbol{1}$, and $\pi(\xi) \propto 1/\xi$, the posterior distributions of log-location and log-scale parameters in the Weibull distribution is proper, which implies that,
	$$\int\prod_{i}^m f(t_{i}|w_{i}, \bPsi)f(\bPsi) d\bPsi <\infty, \textrm{ if } m > p + 1. $$
\end{proposition}
The proof of Proposition~\ref{prop1} is given Appendix~\ref{apd0}. Proposition~\ref{prop1} provides part of the theoretical basis for the inference using the posterior distribution.

In the spatially correlated failure times application, we jointly sample $f(\btheta, \bw|\bt)$ using Markov chain Monte Carlo (MCMC). We transform all parameters to have a range $(-\infty, \infty)$. We have seen that doing this improves the Stan NUTS sampler. Let $\widetilde{\btheta}$ be the transformed unrestricted parameters, then,
{\small
$$\widetilde{\btheta}= \left(\bbeta_1',\bbeta_2', \mu_1,\mu_2,\log(\xi_1),\log(\xi_{21}),
\log(\xi_{22}),\log(\eta), \textrm{logit}(\lambda),\textrm{logit}\left(\frac{ \rho_{12} + 1}{2}\right), \log(\nu), \textrm{logit}\left(\frac{\kappa}{2}\right)\right)'.$$
}
Also, we transform the random effect matrix to avoid linear dependency on $\mu_k$. Let $A = n^{-1}(I_n - J_n)$ and $A\bw_k$ is the transformed spatial random vector for $k = 1,2$. We take the first $n-1$ elements of $A\bw_k$ as the spatial random effect parameters for failure mode $k$, and only use these $n-1$ elements in MCMC sampling. To make the MCMC sampling more numerically stable, we also constrain the range of $\rho_{12}$ to be between $-0.95$ and $0.95$.

After we obtain the draws from posterior distribution, we use the posterior means to obtain point estimates. Also, $95\%$ equal tail credible intervals (CI) are computed from the 0.025 and 0.975 empirical quantiles of the marginal posterior draws for quantities of interest.

	%%%%%%%%%%%%%%%%%%%%%%%%%%%%%%%%%%%%%%%%%%%%%%%%%%%%%%%%%%%%%%%%%%%%%%%%%%%%%%%%%%%%%%%%%%%%%%%%%%%%%%%%%%%%%%%%%%%%%
\section{A Simulation Study}\label{sec:simulation.study}	%%%%%%%%%%%%%%%%%%%%%%%%%%%%%%%%%%%%%%%%%%%%%%%%%%%%%%%%%%%%%%%%%%%%%%%%%%%%%%%%%%%%%%%%%%%%%%%%%%%%%%%%%%%%%%%%%%%%%
In this section, we use simulation to study model estimation performance. We evaluate various metrics such as the root relative mean squared error (RRMSE), relative bias, estimated standard deviation, coverage probability of $95\%$ CI, and mean length of $95\%$ CI for all parameter estimators for $\btheta$.

%%%%%%%%%%%%%%%%%%%%%%%%%%%%%%%%%%%%%%%%%%%%%%%%%%%%%%%%%%%%%%%%%%%%%%%%%%%%%%%%%%%%%%%%%%%%%%%%%%%%%%%%%%%%%%%%%%%%%
\subsection{Simulation Setting} %%%%%%%%%%%%%%%%%%%%%%%%%%%%%%%%%%%%%%%%%%%%%%%%%%%%%%%%%%%%%%%%%%%%%%%%%%%%%%%%%%%%%%%%%%%%%%%%%%%%%%%%%%%%%%%%%%%%%
We study the influence of two factors: the number of units $N$, and the number of spatial locations $d\times d$, on model estimation. We carefully choose value of true parameters to mimic the real GPU dataset. Because the model used for the GPU data is quite sophisticated, to simplify the setting and reduce the computing time, we use one inside cabinet position covariate. In particular, we use the cage factor with three different levels; thus the column rank of the design matrix is $p=2$, and $p$ is the number of covariates. To simplify the setting, we use the same number of rows and columns (i.e., both of number of columns and rows are $d$) for the location setting. We chose $N$ to be 5{,}000, 7{,}000, and 10{,}000, and we chose $d$ to be 5, 7, and 10. In practice, we learned that when the number of failed GPUs is small (i.e., zero failures or one failure) on multiple locations, the posterior samples of the random effects variances stay at the same value in some MCMC chains. Therefore, in simulating the data, we require the proportion of locations that have at least 1 failure is greater than $5\%$, and the proportion of locations have at least 2 failures is greater than $10\%$. We generate the simulation dataset 300 times under each factor-level combination. The value of true parameters are shown in Table~\ref{tab:sim.true}.

\begin{figure}
\begin{center}
	\begin{tabular}{cc}
 	    \includegraphics[width = 0.35\textwidth]{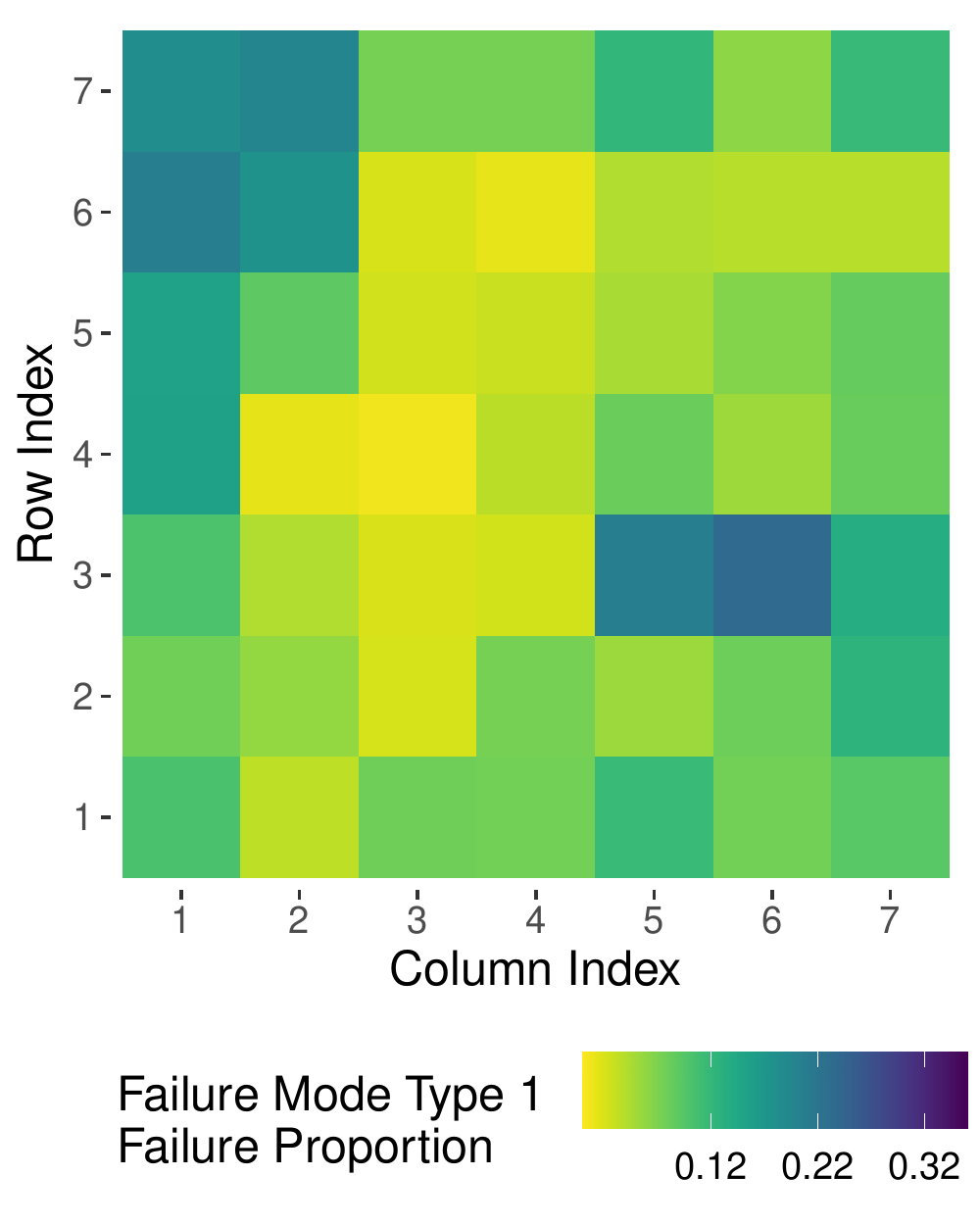} &
		\includegraphics[width = 0.35\textwidth]{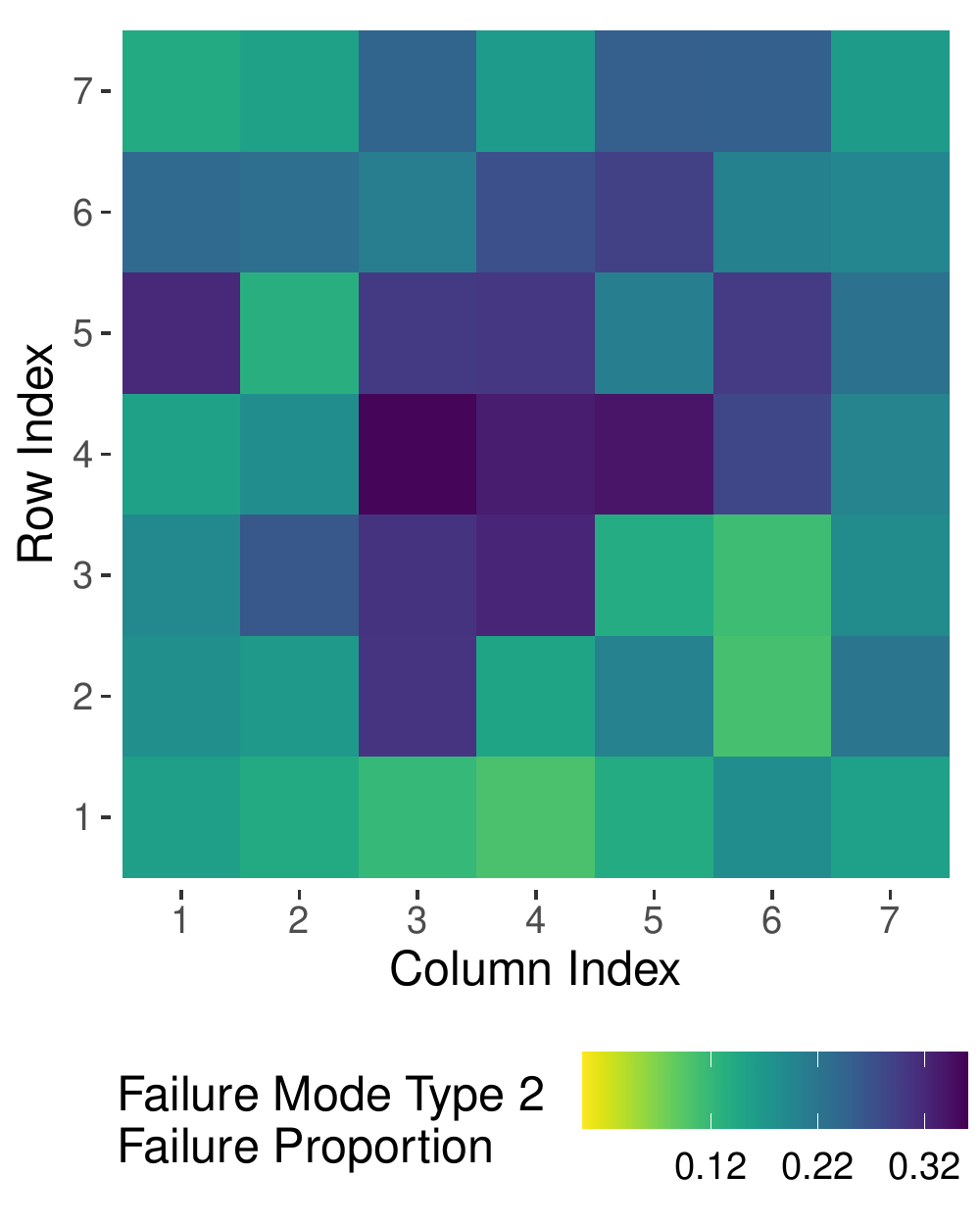}\\
		(a) Failure Mode Type 1 Failure Proportion  & (b) Failure Mode Type 2 Failure Proportion\\
	\end{tabular}
	\caption{Visualization of failure proportions for simulation on $7 \times 7$ locations for failure mode type 1 and failure mode type 2. }\label{fig:sim.data}
\end{center}
\end{figure}

Under each factor-level combination, for each of the $N$ units, we first sample from Binom($3$, $0.5$), and then remove the last column of the $N \times 3$ matrix to obtain the model matrix. For the cabinet location index, we uniformly sample row and column index from $1$ to $d$. We then sample failure times for failure mode type 1 and failure mode type 2 of each unit from two independent Weibull distributions based on the covariates and random effects. Using a simplified setting, we generate beginning date for each GPU from a uniform distribution with the lower bound as 01 January 2012, and the upper bound as 01 January 2018. We set the end date as 01 January 2019. Units that have not failed before 01 January 2019 is categorized as censored. Figure~\ref{fig:sim.data} shows the failure proportions of failure mode types $1$ and $2$ from one dataset generated using $7 \times 7 $ spatial random effects and 7{,}000 units.

\begin{table}
\begin{center}
\caption{The true parameter values used in generating simulated data in all situations.}\label{tab:sim.true}
\begin{tabular}{c|cccccccc}
		\hline\hline
		& $\mu_k$ & $\beta_{k1}$ & $\beta_{k2}$ & $\xi_k$        & $\sigma_k^2$ & $\rho$             & $\nu$                 & $\kappa$              \\\hline
		Failure Mode Type 1 $(k=1)$ & 1.70  & 0.67        & 0.27        & 0.19 & $0.02$     & \multirow{2}{*}{0.00} & \multirow{2}{*}{0.25} & \multirow{2}{*}{1.52} \\ \cline{1-6}
		Failure Mode Type 2 $(k=2)$ & 1.55  & 0.57        & 0.23        & 0.14 & $0.01$     &                    &                       &                       \\ \hline\hline
\end{tabular}
\end{center}
\end{table}

For each factor-level combination and for each simulation trial, we run $3$ chains with 2{,}000 draws after warmup per chain, providing 6{,}000 draws in total. We use random starting values picked by Stan. We check the convergence using Rhat introduced in \shortciteN{vehtari2021rank} and use cutoff $1.1$ to decide whether a chain converged or not. We noticed that there are some MCMC chains in which the samples of random effect variances were not moving. However, the proportion of such chains is relatively small (i.e., less than 1\%) for each factor-level combination. For those chains, we use the same data and run the MCMC again with different starting values picked by Stan. If the chains still do not converge, we exclude the dataset and generate a new dataset. Among all 27{,}000 simulation runs, only four such datasets were excluded.

%%%%%%%%%%%%%%%%%%%%%%%%%%%%%%%%%%%%%%%%%%%%%%%%%%%%%%%%%%%%%%%%%%%%%%%%%%%%%%%%%%%%%%%%%%%%%%%%%%%%%%%%%%%%%%%%%%%%%
\subsection{Results}
%%%%%%%%%%%%%%%%%%%%%%%%%%%%%%%%%%%%%%%%%%%%%%%%%%%%%%%%%%%%%%%%%%%%%%%%%%%%%%%%%%%%%%%%%%%%%%%%%%%%%%%%%%%%%%%%%%%%%

\begin{figure}
\begin{center}
	\includegraphics[width = 0.95\textwidth]{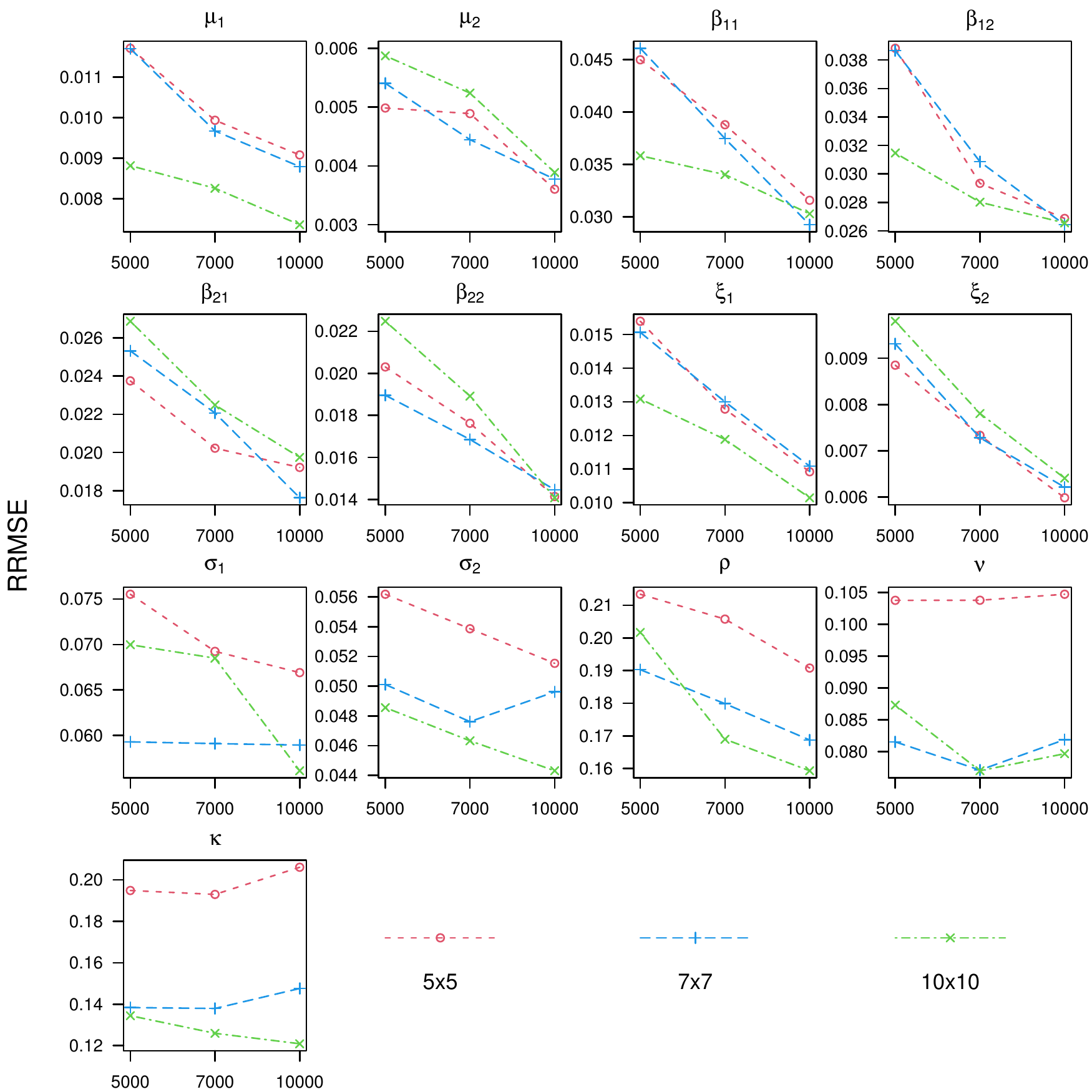}
	\caption{Plots of RRMSE for different number of units and spatial location combinations. The $x$-axis is the number of units $N$.}\label{MSE}
\end{center}
\end{figure}

Figure~\ref{MSE} shows the RRMSE for parameter estimators for $\btheta$. More results about the relative bias, estimated standard deviation, coverage probability, and mean CI length are in Appendix~\ref{apd2}. The RRMSEs are small for all the situations. For most parameters, the RRMSE decreases as the number of units increases. There is no evident trend in relative bias or standard deviation as $N$ in contrast to the observed trend of RRMSE. In addition, when the number of locations increases from $5 \times 5$ to $ 7 \times 7$, the relative bias of both $\nu$ and $\kappa$ decreases for all three levels of $N$. The coverage probability of CI is close to the nominal $0.95$ level for all the cases. The mean length of CIs decreases as the number of units increases for all parameters. When the number of locations increases, the mean length of CIs for $\nu$ and $\kappa$ decreases.

%%%%%%%%%%%%%%%%%%%%%%%%%%%%%%%%%%%%%%%%%%%%%%%%%%%%%%%%%%%%%%%%%%%%%%%%%%%%%%%%%%%%%%%%%%%%%%%%%%%%%%%%%%%%%%%%%%%%%
\section{Analysis of Titan GPU Data}\label{sec:data.analysis}
%%%%%%%%%%%%%%%%%%%%%%%%%%%%%%%%%%%%%%%%%%%%%%%%%%%%%%%%%%%%%%%%%%%%%%%%%%%%%%%%%%%%%%%%%%%%%%%%%%%%%%%%%%%%%%%%%%%%%

%%%%%%%%%%%%%%%%%%%%%%%%%%%%%%%%%%%%%%%%%%%%%%%%%%%%%%%%%%%%%%%%%%%%%%%%%%%%%%%%%%%%%%%%%%%%%%%%%%%%%%%%%%%%%%%%%%%
\subsection{Model Fitting and Comparisons on Different Time-to-Event Distributions and Spatial Correlation Functions}
%%%%%%%%%%%%%%%%%%%%%%%%%%%%%%%%%%%%%%%%%%%%%%%%%%%%%%%%%%%%%%%%%%%%%%%%%%%%%%%%%%%%%%%%%%%%%%%%%%%%%%%%%%%%%%%%%%%%%
In this section, we fit the models proposed in Section~\ref{sec:sm} to the GPU data. We ran $4$ chains, and we generated 2{,}000 draws from each chain after an initial 6{,}000 draws. Because the regular residuals are not defined for a mixture distribution time-to-event model, we define our residuals as $r_{ijk}=  -\log(1- \widehat{F}(t_{ij}|w_{ik})),$ where $\widehat{F}(t_{ij}|w_{ik})$ is obtained by substituting the estimates of parameters $\btheta$. Because the residuals should have WEIB$(0,1)$ distribution (if the fitted model is correctly specified), we assess their distributions by using Weibull probability plots.

\begin{figure}
\begin{center}
	\begin{tabular}{cc}
  		\includegraphics[width = 0.48\textwidth]{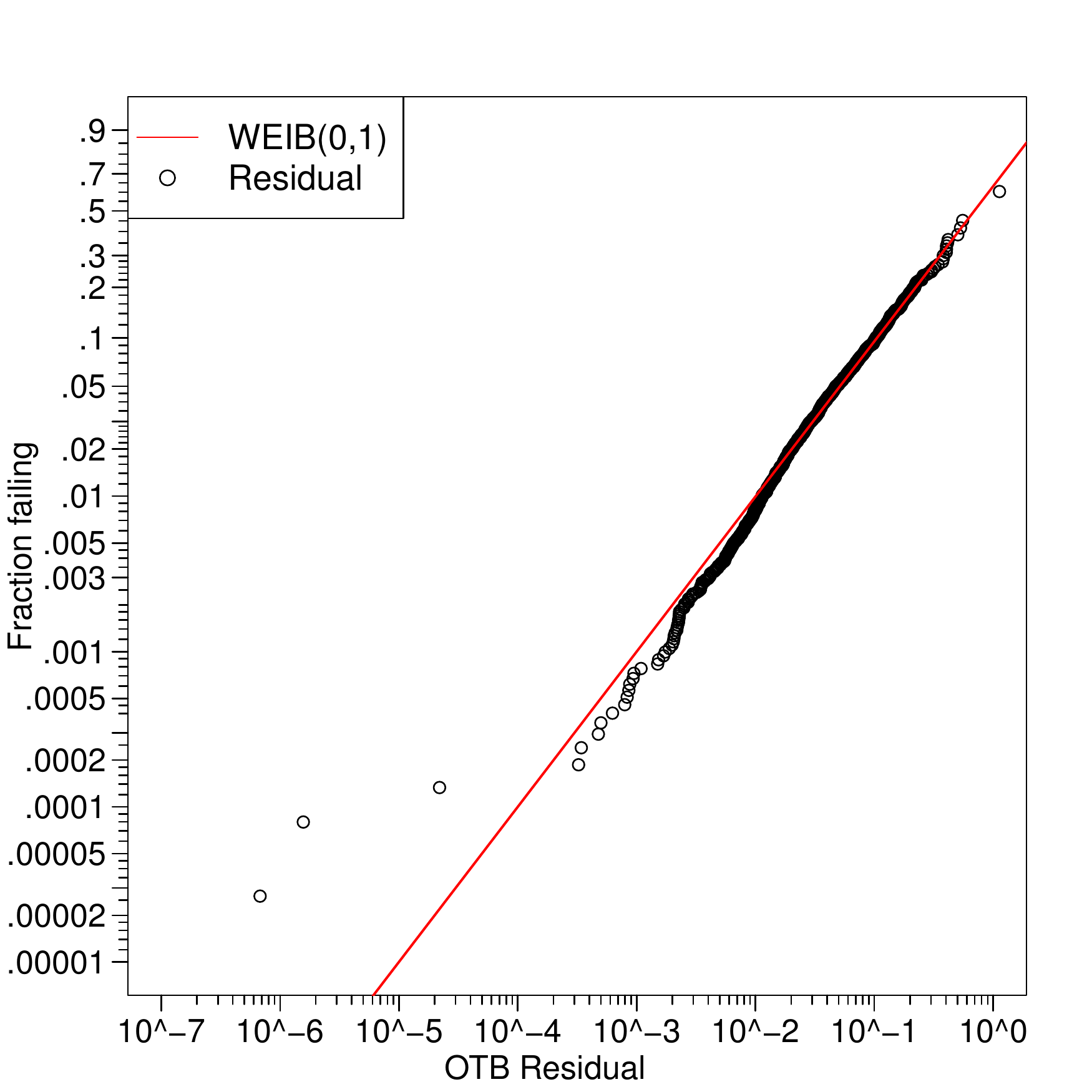} &
		\includegraphics[width = 0.48\textwidth]{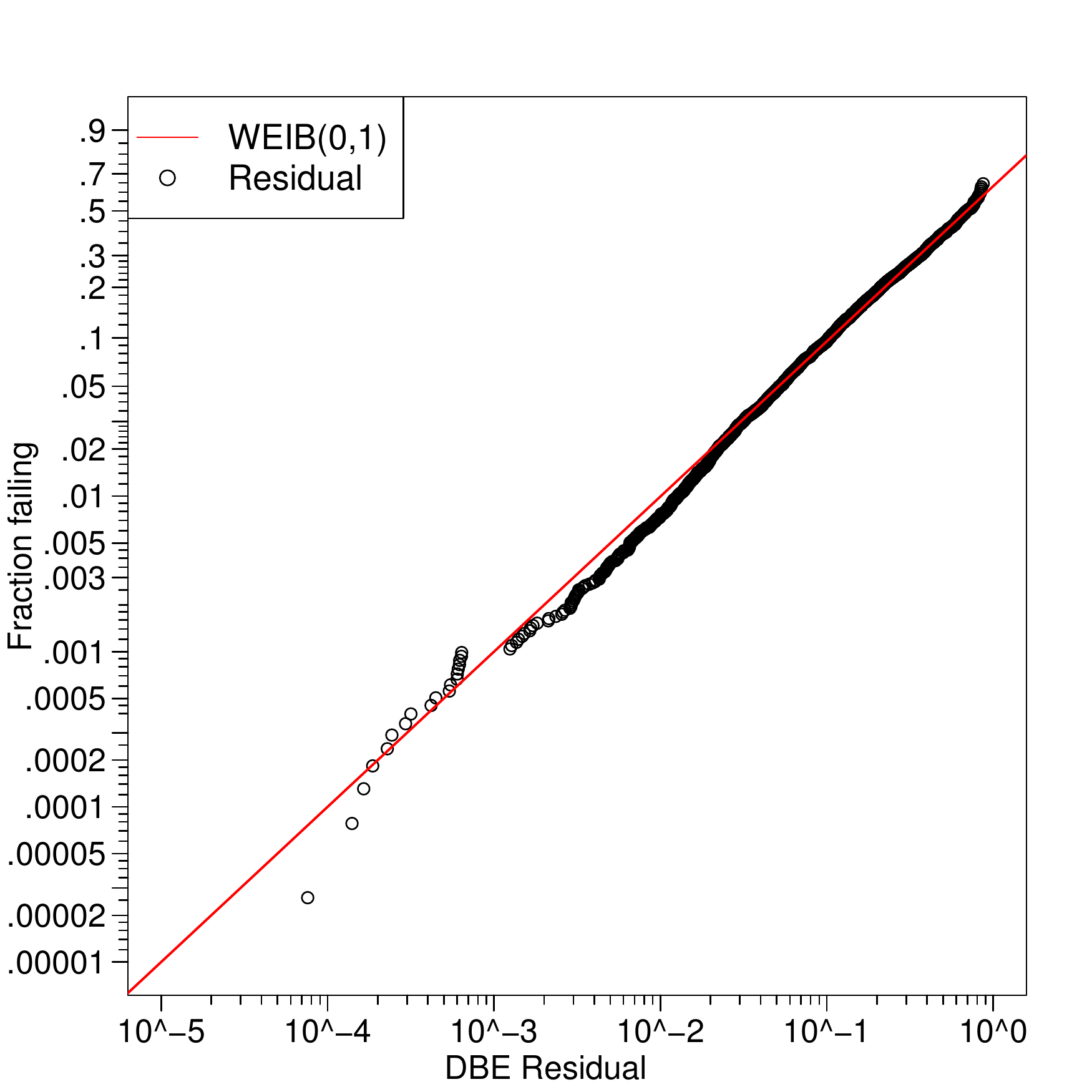}\\[-1.5ex]
	\end{tabular}
	\caption{Weibull probability plots for the OTB  and DBE residuals.}\label{fig:redisual}
\end{center}
\end{figure}

We first fit the model with a Weibull time-to-event distribution and a power exponential spatial correlation function. Figure~\ref{fig:redisual} shows the result for the estimated residuals of the model. The red lines are \textrm{WEIB}$(0,1)$ distributions and the black dots are estimated points. Except for the three data points in Figure~\ref{fig:redisual}(a), the residuals agree well with the $\textrm{WEIB}(0,1)$ reference line, suggesting that the model is fitting reasonably well.

\begin{figure}
	\begin{center}
		\begin{tabular}{ccc}
		    \includegraphics[width=0.176\textwidth]{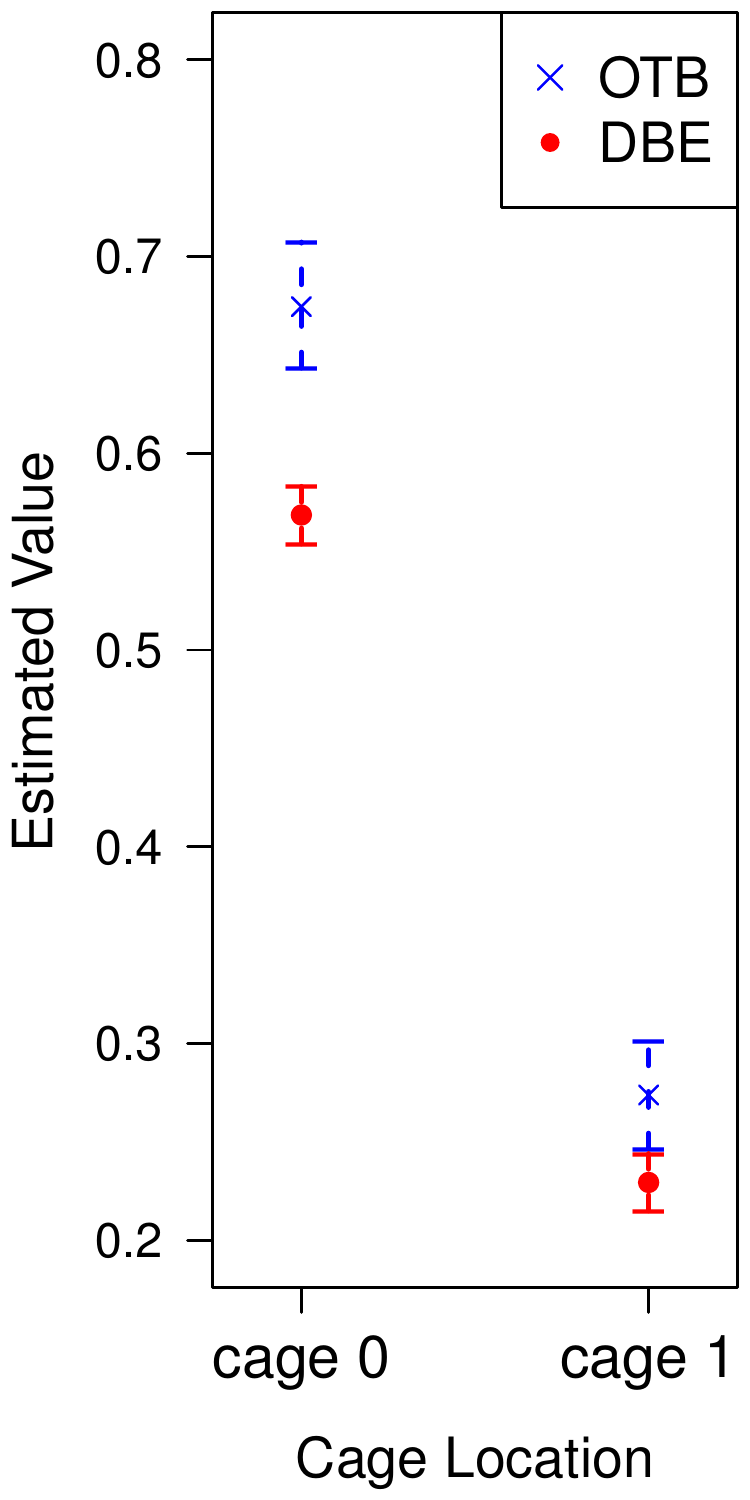} &
			\includegraphics[width=0.50\textwidth]{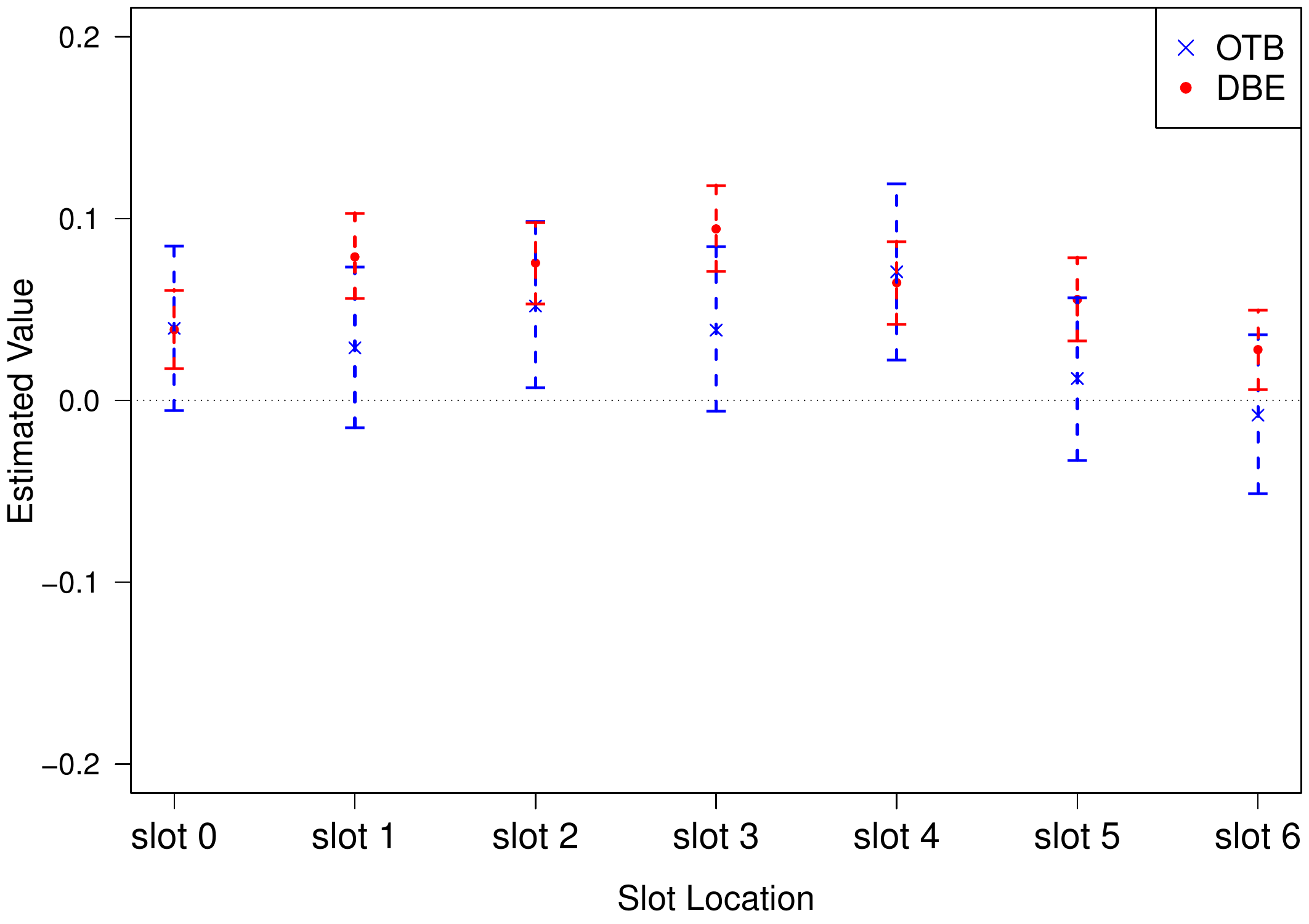} &
			\includegraphics[width=0.235\textwidth]{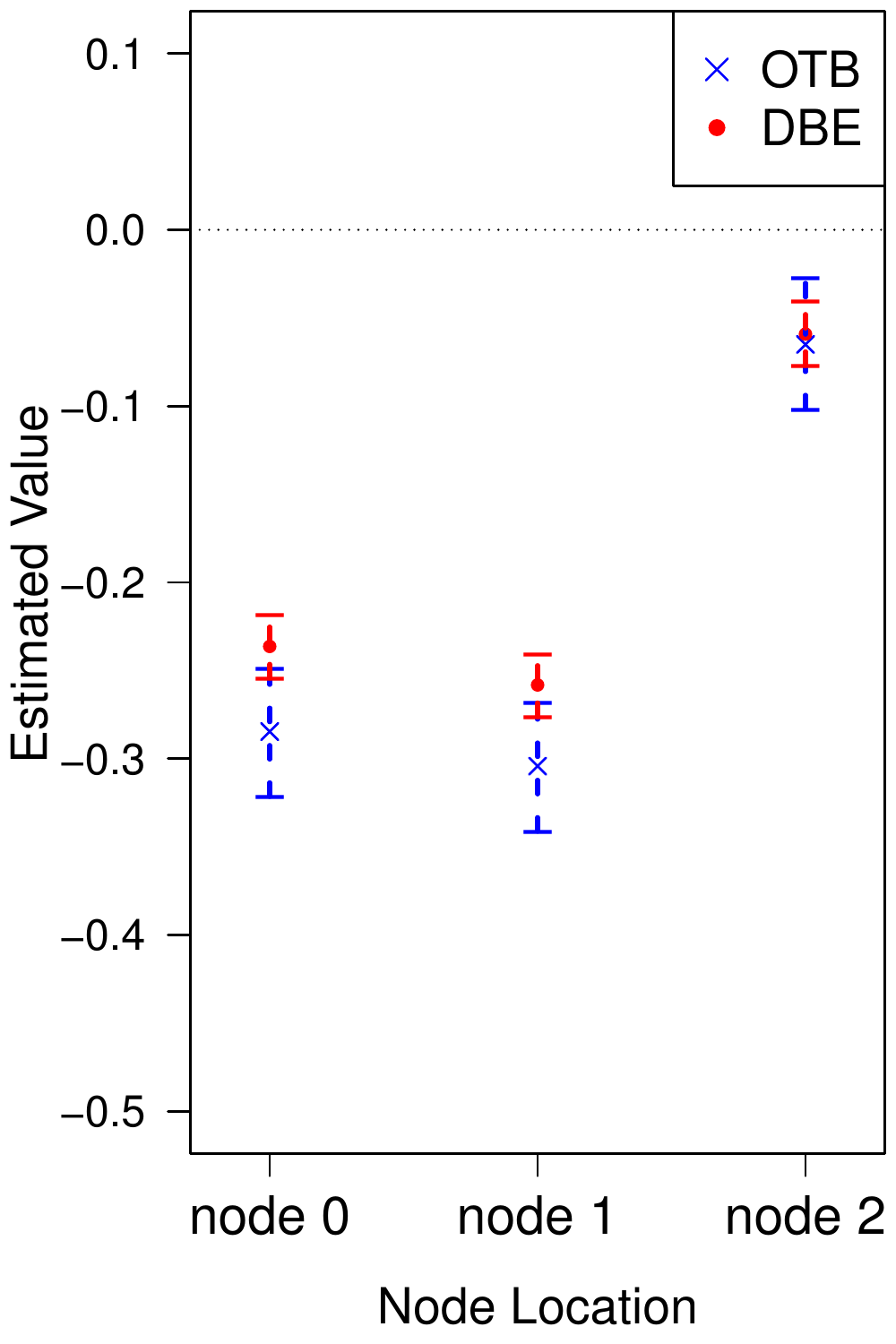}\\[-2ex]
		\end{tabular}
		\caption{Estimated posterior means and $95\%$ CIs of covariate coefficients corresponding to cage, slot, and node positions.} \label{fig:covariate}
	\end{center}
\end{figure}

\begin{figure}
\begin{center}
	\begin{tabular}{c}
		\includegraphics[width = 0.8\textwidth]{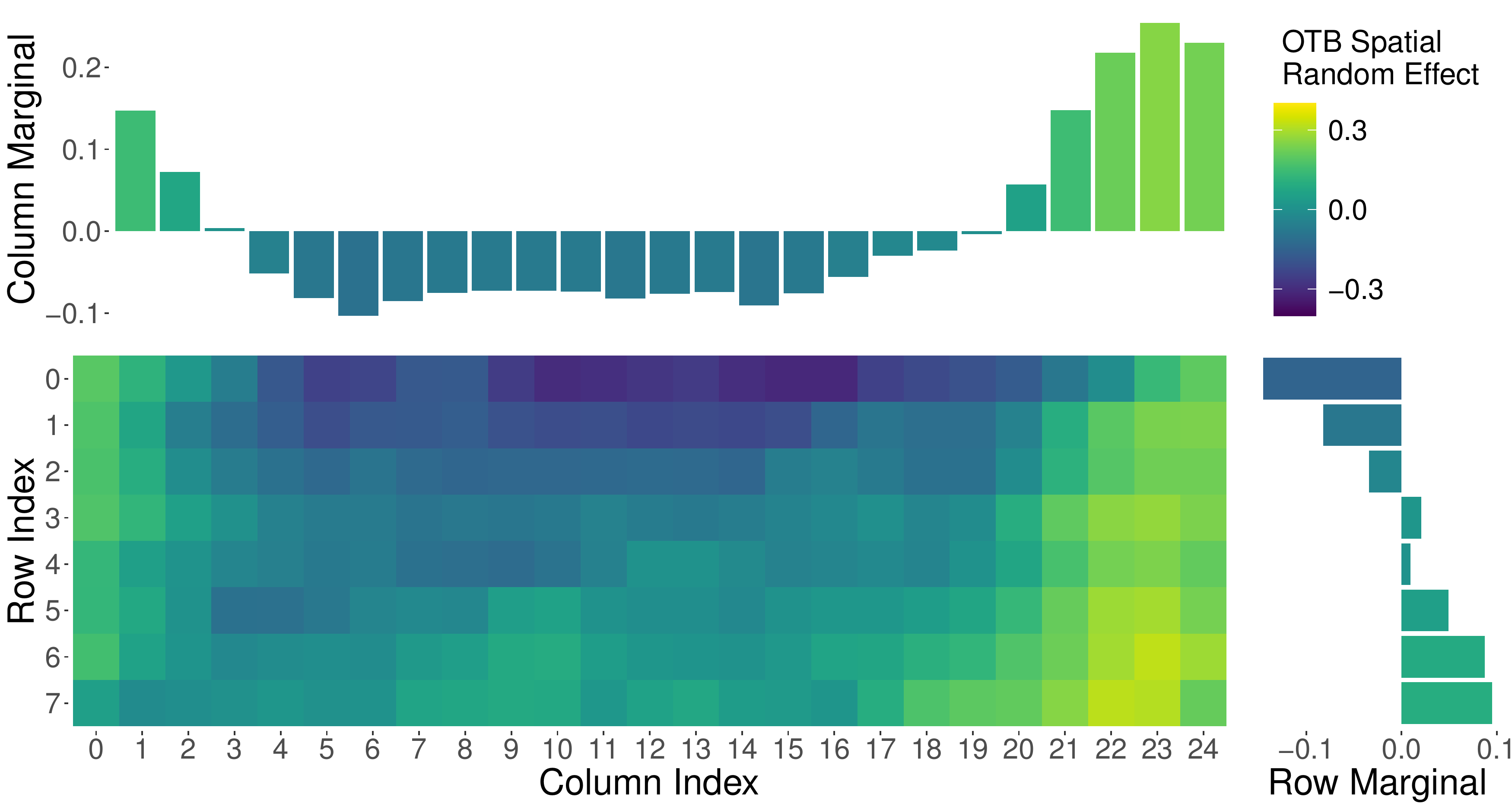}\\[-2ex]
	 \\
\includegraphics[width = 0.8\textwidth]{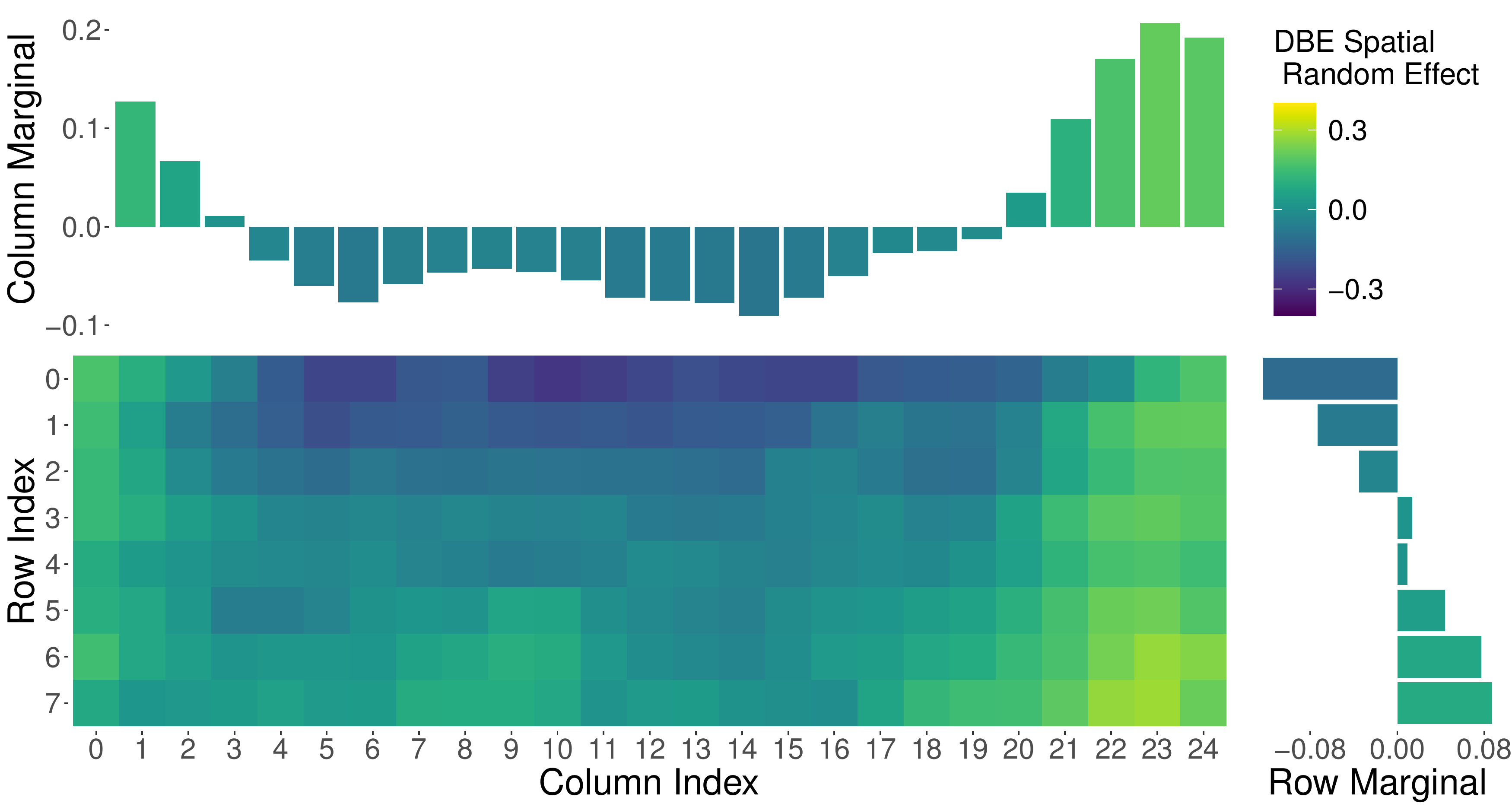}\\[-2ex]
		  \\
	\end{tabular}
	\caption{Heatmap for posterior means of estimated spatially correlated random effects $\bw_1$ and $\bw_2$ for the OTB and DBE failures. The histograms on the top and right of each heatmap show the column marginal and row marginal posterior means of estimated random effects.}\label{fig:randomeff}
\end{center}
\end{figure}

\begin{figure}
\begin{center}
	\includegraphics[width = 0.5\textwidth]{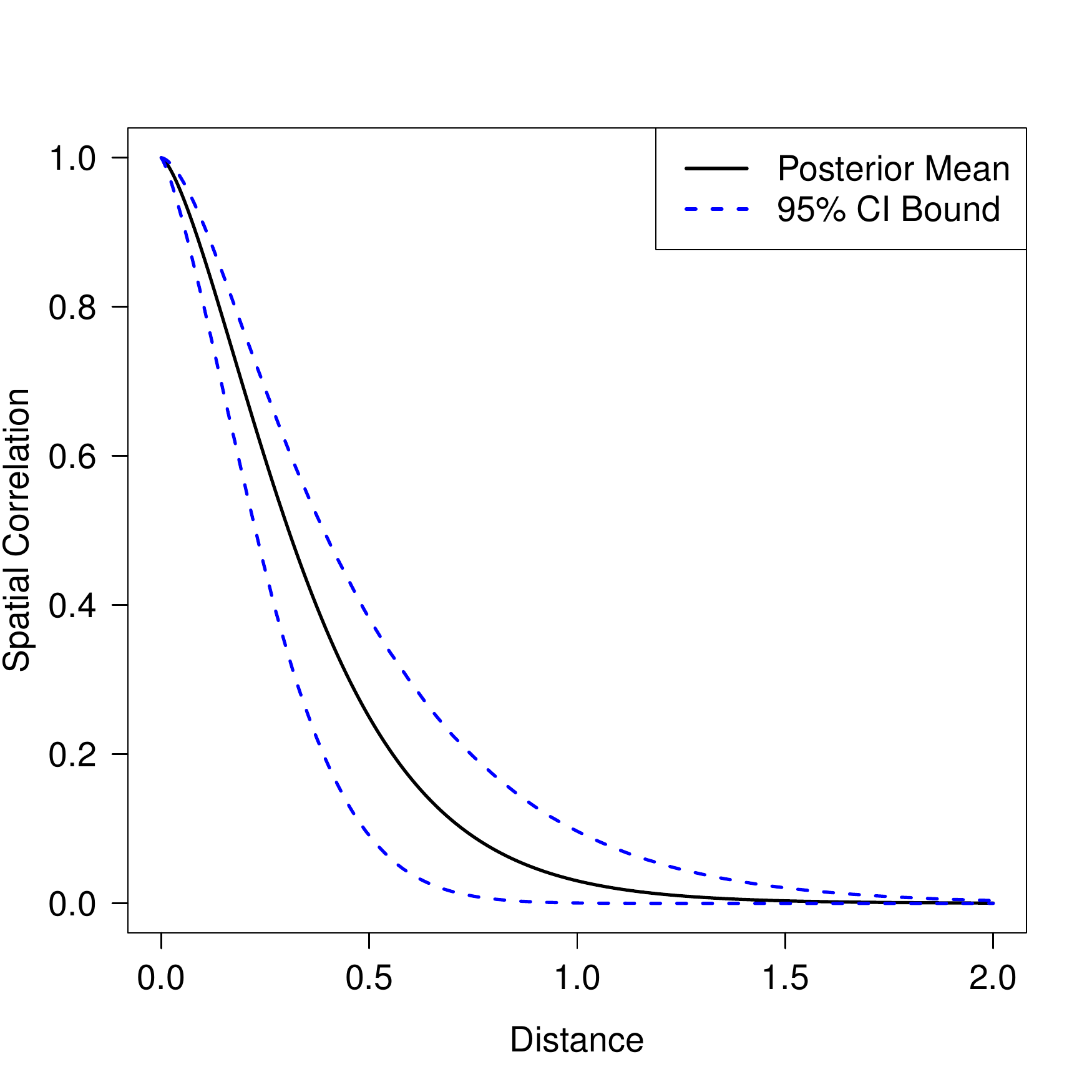}
	\caption{The  95\% CI of estimated spatial correlation function.}\label{fig:spatialcorr}
\end{center}
\end{figure}

\begin{table}
	\begin{center}
		\caption{The estimated posterior means and $95\%$ CIs for the time-to-event distribution scale parameters, mixture proportion, the difference between two DBE distribution modes, and spatial correlation parameters.}
		\begin{tabular}{c|cccc|c|ccc}
			\hline\hline
			\multirow{2}{*}{Parameter}   & \multirow{2}{*}{Estimate}  &  \multicolumn{2}{c}{CI} & & \multirow{2}{*}{Parameter}& \multirow{2}{*}{Estimate}  &  \multicolumn{2}{c}{CI} \\\cline{3-4} \cline{8-9}
			&  & Lower  & Upper  && & & Lower  & Upper  \\ \hline
			$\xi_{1}$ & 0.20 & 0.19 & 0.21   &&  $\sigma_{1}$ & 0.13 & 0.10 & 0.17 \\ \hline
			$\xi_{21}$ & 0.14 & 0.14 & 0.15 &&  $\sigma_{2}$ &0.11 & 0.09 & 0.15 \\\hline
			$\xi_{22}$ & 1.71 & 0.84 & 3.58 &&  $\rho_{12}$ & 0.92 & 0.83 & 0.95 \\ \hline
			$\eta$ & 10.56 & 4.55 & 24.06    &&  $\nu$ & 0.40 & 0.29 & 0.52 \\ \hline
			$\lambda$ & 0.60 & 0.55 & 0.65       &&  $\kappa$ & 1.46 & 1.21 & 1.71\\\hline\hline
		\end{tabular}\label{tab:estimat}
	\end{center}
\end{table}

As a quick overview of estimation results, Figure~\ref{fig:covariate} shows the estimates and $95\%$ CIs for the inside cabin locations. Table~\ref{tab:estimat} shows the estimate and $95\%$ credible interval for all of the other parameters.  Figure~\ref{fig:randomeff} shows the estimated random effects at different cabin locations.

Because the Weibull and lognormal are two commonly used parametric distributions for modeling time-to-event data, we also compare the model performance for these two distributions. In addition, we consider five correlation models: the power exponential (PEXP), the exponential (EXP), and the Gaussian (GAU) correlation functions, the AFT models with no spatial correlated random effects (AFT INDEP), and the AFT models with no random effects. All the models we consider are listed in Table~\ref{tab:dictab}.

We compare all the models using the leave-one-out cross-validation information criterion (LOOIC) in \shortciteN{vehtari2017practical}. Table~\ref{tab:dictab} shows the LOOIC for all the models. The comparison indicates that the Weibull PEXP model is the best one in terms of LOOIC. The comparison also shows that adding spatial random effects to the model is important. The estimates and CIs of the parameters for those models are given in Appendix~\ref{apd5}.

\begin{table}
\begin{center}
	\caption{LOOIC values for all models under comparison (bold font indicates the smallest). }\label{tab:dictab}
	\begin{tabular}{c|ccccc}
		\hline\hline
		& PEXP    & EXP     & GAU                 & AFT INDEP & AFT     \\ \hline
		Weibull   & $\textbf{26272.8}$ & 26280.7 & 26279.4   & 26390.7 &28138.2 \\ \hline
		Lognormal & 26289.1 & 26294.8 & 26307.5 & 26400.9  &28072.7 \\\hline\hline
\end{tabular}
\end{center}
\end{table}

%%%%%%%%%%%%%%%%%%%%%%%%%%%%%%%%%%%%%%%%%%%%%%%%%%%%%%%%%%%%%%%%%%%%%%%%%%%%%%%%%%%%%%%%%%%%%%%%%%%%%%%%%%%%%%%%%%%%%
\subsection{Alternative Methods Comparison}
%%%%%%%%%%%%%%%%%%%%%%%%%%%%%%%%%%%%%%%%%%%%%%%%%%%%%%%%%%%%%%%%%%%%%%%%%%%%%%%%%%%%%%%%%%%%%%%%%%%%%%%%%%%%%%%%%%%%%

There are other estimation methods for fitting a time-to-event model with random effects. When using a Bayesian model, the integrated nested Laplace approximation (INLA) \shortcite{rue2009approximate} can be used for approximate Bayesian inference. For frequentist inference, the EM algorithm can be used to obtain maximum likelihood estimators. However, INLA can not be directly implemented in our case because of the mixture distribution for the DBE failure mode. Therefore, we compare our Bayesian results (i.e., the Weibull PEXP model with the best LOOIC) with EM algorithm results. Details about the formulas and implementation of EM algorithm are in Appendix~\ref{apd3}. The EM estimates are also available in Appendix~\ref{apd3}. All EM estimates for parameters are within the CIs from the Bayesian method, indicating good agreement from different methods. However, we do notice that the EM algorithm is much slower due to the use of MCMC in the E step for each iteration.

%%%%%%%%%%%%%%%%%%%%%%%%%%%%%%%%%%%%%%%%%%%%%%%%%%%%%%%%%%%%%%%%%%%%%%%%%%%%%%%%%%%%%%%%%%%%%%%%%%%%%%%%%%%%%%%%%%%%%
\subsection{Results Interpretation}
%%%%%%%%%%%%%%%%%%%%%%%%%%%%%%%%%%%%%%%%%%%%%%%%%%%%%%%%%%%%%%%%%%%%%%%%%%%%%%%%%%%%%%%%%%%%%%%%%%%%%%%%%%%%%%%%%%%%%
In this section, we show the estimation results for the Weibull PEXP model.	
Figure~\ref{fig:randomeff} shows the estimated random effects at different cabinet locations, suggesting that the mode of failure times varies among cabinet locations. A larger random effect indicates that the GPUs at those locations tend to have longer lifetimes. From Figure~\ref{fig:randomeff}, we can see that the column marginals transition from positive values, to negative values and then to positive values, by using the labels from the logical connection. We also can see that the row marginals transition from negative values to positive values (shorter lifetimes to longer lifetimes) from row 0 to row 7, which may be caused by physical cooling heterogeneity from row 0 to row 7. As indicated in Figure~\ref{fig:tts}(a), heat dissipation near a wall (near row 0) may not be as good as in an open space (near row 7). Figure~\ref{fig:covariate} shows the estimates and $95\%$ CIs for the inside cabinet positions. Table~\ref{tab:estimat} shows the estimate and $95\%$ CI for all other parameters. The baseline levels for categorical variables cage, slot, and node are cage $2$, slot $7$, and node $3$. The estimated coefficients in Figure~\ref{fig:covariate} shows the difference on the log scale of the mode of the failure times between current cage, slot, node levels and the baseline levels, holding other parameters fixed. The positive estimates in Figure~\ref{fig:covariate}(a) indicate that the failure time for GPUs inside cage $0$ and cage $1$ tend to be longer than those for the GPUs inside cage $2$. The difference between cage $0$ and cage $1$ is also significant. The cage effect is likely due to differences in the temperature in the supercomputer. The temperature around cages at the bottom floors of the cabinets is lower than the cages at higher floors of the cabinets, because of the airflow inside the supercomputer. In addition, the influence of the positions of cages within cabinets is stronger for OTB failures than for DBE failures, which shows that the OTB failure mechanism is more sensitive to temperature.

	On the other hand, the negative estimate in Figure~\ref{fig:covariate}(c) shows that GPUs on node $3$ have the longest failure times when compared with GPUs on nodes 0--2. The difference between nodes 0--1 and nodes 2--3 may also be caused by temperature differences similar to the differences among different positions of cage. This is because nodes 0--1 are at a lower vertical level than nodes 2--3, as shown in Figure~\ref{fig:tts}(c). Slot position influence is relatively weak compared with cage and node position influence, and the influence of slot on the OTB and DBE failure-time distributions is similar. An explanation for this is that the cooling condition is relatively uniform for slots across the same cage (see Figure \ref{fig:tts}(b) which depicts the within cabinet layout).

	The estimates of parameters related to the spatial random effects suggest that the row-column cabinet locations also have an effect on the failure-time distributions. Table~\ref{tab:estimat} shows that the posterior means of $\sigma_1$ and $\sigma_2$ are $0.13$ and $0.11$, suggesting that the variation of random effects exists. The estimate of $\rho_{12}$ and its corresponding CI in Table~\ref{tab:estimat} provide evidence of strong positive correlations between OTB spatial random effects and DBE spatial random effects. Figure~\ref{fig:spatialcorr} shows the curve of the spatial correlation function. The black line is the posterior mean, and the two blue lines are lower and upper endpoints of the CI of the correlation function. The correlation decays rapidly, and the value at the largest distance ($\sqrt{2}$) in the dataset is around $0.002$.
	
\begin{figure}
		\begin{center}
			\begin{tabular}{cc}
				\includegraphics[width=0.45\textwidth]{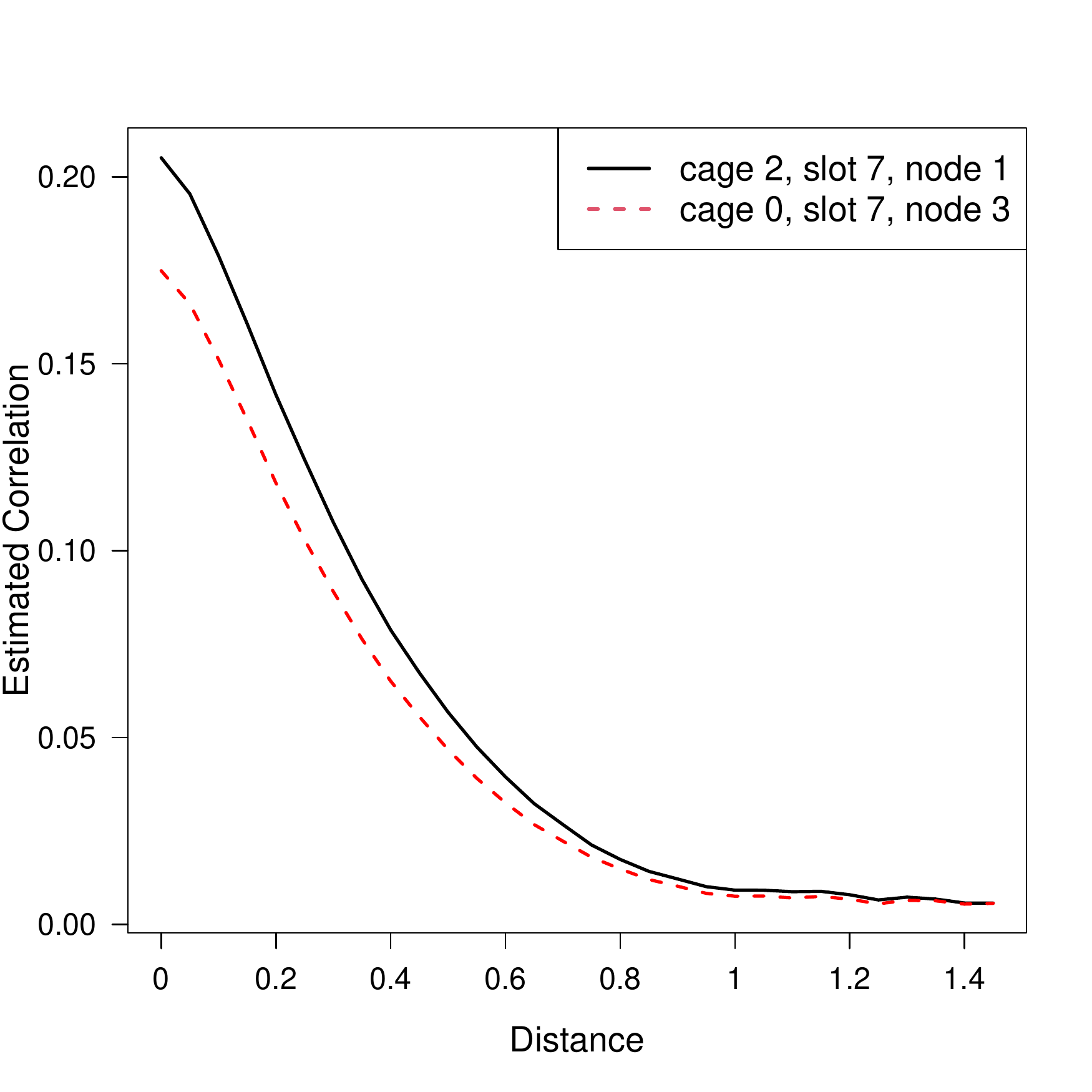} &
				\includegraphics[width=0.45\textwidth]{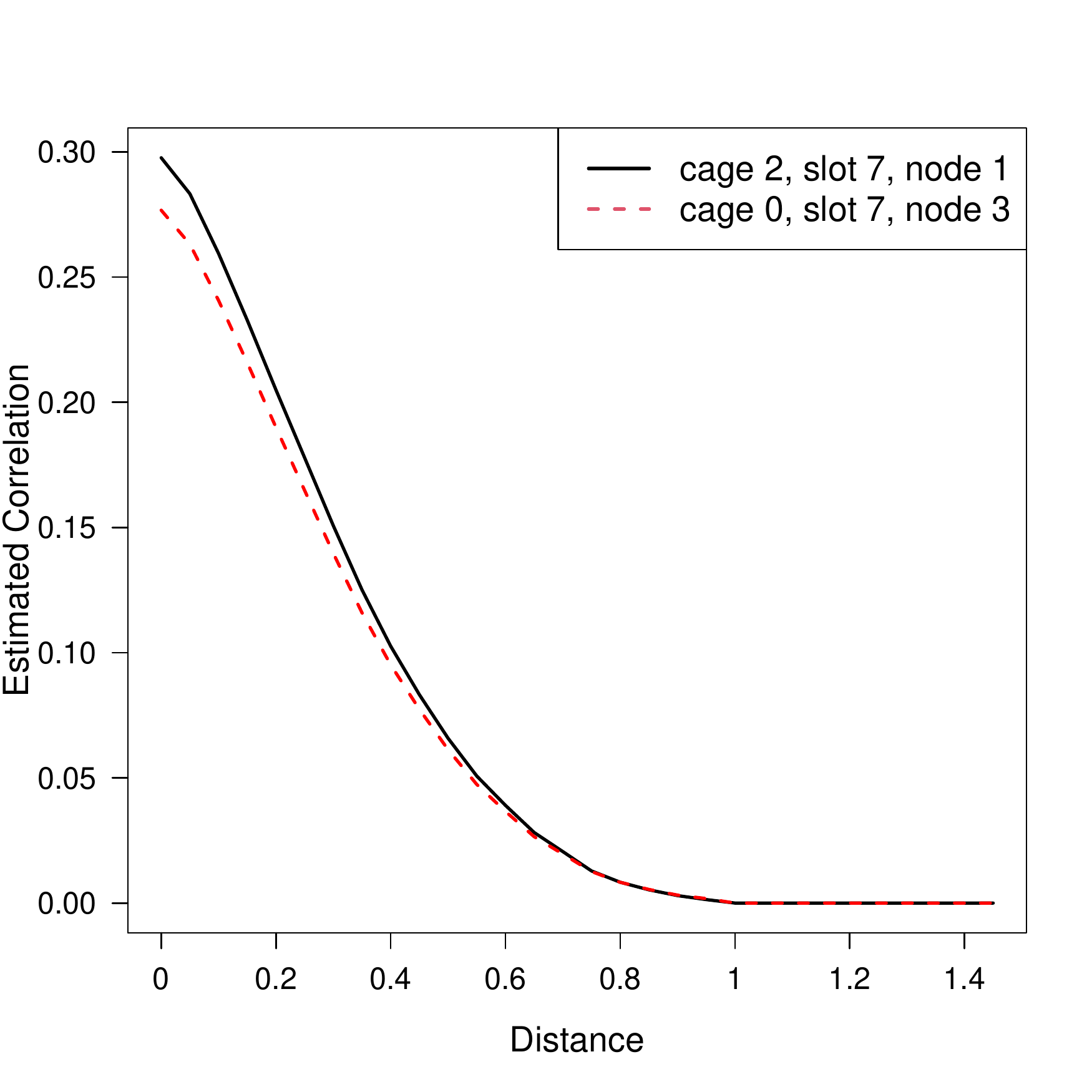}\\
				(a) OTB Failure   & (b)  DBE Failure 	
			\end{tabular}
			\caption{Marginal correlation of $T_{ijk}$ and $T_{i^{\ast}jk}$ at the inside cabinet position cage 2, slot 7, and node 3 for OTB and DBE failures.} \label{fig:spat}
		\end{center}
	\end{figure}

We can also estimate the marginal posterior correlation between the OTB failure time and the DBE failure times at different cabinet locations and inside cabinet positions by integrating out the random effects. The formulas and calculation steps are given in Appendix~\ref{apd6}. The correlation between $T_{ij1}$ and $T_{ij2}$ depends on the value of the covariates $\bx_j$. At inside cabinet position cage 2, slot 7, and node 1, the estimated correlation between $T_{ij1}$ and $T_{ij2}$ is 0.24. At inside cabinet position cage 0, slot 7 and node 3, the estimated correlation is 0.21. Also, for a fixed index $j$ (i.e., fixed inside cabinet position), we compute the marginal correlation between $T_{ijk}$ and $T_{i^{\ast}jk}$, $i \neq i^{\ast}$. Figure~\ref{fig:spat} shows the correlation curve for $T_{ijk}$ and $T_{i^{\ast}jk}$. To show the trend of the correlation curve, the $x$-axis is the distance value. The small fluctuation when distance is large comes from estimation error.
	
	%%%%%%%%%%%%%%%%%%%%%%%%%%%%%%%%%%%%%%%%%%%%%%%%%%%%%%%%%%%%%%%%%%%%%%%%%%%%%%%%%%%%%%%%%%%%%%%%%%%%%%%%%%%%%%%%%%%%%
\section{Conclusions and Areas for Future Research}\label{sec:conclusoin}	%%%%%%%%%%%%%%%%%%%%%%%%%%%%%%%%%%%%%%%%%%%%%%%%%%%%%%%%%%%%%%%%%%%%%%%%%%%%%%%%%%%%%%%%%%%%%%%%%%%%%%%%%%%%%%%%%%%%%
This paper focuses on modeling GPU failure times under competing risks taking into account the GPU locations inside the supercomputer based on Titan GPU dataset. We propose parametric models with spatially correlated random effects. We use lognormal and Weibull distributions to model the time-to-event distribution, and use exponential, Gaussian, and power exponential correlation covariance functions for spatial random effects. Bayesian methods are used in estimation and inference. We show that the DBE failures and OTB failures interact with each other, and the OTB and DBE random effects are highly correlated with each other. The location of GPUs in the Titan supercomputer has a strong influence on the failure-time distributions.
	
Our simulation study shows that the proposed method works well and provides accurate estimates. The variances of estimators related to the power exponential function and the correlation between two failure modes are larger when compared with other estimators.

Although the Titan GPU dataset we used records only the OTB and DBE failures, there are other GPU failure types in Titan that were not provided in the available dataset. It would be interesting to take more failure types into consideration, and explore the influence of different GPU locations on other failure modes.  It will also be interesting to use a different distance function for each failure mode, which, however, will increase the complexity of the model and may impose challenges for inference. In addition, combining a nonparametric AFT model with spatially correlated random effects is a possible future research topic. A more flexible nonparametric model combined with the random effects may cause additional estimation difficulty and identifiability problems, especially when the number of failures at some locations are small. However, it is interesting to study this in the further. Using INLA to speed up the estimation is another potential future topic. To use INLA, the label of each GPU for mixture distribution needs to be sampled by MCMC, which can cause difficulty in setting priors of the labels and the convergence of the chains. Another future research topic is considering the generalized limited failure population (GLFP) model as introduced in Section~23.1 of \citeN{meeker2022statistical} to fit the data, which can make the model more flexible. However, considering GLFP under spatial correlations would be challenging.

%%%%%%%%%%%%%%%%%%%%%%%%%%%%%%%%%%%%%%%%%%%%%%%%%%%%%%%%%%%%%%%%%%%%%%%%%%%%%%%%%%%%%%%%%%%%%%%%%%%%%%%%%%%%%%%%%
	\section*{Acknowledgments}
	The authors acknowledge the Advanced Research Computing program at Virginia Tech for providing computational resources.
	%%%%%%%%%%%%%%%%%%%%%%%%%%%%%%%%%%%%%%%%%%%%%%%%%%%%%%%%%%%%%%%%%%%%%%%%%%%%%%%%%%%%%%%%%%%%%%%%%%%%%%%%%%%%%%%%%
	\appendix

	\section{Proof of Proposition 1}\label{apd0}

In this appendix, we provide a proof for Proposition~1 in Section~\ref{sec:model.est.inf}.
Let $z_i=(a_i-\mu)/\xi$, where $a_i = \tlog(t_i) - \bx_i\bbeta - w_i$. We have,	
	\begin{align*}
\int\prod_{i}f(t_{i}|w_{i}, \bPsi)f(\bPsi) d\bPsi
\propto \int_{0}^\infty\int_{-\infty}^\infty \cdots \int_{-\infty}^\infty\frac{1}{\xi } \prod_{i} \frac{c}{\xi t_{i}}
		d\mu\, d\beta_1 \cdots d\beta_p\, d\xi \notag,
	\end{align*}
where $c=\exp \left[ \sum_{i} z_i   -\sum_i \exp(z_i) \right]$. Let $u = \sum_{i} \exp(z_i)$. Then,
	\begin{align*}
\int\prod_{i}f(t_{i}|w_{i}, \bPsi)f(\bPsi) d\bPsi
\propto  \int_{-\infty}^\infty \cdots \int_{0}^\infty  \frac{1}{\xi} \prod_{i} \frac{1}{\xi t_i} \cdot
		\int_{ -\infty}^ \infty \exp \left\{  \sum_i  \frac{b_i}{\xi} - u  \right\}  \frac{\xi}{u}\, du\,
		 d\xi\, d\beta_1 \cdots d\beta_p,
	\end{align*}
where
$b_i=a_i - \xi \tlog \left[ \sum_i  \exp(a_i/\xi)\right]+ \xi \tlog(u).$ The integral related to the $u$ part is
	\begin{align*}
		\int_{-\infty} ^ \infty \exp \left[m \tlog(u) -u \right] u^{-1} du = \int_{-\infty} ^{\infty}  u^{m-1} \exp(-u) du = \Gamma(m-1).
	\end{align*}
The integral related to the $\xi$ part is
	\begin{align}\label{form:1}
		\int_{0}^{\infty} \xi^{-m} \exp \left\{ \sum_i \left(a_i/\xi  - \tlog \left[ \sum_i \exp \left(a_i/\xi\right)\right]\right)\right\} d\xi.
	\end{align}
	Let $a_{\tmax} = \tmax a_i$.  By Proposition A.3 in \shortciteN{ramos2020posterior}, because
$$\lim_{\xi \rightarrow 0} \frac{\sum_i \exp \left(a_i/\xi\right)}{\exp \left(a_{\tmax}/\xi \right) } = \lim_{\xi \rightarrow 0} \sum_{i} \exp \left(\frac{a_i - a_{\tmax}}{\xi}\right) = 0,$$
so $\lim_{\xi \rightarrow \infty} \sum_i \exp \left(a_i/\xi\right)/\exp \left(a_{\tmax}/\xi \right)= m$, and $\sum_i \exp \left(a_i/\xi\right) \propto \exp \left(a_{\tmax}/\xi \right)$.
	\eqref{form:1} becomes
	\begin{align*}
		\int_{0}^{\infty} \frac{1}{\xi^m} \prod_{i=1}^m \exp (a_i - a_{\tmax}) ^{1/\xi} d\xi = \int_{0}^\infty \frac{1}{\xi^m} \exp \left\{\tlog \left[\prod_{i} \exp (a_i - a_{\tmax})\right]\frac{1}{\xi} \right\}d\xi.
	\end{align*}
	We have that \eqref{form:1} is equal to $\Gamma(-m + 1)/\left(\tlog \left[\prod_{i} \exp (a_{\tmax} - a_i)\right] \right)^{m-1}$, because $a_i - a_{\tmax} \leq 0$, and $\log \left[\prod_i \exp(a_i - a_{\tmax}) \right] \leq 0$. Then,
	\begin{align*}
		\int\prod_{i}f(t_{i}|w_{i}, \bPsi) f(\bPsi)d\bPsi \propto \int _{-\infty}^\infty \cdots \int_{ -\infty}^{\infty} \frac{1}{ \left(\tlog \left[\prod_{i} \exp(a_{\tmax} - a_i)\right] \right)^{m-1} }d\beta_1 \cdots d\beta_p.
	\end{align*}
	Let $b_{i1} = \tlog (t_i) - w_i - \sum_{l = 2}^p x_{il}\beta_l$, and $c_{1} = \textrm{max}(b_{i1}) - b_{i1}$. We know $a_{\tmax} = \max( b_{i1} ) - \max(x_{i1})\beta_1$, if $\beta_1 \leq 0$ and $a_{\tmax} =\max( b_{i1}) - \min(x_{i1})\beta_1$, if $\beta_1 > 0.$
	Then,
	\begin{align*}
		\int_{ -\infty}^{\infty} \frac{1}{ \left(\tlog \left[\prod_{i} \exp ( a_{max} - a_i)\right] \right)^{m-1} }d\beta_1  = &
		\int_{-\infty}^{ 0 }\left\{ mc_1 + \sum_i \left[ x_{i1} - \tmax(x_{i1})\right]\beta_1 \right \}^{-m+1}d\beta_1 \\
		&+ \int_{0}^{ \infty } \left\{mc_1 + \sum_i \left[ x_{i1} - \tmin(x_{i1})\right]\beta_1\right\}^{-m+1}d\beta_1	.
	\end{align*}
	When $m > 2$,
	\begin{align*}
		&\int_{ -\infty}^{\infty} \frac{1}{ \left(\tlog \left[\prod_{i} \exp ( a_{\tmax} - a_i)\right] \right)^{m-1} }d\beta_1  \\
		& =\frac{1}{-m + 2} \left(\frac{1}{\sum_i (x_{i1} - \tmin(x_{i1}))} - \frac{1}{\sum_i (x_{i1} - \tmax(x_{i1}))}\right)(mc_1)^{-m+2}.
	\end{align*}
	
	Similarly, let $b_{i2} = \tlog (t_i) - w_i - \sum_{l = 3}^p x_{il}\beta_l$, $c_2 = \tmax(b_{i2}) - b_{i2}$, $\ldots$, $b_{ip-1} = \tlog (t_i) - w_i -  x_{ip}\beta_p$, $c_{p-1} = \tmax(b_{ip-1}) - b_{ip-1}$, $b_{ip} = \log(t_i) - w_i$.  For $j=2, \ldots, p-1$, we have $\tmax(b_{ij}) = \max( b_{ij} ) - \max(x_{ij})\beta_j$, if $\beta_j \leq 0$ and
$\tmax(b_{ij})=\max( b_{ij}) - \min(x_{ij})\beta_j$, if $\beta_j > 0$.
Then, it is clear that $\int _{-\infty}^\infty c_{p-1}^{-m+p} d\beta_p < \infty$ if $m > (p + 1)$, thus
$ \int\prod_{i}^m f(t_{i}|w_{i}, \bPsi)f(\bPsi) d\bPsi <\infty,
\textrm{ if }  m > (p + 1).$
	
%%%%%%%%%%%%%%

%%%%%%%%%%%%%%%%%%%%%%%%%%%%%%%%%%%%%%%%%%%%%%%%%%%%%%%%%%%%%%%%%%%%%%%%%%%%%%%%%%%%%%%%%%%%%%%%%%%%%%%%%%%%%%%%%%%%%%%%%%%%%%%%%%%%%%%
\section{Additional Data Visualization} \label{apd1}
%%%%%%%%%%%%%%%%%%%%%%%%%%%%%%%%%%%%%%%%%%%%%%%%%%%%%%%%%%%%%%%%%%%%%%%%%%%%%%%%%%%%%%%%%%%%%%%%%%%%%%%%%%%%%%%%%%%%%%%%%%%%%%%%%%%%%%%

Under the original labeling, column indexes follow the physical column location order. However, the internal connection of columns does not follow the physical location order. For example, column 1 under the original labeling is connected with column 3, and column 3 is connected with column 5. Therefore, we relabel the columns using the connection index. That is, column~3 is relabeled as column 2, and column 5 is relabeled as column 3.

\begin{figure}
	\begin{center}
		\begin{tabular}{c}
			\includegraphics[width = 0.8\textwidth]{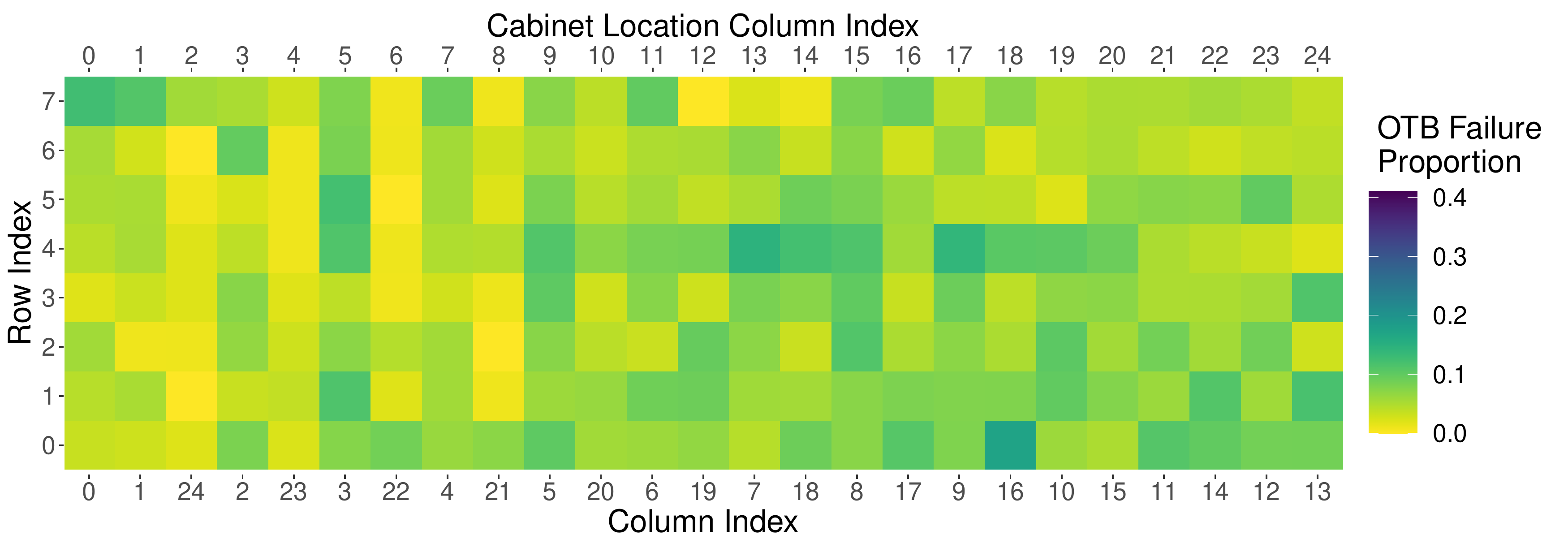}\\
			%(a) OTB \\
			\includegraphics[width = 0.8\textwidth]{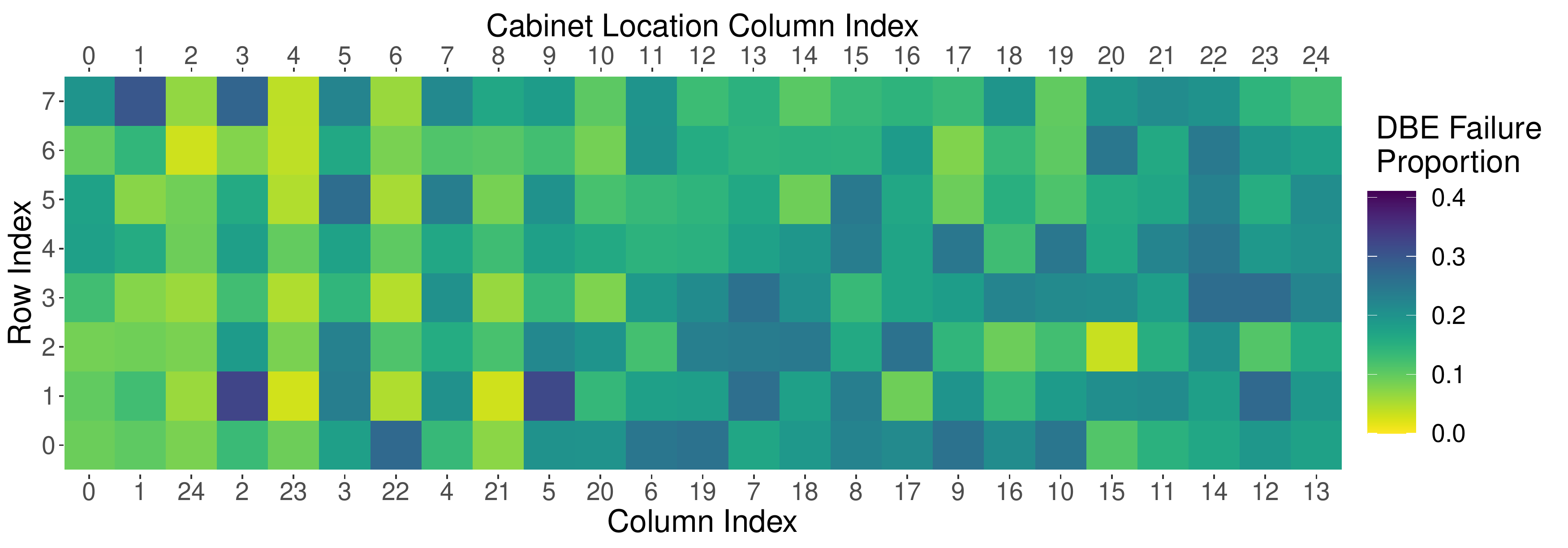}%\\
			%(b) DBE\\
		\end{tabular}
		\caption{Visualization of  OTB and DBE failure proportions on 200 different cabinet locations. The top $x$-axis label shows the cabinet location column index, and the bottom $x$-axis label shows the column index based on cabinet connection. }\label{fig:visual.heat.map.ori}
	\end{center}
\end{figure}

\begin{figure}
	\begin{center}
		\begin{tabular}{c}
			\includegraphics[width = 0.8\textwidth]{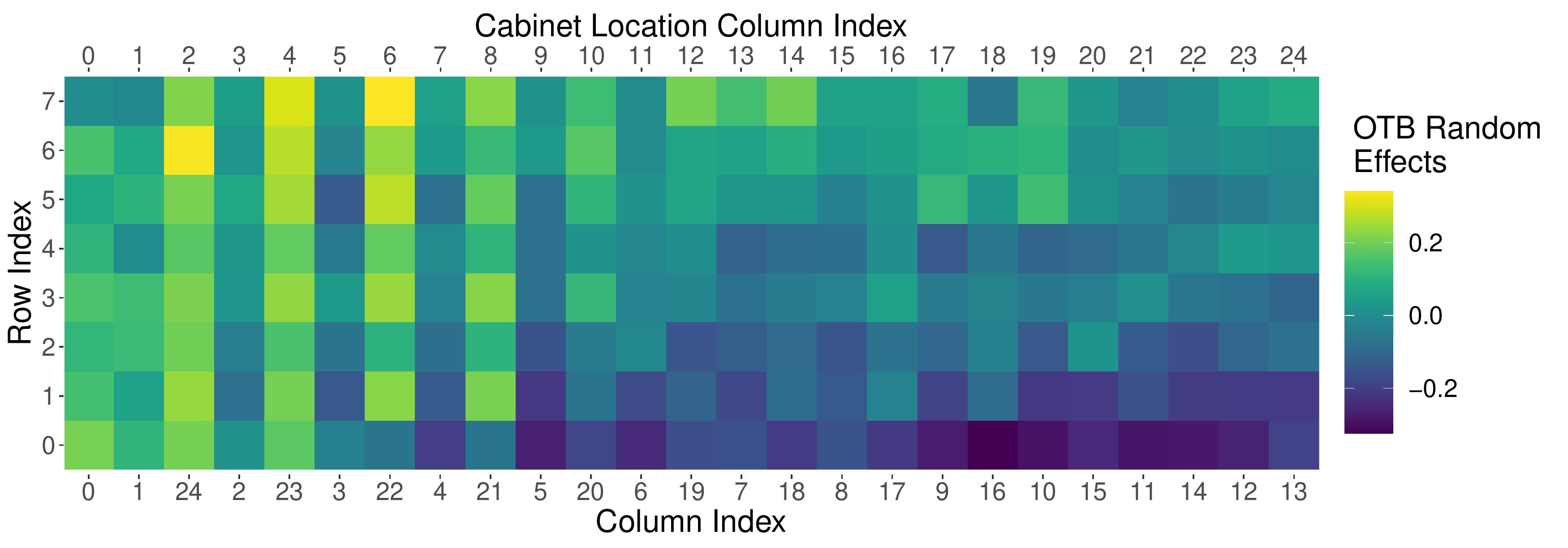}\\
			%(a) OTB \\
			\includegraphics[width = 0.8\textwidth]{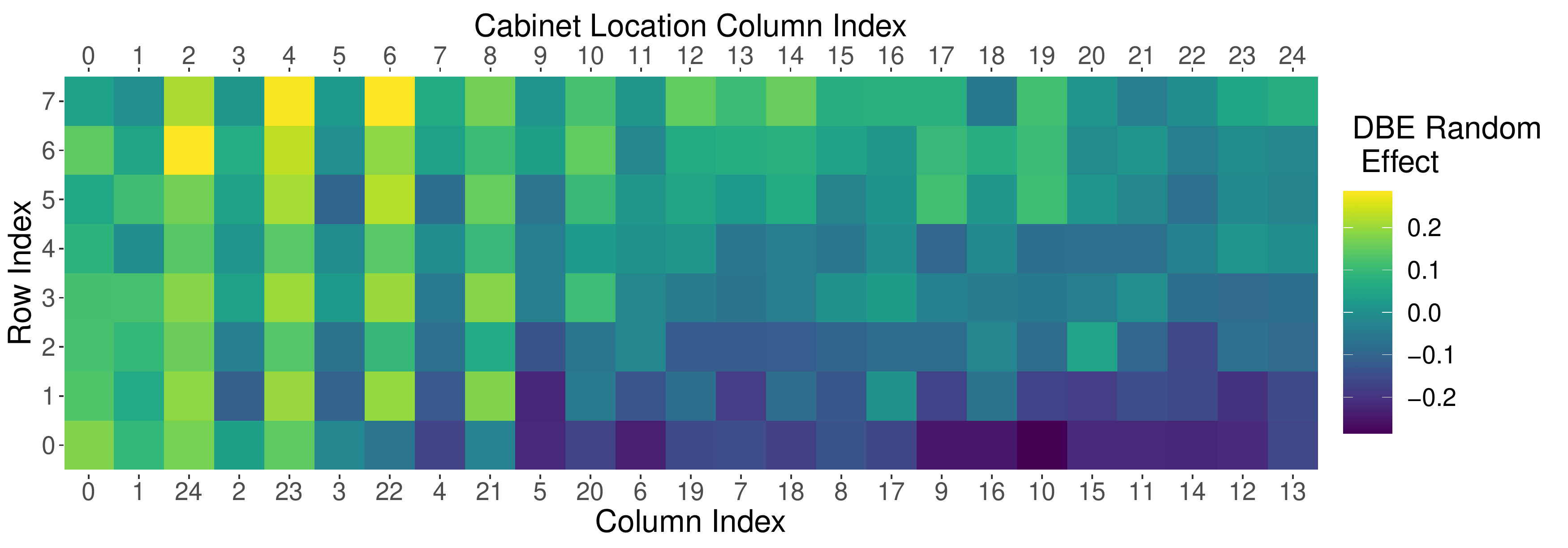}%\\
			%(b) DBE\\
		\end{tabular}
		\caption{Visualization of estimated OTB and DBE random effects using the Weibull AFT INDEP model on 200 different cabinet locations. The top $x$-axis label shows the cabinet location column index, and the bottom $x$-axis label shows the column index based on cabinet connection. }\label{fig:random.heat.map.ori}
	\end{center}
\end{figure}

Figure~\ref{fig:visual.heat.map.ori} shows the heatmaps of OTB and DBE failure proportion on 200 cabinet locations. The $x$-axis show both the cabinet location column index (i.e., the original labeling, marked on the top of the panel in the figure) and the column index built based on connection (i.e., the label used in modeling, marked on the bottom of the panel in the figure). Figure~\ref{fig:random.heat.map.ori} shows similar heatmaps, while the color represents estimated random effects' means using the Weibull AFT INDEP model. Under this model, the order of columns does not influence the model fitting.

Both Figures~\ref{fig:visual.heat.map.ori} and~\ref{fig:random.heat.map.ori} show that using the original cabinet location as column index makes the random effects on different locations less ordered compared with using connectivity based the column index.

%%%%%%%%%%%%%%%%%%%%%%%%%%%%%%%%%%%%%%%%%%%%%%%%%%%%%%%%%%%%%%%%%%%%%%%%%%%%%%%%%%%%%%%%%%%%%%%%%%%%%%%%%%%%%%%%%%%%%%%%%%%%%%%%%%%%%%%
\section{Additional Simulation Results}\label{apd2}
%%%%%%%%%%%%%%%%%%%%%%%%%%%%%%%%%%%%%%%%%%%%%%%%%%%%%%%%%%%%%%%%%%%%%%%%%%%%%%%%%%%%%%%%%%%%%%%%%%%%%%%%%%%%%%%%%%%%%%%%%%%%%%%%%%%%%%%

\begin{figure}
 	\begin{center}
 		\begin{tabular}{cc}
 		  \includegraphics[width = 0.48\textwidth]{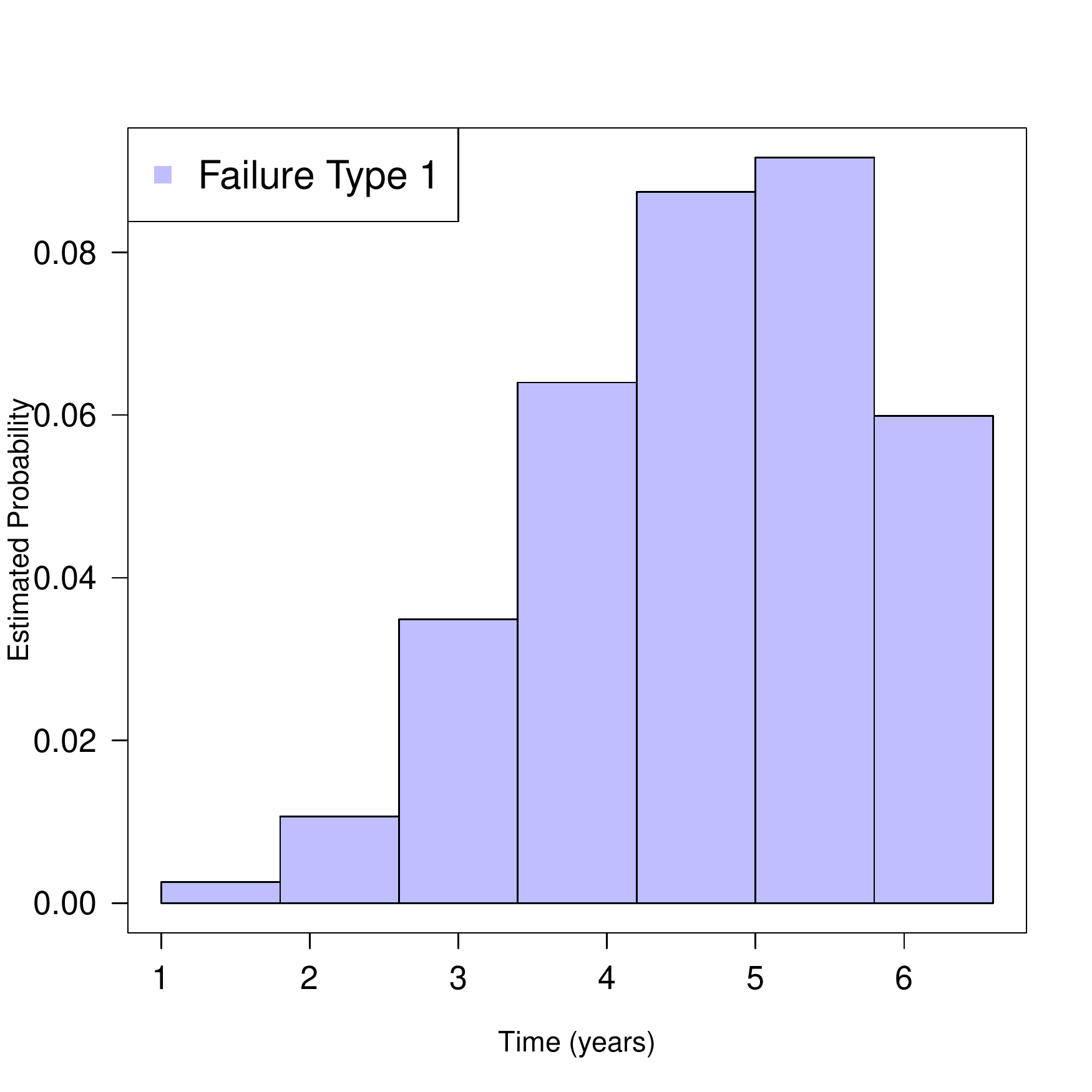}&  \includegraphics[width = 0.48\textwidth]{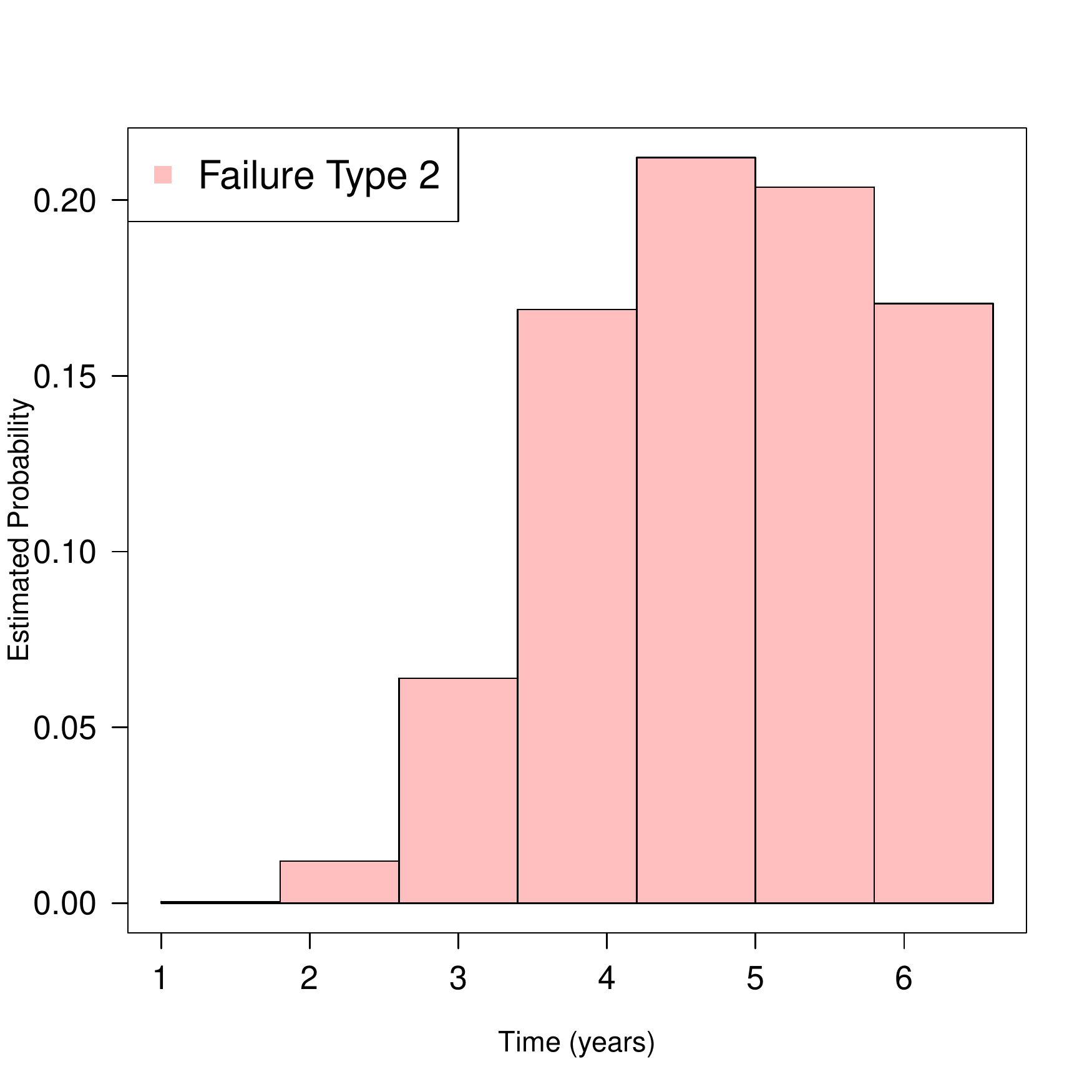}
 		\end{tabular}
 		\caption{Barplots of discretized estimated probability mass function of the two failure types using Kaplan-Meier estimator based on the simulation dataset.}\label{censdat}
 	\end{center}
 \end{figure}

In this section, we present additional simulation results. Figure \ref{censdat} shows the barplots of discretized estimated probability mass function of the two failure types in a simulation data set generated using $ 7 \times 7$ spatial locations and $7{,}000$ units. The figures are similar to the barplots drawn using the real GPU data, indicating the simulated data is close to the real GPU data.

\begin{figure}
	\begin{center}
		\includegraphics[width = 0.95\textwidth]{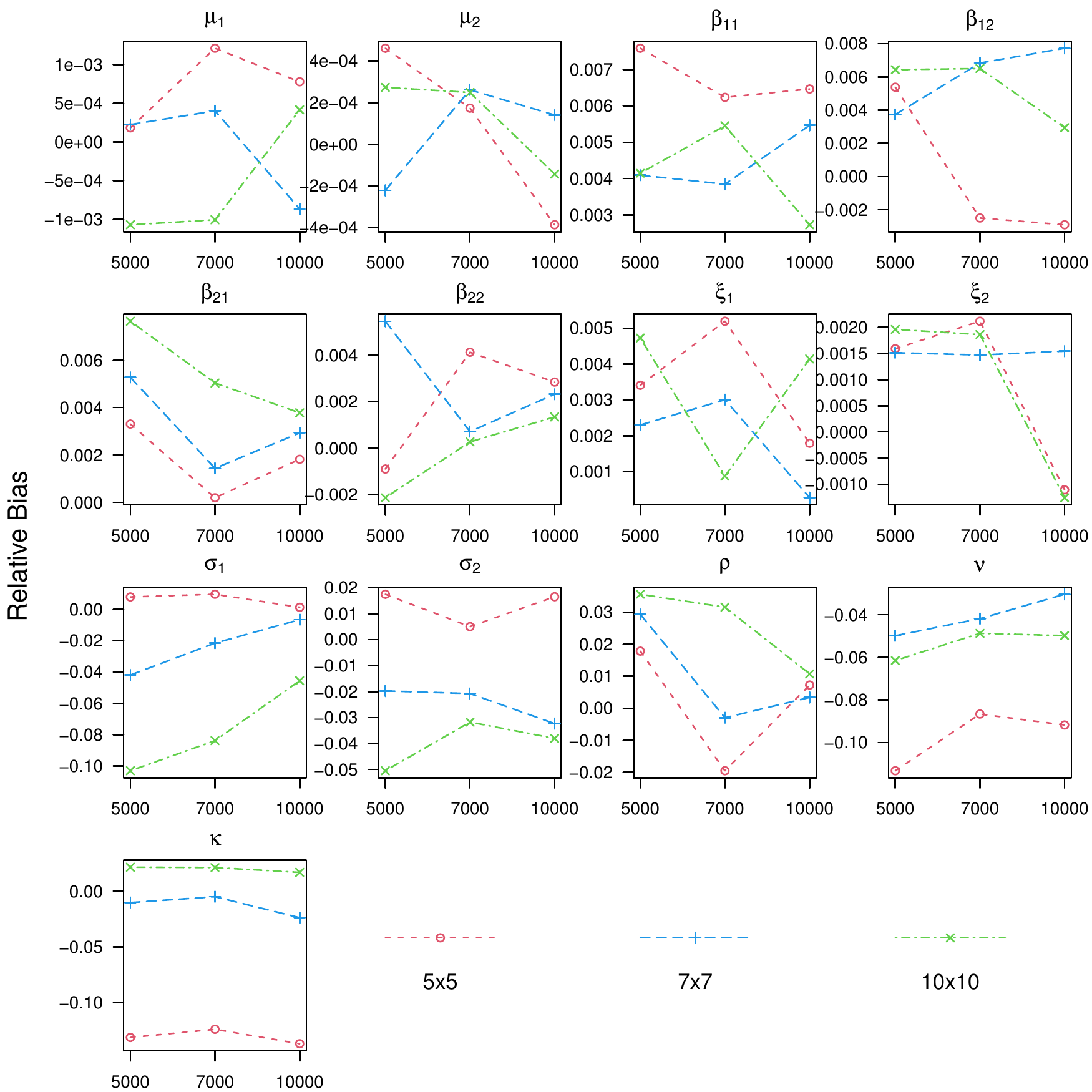}
		\caption{Plot of root relative bias as a function of the number of units for three different number of spatial location settings.}\label{relbias}
	\end{center}
\end{figure}

\begin{figure}
	\begin{center}
		\includegraphics[width = 0.95\textwidth]{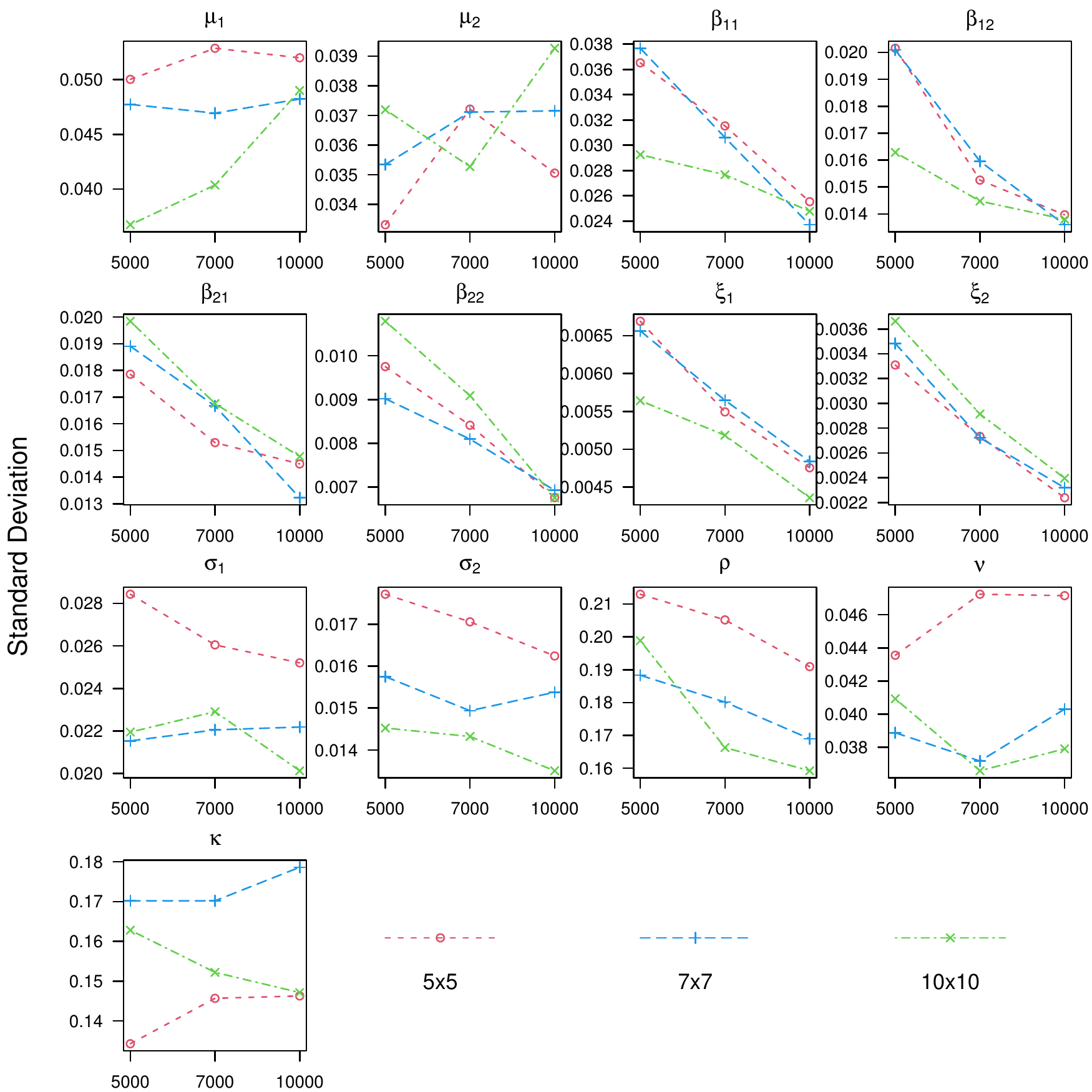}
		\caption{Plots of estimated standard deviation for different number of units and spatial location combinations. The $x$-axis is the number of units $N$.}\label{SD}
	\end{center}
\end{figure}

\begin{figure}
	\begin{center}
		\includegraphics[width = 0.95\textwidth]{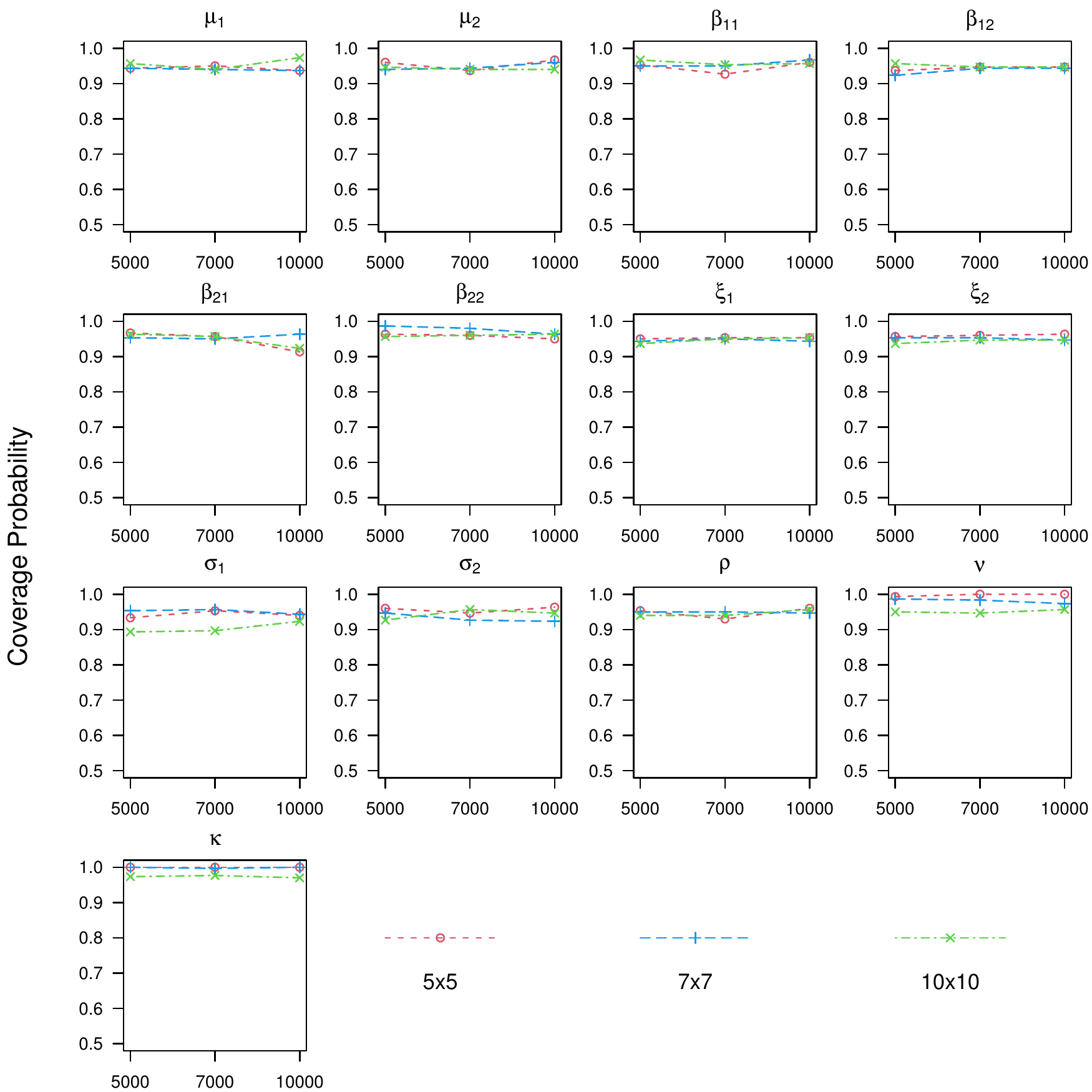}
		\caption{Plots of coverage probability for different number of units and spatial location combinations. The $x$-axis is the number of units $N$.}\label{CP}
	\end{center}
\end{figure}

\begin{figure}
	\begin{center}
		\includegraphics[width = 0.95\textwidth]{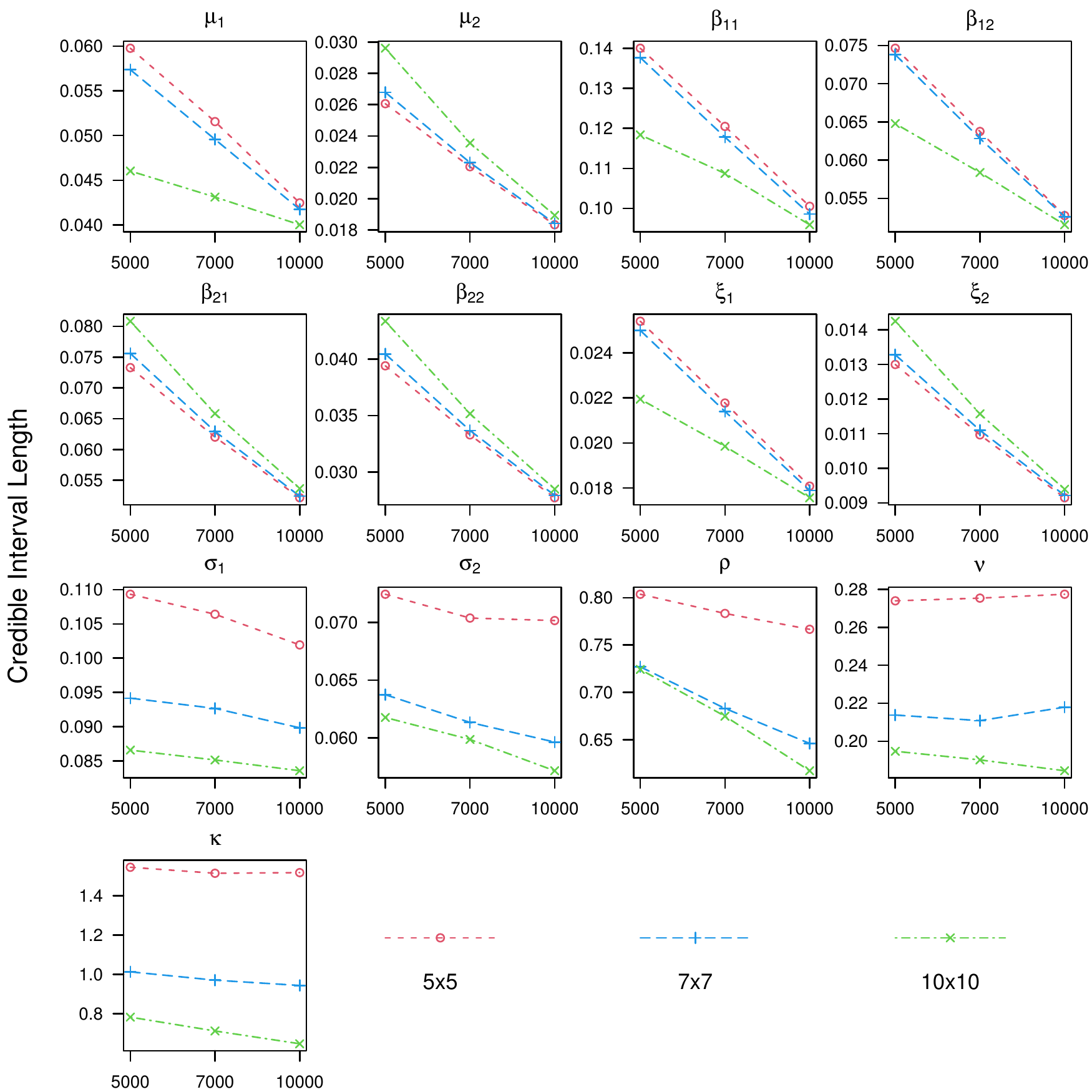}
		\caption{Plots of mean credible interval length for different number of units and spatial location combinations. The $x$-axis is the number of units $N$.}\label{CIL}
	\end{center}
\end{figure}

Figures~\ref{relbias}--\ref{CIL} show the relative bias, estimated standard deviation of posterior means, coverage probability of CIs and mean CI lengths of all the estimators. The relative bias and estimated standard deviations are small for all parameters. When the number of location increases from $5 \times 5$ to $ 7 \times 7$, the relative bias of both $\nu$ and $\kappa$ decreases. The coverage probability of CI is close to $0.95$ for all the cases. The mean length of CI decreases as the number of units increases for all parameters.

%%%%%%%%%%%%%%%%%%%%%%%%%%%%%%%%%%%%%%%%%%%%%%%%%%%%%%%%%%%%%%%%%%%%%%%%%%%%%%%%%%%%%%%%%%%%%%%%%%%%%%%%%%%%%%%%%%%%%%%%%%%%%%%%%%%%%%%
\section{Details for EM Algorithm and Results}\label{apd3}
%%%%%%%%%%%%%%%%%%%%%%%%%%%%%%%%%%%%%%%%%%%%%%%%%%%%%%%%%%%%%%%%%%%%%%%%%%%%%%%%%%%%%%%%%%%%%%%%%%%%%%%%%%%%%%%%%%%%%%%%%%%%%%%%%%%%%%%
In this section, we give the detailed formulas of the EM algorithm, which is used in the alternative methods comparison. For the OTB failure times, the pdf is
\begin{align*}
	f_1(t_{ij}|w_{ik})= \frac{1}{\xi_{k} t_{ij}}\phi\left [\frac{\log (t_{ij})-\mu_{ijk}}{\xi_{k}}\right],
\end{align*}
 the cdf is
\begin{align*}
F_1(t_{ij}|w_{ik})= \Phi\left[\frac{\log(t_{ij})- \mu_{ijk}}{\xi_{k}} \right],
\end{align*}
and
\begin{align*}
\mu_{ijk} = &\mu_k+\bx_j' \bbeta_k+ w_{ik}=\bx_j' \bbeta_k+ u_{ik}, k=1,
\end{align*}
where $u_{ik}=\mu_k+w_{ik}$. For DBE failure times, the pdf is
\begin{align*}
		f_2(t_{ij}|w_{ik}) = \lambda\left[\frac{1}{\xi_{k1} t_{ij}}\phi\left (\frac{\log (t_{ij})-\mu_{ijk1}}{\xi_{k1}}\right)\right] + (1- \lambda) \left[\frac{1}{\xi_{k2} t_{ij}}\phi\left (\frac{\log (t_{ij})-\mu_{ijk2}}{\xi_{k2}}\right) \right],
\end{align*}
the cdf is
\begin{align*}
F_2(t_{ij}|w_{ik}) = \lambda\Phi\left (\frac{\log (t_{ij})-\mu_{ijk1}}{\xi_{k1}}\right) + (1- \lambda) \Phi\left (\frac{\log (t_{ij})-\mu_{ijk2}}{\xi_{k2}}\right),
\end{align*}
and
\begin{align*}
	\mu_{ijk1} = &\mu_k+\bx_j' \bbeta_k+ w_{ik}=\bx_j' \bbeta_k+ u_{ik}\\
	\mu_{ijk2} = &\mu_k+ \eta + \bx_j' \bbeta_k+ w_{ik}=\eta + \bx_j' \bbeta_k+ u_{ik}, \quad k = 2.
\end{align*}

We sort $u_{ik}$ as $\bu=(u_{11}, u_{12},\ldots, u_{n1}, u_{n2})'$ and $\muvec=(\mu_{1}, \mu_{2},\ldots, \mu_{1}, \mu_{2})'$.  Note that the $\bu$ vector is different from the $\bw$ in the paper.

The covariance matrix is
	\begin{equation*}
		\Sigma_{\bw} =  \Omega \otimes \Sigma_f,
	\end{equation*}
where
\begin{align*}
		\Sigma_f =  \begin{pmatrix}
			\sigma_{1} & \rho_{12}\sqrt{\sigma_{1} \sigma_{2}}\\
			\rho_{12}\sqrt{\sigma_{1} \sigma_{2}} & \sigma_{2}
		\end{pmatrix}.
\end{align*}

The log-likelihood conditioning on $\bu$ is:
\begin{align*}
\loglik(\btheta_{\bt}|\bt,\bu)&= \sum_{k=1}^{2}\sum_{i,j}\delta_{ijk}\log[f_k(t_{ij}|u_{ik})]+(1-\delta_{ijk})
\log[1-F_k(t_{ij}|u_{ik})].
\end{align*}
The log joint density for $\bu$ is:
\begin{align*}
&\log[f_{\bu}(\bu|\btheta_{\bu})]\\
&=-\frac{2n}{2}\log(2\pi) -\frac{1}{2}\log(|\Sigma_{\bw}|)-\frac{1}{2} (\bu-\muvec)' \Sigma_{\bw}^{-1}(\bu-\muvec)\\
&=-\frac{2n}{2}\log(2\pi)-\frac{n}{2}\log(|\Sigma_f|)-\frac{2}{2}\log(|\Omega|)-\frac{1}{2} \trace\left[(\Omega^{-1}\otimes \Sigma_f^{-1})(\bu-\muvec)(\bu-\muvec)'\right].
\end{align*}
Then, the log likelihood for $\btheta=(\btheta_{\bt}',\btheta_{\bu}')'$ is:
\begin{align*}
\loglik(\btheta|\bt,\bu)= \loglik(\btheta_{\bt}|\bt,\bu)+\log[f_{\bu}(\bu|\btheta_{\bu})].
\end{align*}
For the E step, the conditional expectation of the first term is
$\E_{\bu|\bt}\loglik(\btheta_{\bt};\bt,\bu)$, which needs to be evaluated numerically. The conditional expectation of the second term is
$$
-\frac{2n}{2}\log(2\pi)-\frac{n}{2}\log(|\Sigma_f|)-\frac{2}{2}\log(|\Omega|)-\frac{1}{2} \trace\left[(\Omega^{-1}\otimes \Sigma_f^{-1})\E_{\bu|\bt}(\bu-\muvec)(\bu-\muvec)'\right].$$

We use the Gibbs sampler to obtain samples of $\bu$ to evaluate $\E_{\bu|\bt}\loglik(\btheta_{\bt};\bt,\bu)$. In each MCMC step, let $y$ be the proposed step, $B=(\Omega^{-1}\otimes \Sigma_f^{-1})=(b_{sl})$, and $\deltavec_{ik}$ be a 0/1 vector that only the corresponding position of $u_{ik}$ in $\bu$ is one. The log of the probability for updating $u_{ik}$ is $a_1+a_2$, where
\begin{align*}
a_1=&\left\{\sum_{j}\delta_{ijk}\log[f_k(t_{ij}|u_{ik}+y)]+(1-\delta_{ijk})
\log[1-F_k(t_{ij}|u_{ik}+y)]\right\}\\
&-\left\{\sum_{j}\delta_{ijk}\log[f_k(t_{ij}|u_{ik})]+(1-\delta_{ijk})
\log[1-F_k(t_{ij}|u_{ik})]\right\},
\end{align*}
and
\begin{align*}
a_2&=-\frac{1}{2}(\bu+y\deltavec_{ik}-\muvec)'B(\bu+y\deltavec_{ik}-\muvec)+\frac{1}{2} (\bu-\muvec)'B(\bu-\muvec)\\
&=-y(B\deltavec)'(\bu-\muvec)-\frac{y^2}{2}\deltavec'B\deltavec.
\end{align*}

Because certain $\nu$ and $\kappa$ can cause negative definite $\Omega$, we add an additional penalty term in the M step to constrain the set of $\nu$ and $\kappa$ estimates. To do this, we use a grid search method. We set different $\nu$ and $\kappa$ values and calculate the smallest eigenvalue of $\Omega$. The border of the smallest positive eigenvalue of $\Omega$ can be seen in the heatmap in Figure~5 of the paper. For $\nu$ and $\kappa$ combination that passes the border, we add a large penalty (e.g., 10{,}000) to the optimization function. By doing this, the estimated $\Omega$ matrix is always positive definite in the EM algorithm. Table~\ref{tab:emmat} shows the estimates based on the EM algorithm.

\begin{table}
	\begin{center}
		\caption{Estimates of all parameters (denoted as ``para.'' in the table) in the Weibull model with the powered exponential correlation function using the EM algorithm.}
		\begin{tabular}{c|c|c|c|c|c|c|c}
			\hline\hline
Para. & Estimate & Para. & Estimate &Para. & Estimate & Para. & Estimate\\
\hline
$\mu_1$
& 1.90
& $\xi_1$
& 0.20
& $\beta_{11}$
& 0.67
& $\beta_{12}$
& 0.27
\\
\hline
$\beta_{13}$
& 0.04
& $\beta_{14}$
& 0.03
& $\beta_{15}$
& 0.05
& $\beta_{16}$
& 0.04
\\
\hline
$\beta_{17}$
& 0.07
& $\beta_{18}$
& 0.01
& $\beta_{19}$
& $-$0.01
& $\beta_{1,10}$
& $-$0.28
\\
\hline
$\beta_{1,11}$
& $-$0.30
& $\beta_{1,12}$
& $-$0.07
& $\mu_2$
& 1.50
& $\xi_{21}$
& 0.14
\\
\hline
$\xi_{22}$
& 1.22
& $\eta$
& 7.09
& $\lambda$
& 0.59
& $\beta_{21}$
& 0.57
\\
\hline
$\beta_{22}$
& 0.23
& $\beta_{23}$
& 0.04
& $\beta_{24}$
& 0.08
& $\beta_{25}$
& 0.08
\\
\hline
$\beta_{26}$
& 0.09
& $\beta_{27}$
& 0.06
& $\beta_{28}$
& 0.06
& $\beta_{29}$
& 0.03
\\
\hline
$\beta_{2,10}$
& $-$0.24
& $\beta_{2,11}$
& $-$0.26
& $\beta_{2,12}$
& $-$0.06
& $\sigma_{1}$
& 0.02
\\
\hline
$\sigma_{2}$
& 0.02
& $\rho_{12}$
& 0.95
& $\nu$
& 0.48
& $\kappa$
& 1.46
\\

	\hline
		\hline
	
		\end{tabular}\label{tab:emmat}
	\end{center}
\end{table}
%%%%%%%%%%%%%%%%%%%%%%%%%%%%%%%%%%%%%%%%%%%%%%%%%%%%%%%%%%%%%%%%%%%%%%%%%%%%%%%%%%%%%%%%%%%%%%%%%%%%%%%%%%%%%%%%%%%%%%%%%%%%%%%%%%%%%%%
\section{Details on MCMC Diagnostics}\label{apd4}
%%%%%%%%%%%%%%%%%%%%%%%%%%%%%%%%%%%%%%%%%%%%%%%%%%%%%%%%%%%%%%%%%%%%%%%%%%%%%%%%%%%%%%%%%%%%%%%%%%%%%%%%%%%%%%%%%%%%%%%%%%%%%%%%%%%%%%%

In this section, we provide some details on MCMC diagnostics. Rhat is usually used to evaluate the convergence of MCMC chains. A Rhat value that closes to 1 indicates good mixing. A cut off value of 1.1 is usually used in convergence diagnostics. In addition to Rhat, a proper effective sample size is also needed in convergence diagnostics. \shortciteN{vehtari2021rank} proposed to use bulk effective sample size (ESS) and tail ESS to estimate the effective sample size of MCMC chains in the bulk and tail of the posterior distribution. Bulk ESS and tail ESS larger than $400$ are suggested in \shortciteN{vehtari2021rank} to ensure the Rhat estimation is reliable.

\begin{figure}
	\centering
	\includegraphics[width=0.48\textwidth]{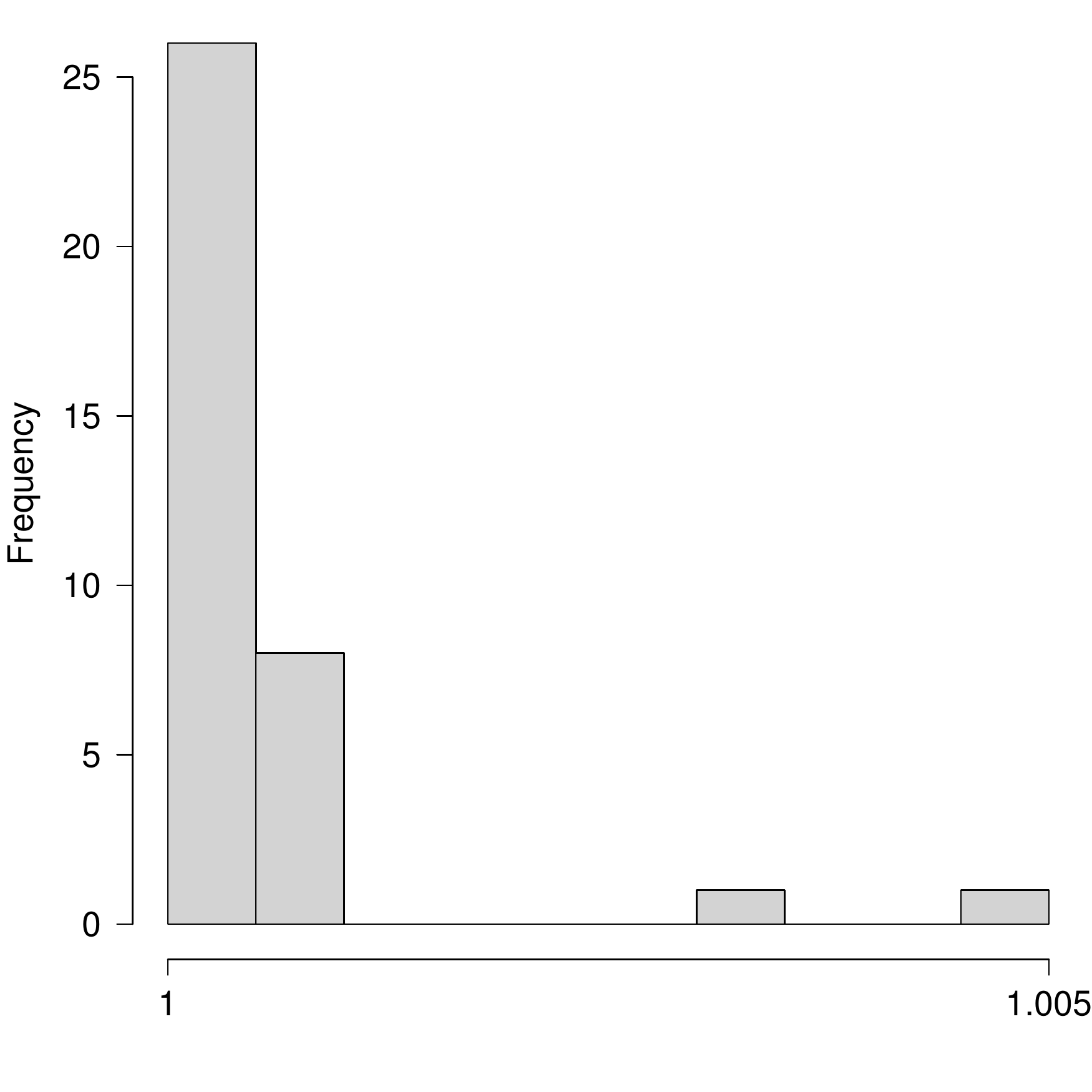}	
	\caption{Histogram of Rhat for all parameters.} \label{fig:Rhat}
\end{figure}

\begin{figure}
	\begin{center}
		\begin{tabular}{c}
			\includegraphics[width=0.6\textwidth]{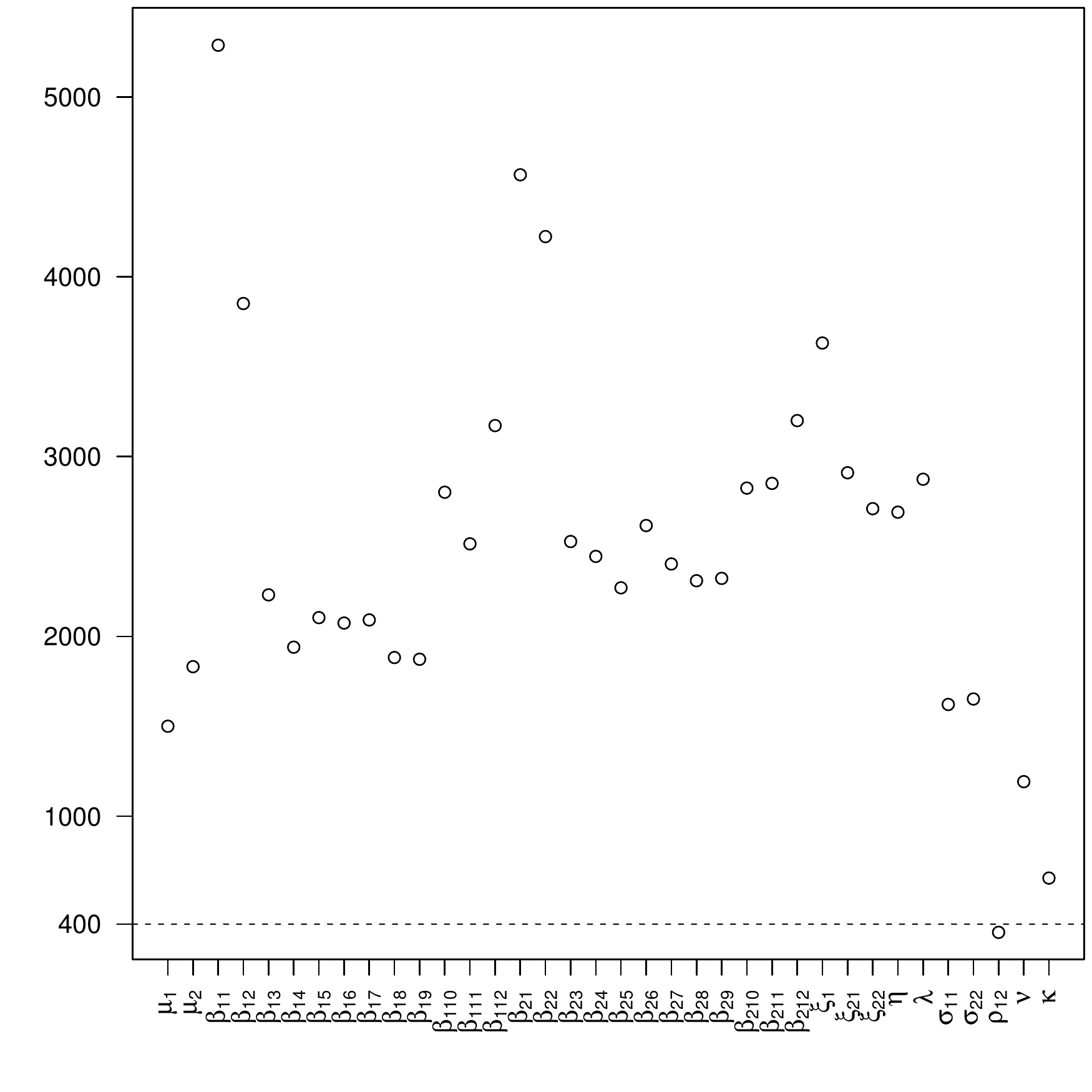} \\[-4ex]
(a)  Bulk ESS  \\[4ex]
			\includegraphics[width=0.6\textwidth]{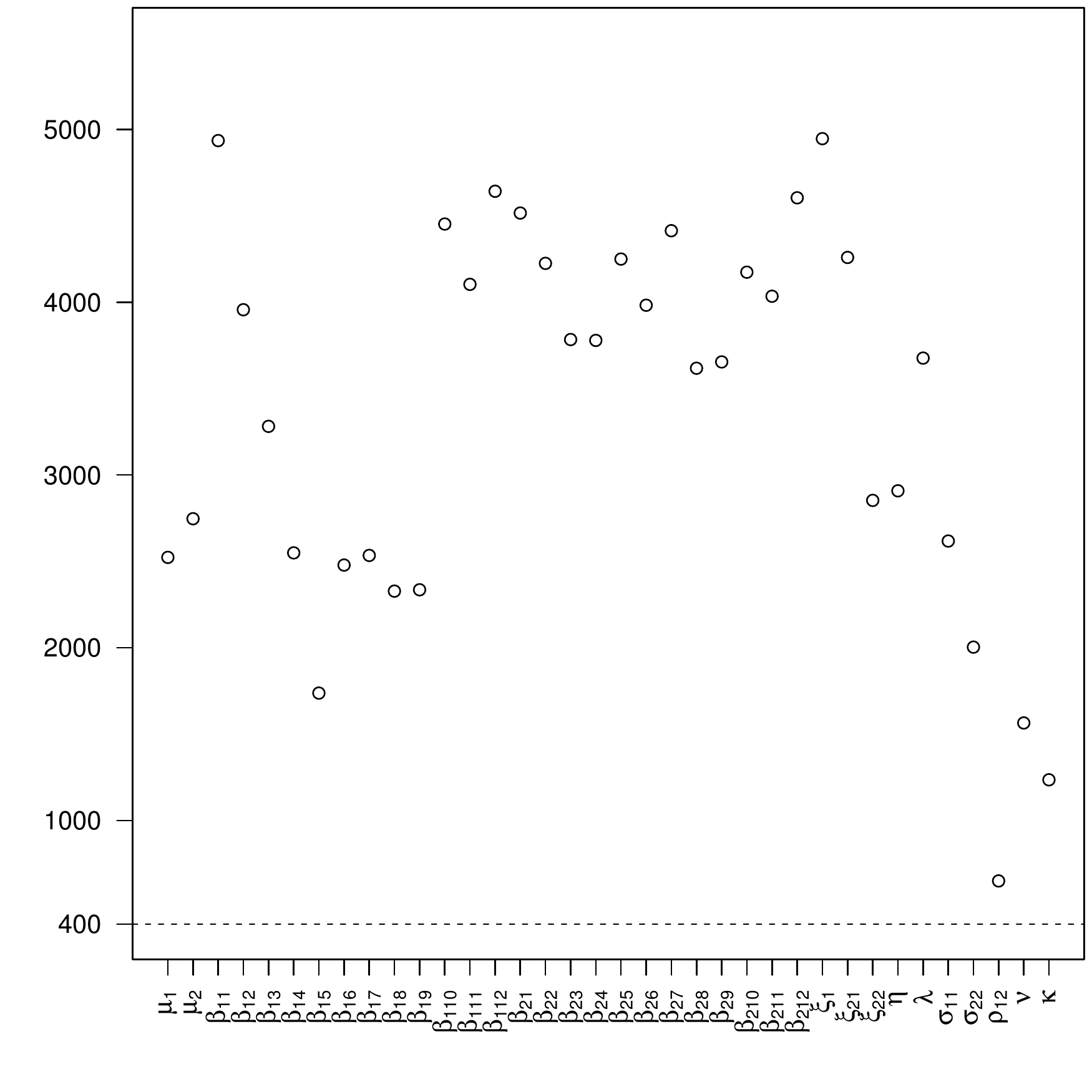}\\[-4ex]
			 (b)  Tail ESS 	
		\end{tabular}
		\caption{Plots of bulk ESS and tail ESS for all parameters.} \label{fig:converge}
	\end{center}
\end{figure}

Figures~\ref{fig:Rhat} and \ref{fig:converge} show the Rhat, bulk ESS, and tail ESS for all parameters. The Rhat, bulk ESS and tail ESS are calculated using Stan based on \shortciteN{vehtari2021rank}. Small Rhat values suggest convergence of the chains. Both the bulk ESS and tail ESS are acceptable for all the parameters. The only bulk ESS value smaller than $400$ is from $\rho_{12}$, which is $350$ and is close to $400$. We also check the behavior of the trace plots. We believe that there is no convergence problem in the MCMC chains.

%%%%%%%%%%%%%%%%%
\section{Estimates and CI of Parameters for Different Models}\label{apd5}

\begin{figure}
	\begin{center}
		\includegraphics[width = 0.95\textwidth]{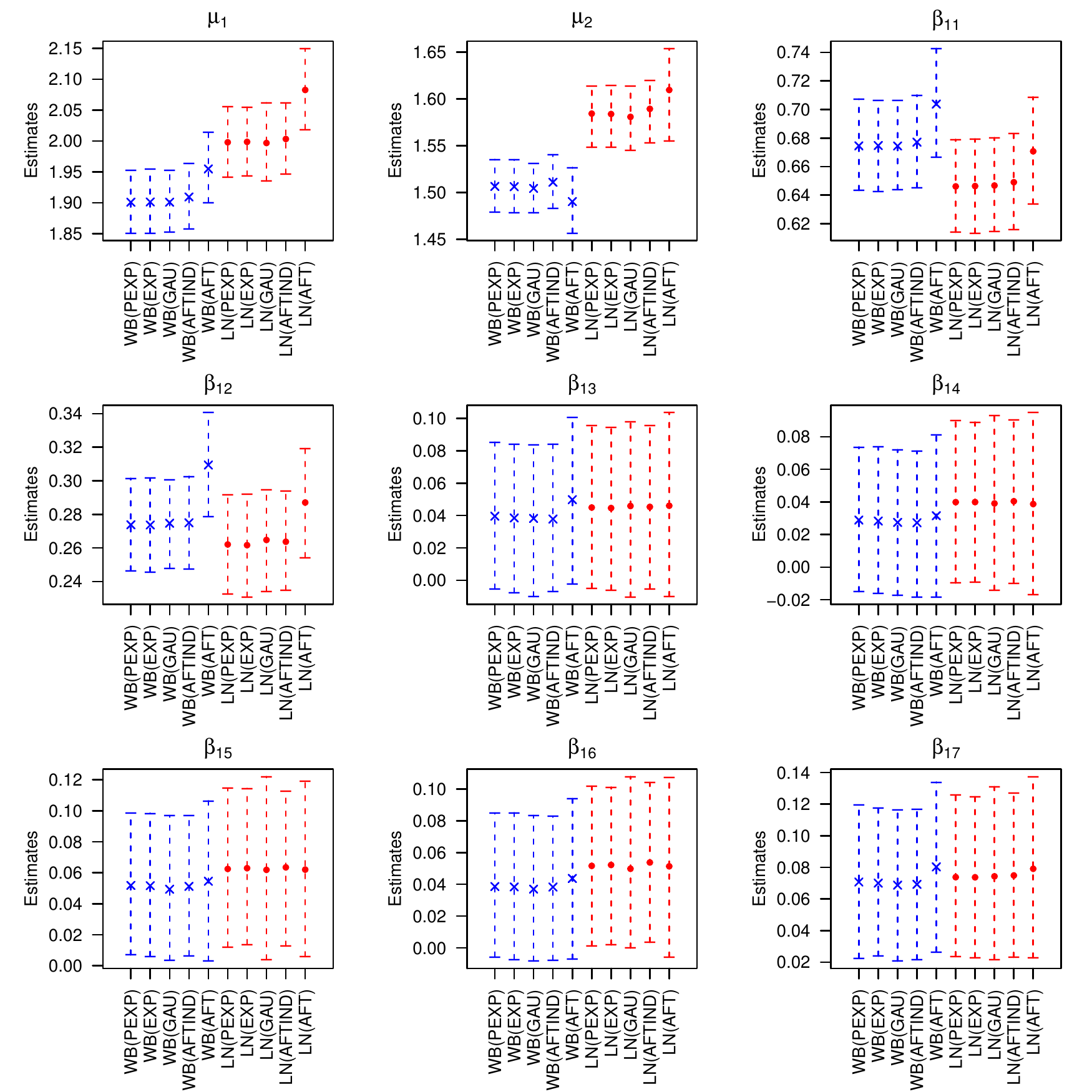}
		\caption{Comparison of CIs for all parameters (set 1) from different models.}\label{CrI1}
	\end{center}
\end{figure}

\begin{figure}
	\begin{center}
		\includegraphics[width = 0.95\textwidth]{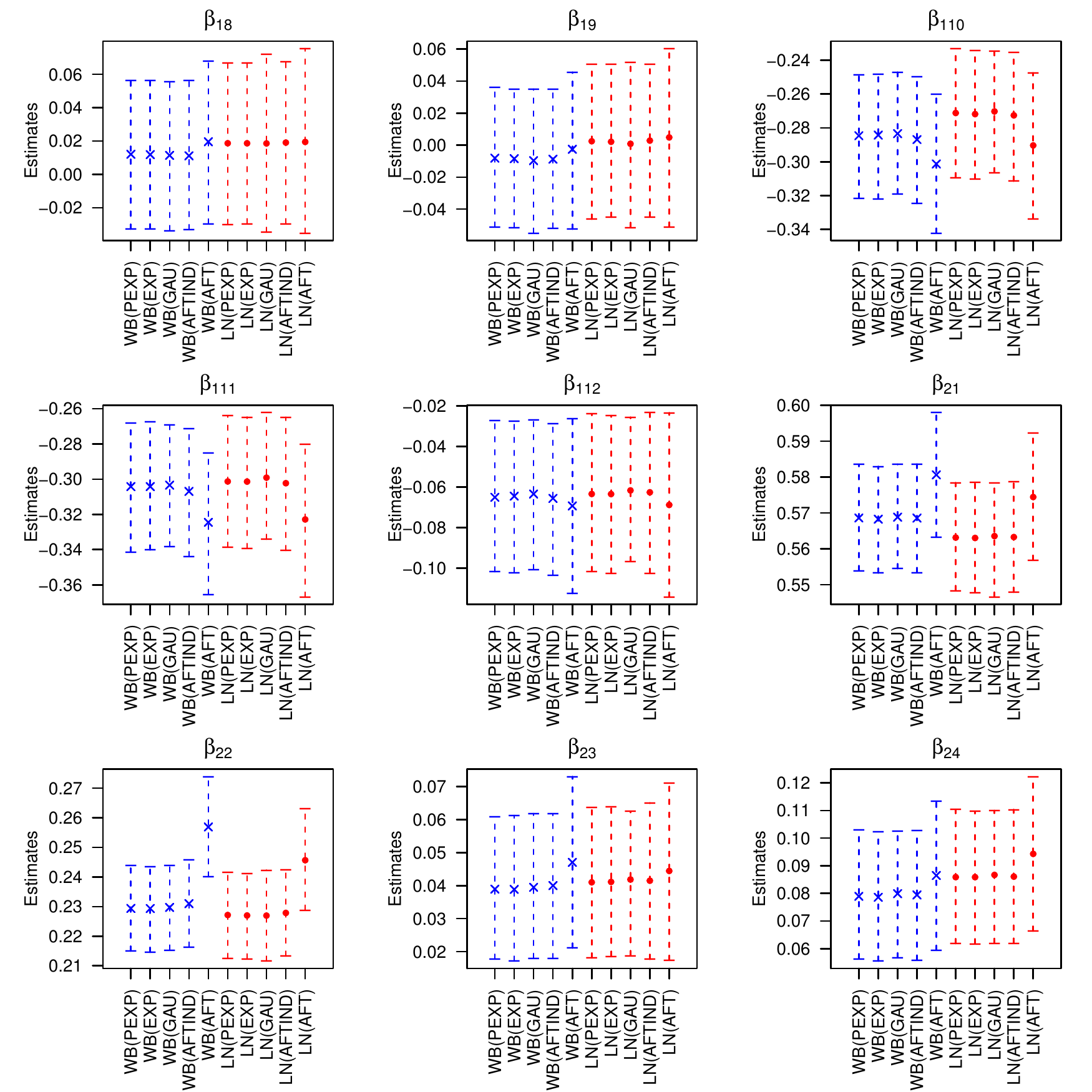}
		\caption{Comparison of CIs for all parameters (set 2) from different models.}\label{CrI2}
	\end{center}
\end{figure}

\begin{figure}
	\begin{center}
		\includegraphics[width = 0.95\textwidth]{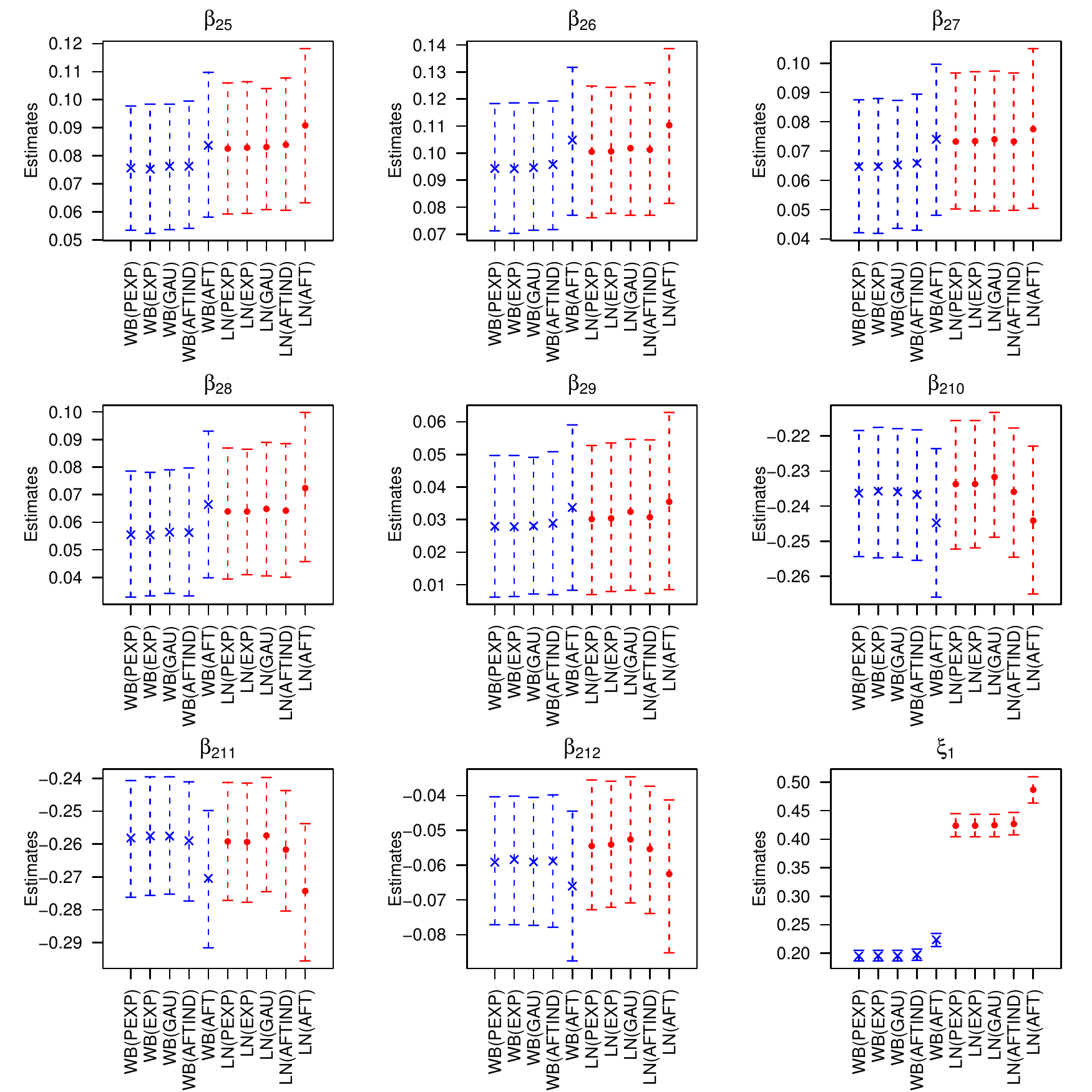}
		\caption{Comparison of CIs for all parameters (set 3) from different models.}\label{CrI3}
	\end{center}
\end{figure}

\begin{figure}
	\begin{center}
		\includegraphics[width = 0.95\textwidth]{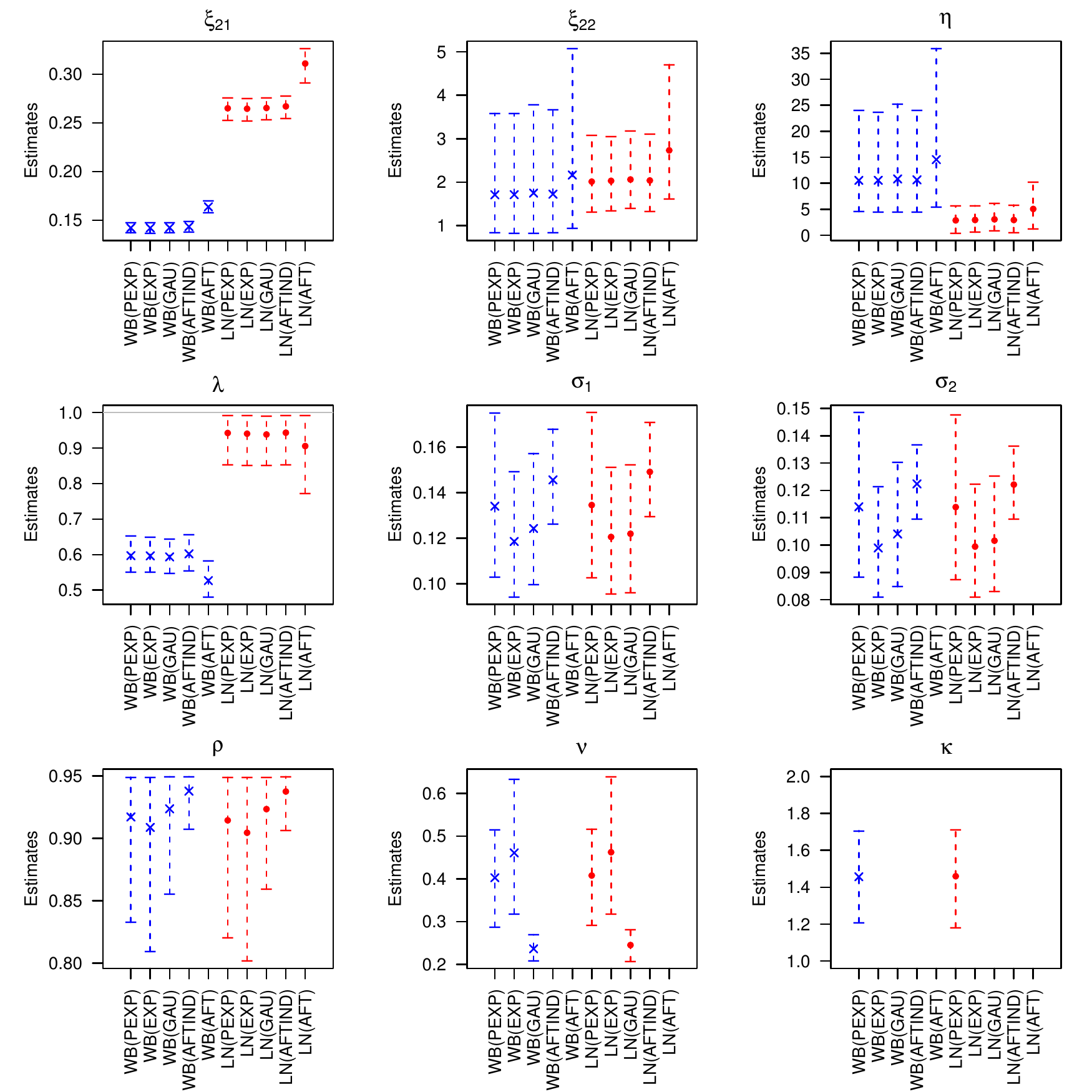}
		\caption{Comparison of CIs for all parameters (set 4) from different models.}\label{CrI4}
	\end{center}
\end{figure}

This section gives the estimates and CIs of all parameters based on all candidate models. Figures \ref{CrI1}--\ref{CrI4} show the estimated posterior means and CIs of parameters from all candidate models. In the figures, WB is short for Weibull, and LN is short for lognormal.

For $\bbeta_1$, $\bbeta_2$, $\sigma_{1}$, $\sigma_{2}$ and $\rho_{12}$, the posterior means and CIs from all models are similar. For $\mu_1$, $\mu_2$, $\xi_1$, $\xi_{21}$ and $\xi_{22}$, the estimates from all Weibull models are smaller than the estimates from lognormal models. The estimates of mixture proportion for Weibull models are close to 0.5, but for lognormal models the estimates are close to 1. This suggests that the lognormal models cannot separate the two modes for DBE failure time distribution. All models using the Gaussian correlation function has smaller estimated $\nu$ compared with models using the exponential or powered exponential correlation functions. Based on the LOOIC table, the best model is the Weibull PEXP model.


%%%%%%%%%%%%%%%%%%%%%%%%%%%%%%%%%%%%%%%%%%%%%%%%%%%%%%%%%%%%%%%%%%%%%%%%%%%%%%%%%%%%%%%%%%%%%%%%%%%%%%%%%%%%%%%%%%%%%%%%%%%%%%%%%%%%%%%
\section{Correlations Among Marginal Failure Times}\label{apd6}
%%%%%%%%%%%%%%%%%%%%%%%%%%%%%%%%%%%%%%%%%%%%%%%%%%%%%%%%%%%%%%%%%%%%%%%%%%%%%%%%%%%%%%%%%%%%%%%%%%%%%%%%%%%%%%%%%%%%%%%%%%%%%%%%%%%%%%%

We assume the failure times are independent conditioning on correlated random effects. For fixed location index $i$ and inside cabinet index $j$, the marginal correlation between $T_{ij1}$ and $T_{ij2}$ can be calculated by integrating out random effects. Similarly, the correlation between $T_{ijk}$ and $T_{i^{\ast}jk}$ can be obtained. Because the Weibull PEXP model has the best performance based on the LOOIC, we calculate the correlation of random failure times based on the Weibull PEXP model. We plug in the estimated posterior means of all the parameters in calculation.

At a particular cabinet location $i$ and inside cabinet position $j$, we have,
\begin{align}
	\textrm{E}(T_{ij1},T_{ij2}) = & \int\int t_{ij1}t_{ij2}f(t_{ij1},t_{ij2})d t_{ij1} d t_{ij2} \notag \\
	 = & \int\int t_{ij1}t_{ij2}\int \int f(t_{ij1},t_{ij2}|w_{i1},w_{i2})f(w_{i1},w_{i2})dw_{i1} dw_{i2} d t_{ij1} d t_{ij2}.\notag
\end{align}
Having $n_l$ random effect draws from the multivariate normal distribution,
$$
\int\int f(t_{ij1},t_{ij2}|w_{i1},w_{i2})f(w_{i1},w_{i2})dw_{i1} dw_{i2} \approx \frac{1}{n_l}\sum_{l = 1}^{n_l} f(t_{ij1},t_{ij2}|w_{i1l},w_{i2l}).
$$
Having $n_r$ uniform draws of $T_{ij1}$ and $T_{ij2}$ from the two dimensional space,
$$
\textrm{E}(T_{ij1}T_{ij2}) \approx \frac{\sum_{r = 1}^{n_r} t_{ij1r}t_{ij2r}\frac{1}{n_l}\sum_{l=1}^{n_l}f(t_{ij1r},t_{ij2r}|w_{i1l},w_{i2l})}{\sum_{r=1}^{n_r} \frac{1}{n_l}\sum_{l=1}^{n_l}f(t_{ij1r},t_{ij2r}|w_{i1l},w_{i2l})}.
$$
Similarly,
\begin{align*}
\textrm{E}(T_{ij1}) \approx \frac{\sum_{r = 1}^{n_r} t_{ij1r} \frac{1}{n_l}\sum_{l = 1}^{n_l} f(t_{ij1r}|w_{i1l})}{\sum_{r=1}^{n_r}\frac{1}{n_l}\sum_{l = 1}^{n_l} f(t_{ij1r}|w_{i1l})},\quad \textrm{E}(T_{ij1}^2) \approx \frac{\sum_{r = 1}^{n_r} t_{ij1r}^2 \frac{1}{n_l}\sum_{l = 1}^{n_l} f(t_{ij1r}|w_{i1l})}{\sum_{r=1}^{n_r}\frac{1}{n_l}\sum_{l = 1}^{n_l} f(t_{ij1r}|w_{i1l})},\\
\textrm{E}(T_{ij2}) \approx \frac{\sum_{r = 1}^{n_r} t_{ij2r} \frac{1}{n_l}\sum_{l = 1}^{n_l} f(t_{ij2r}|w_{i2l})}{\sum_{r=1}^{n_r}\frac{1}{n_l}\sum_{l = 1}^{n_l} f(t_{ij2r}|w_{i2l})}, \quad \textrm{E}(T_{ij2}^2) \approx \frac{\sum_{r = 1}^{n_r} t_{ij2r}^2 \frac{1}{n_l}\sum_{l = 1}^{n_l} f(t_{ij2r}|w_{i2l})}{\sum_{r=1}^{n_r}\frac{1}{n_l}\sum_{l = 1}^{n_l} f(t_{ij2r}|w_{i2l})}.
\end{align*}
Then, we can calculate the correlation $$\textrm{Cor}(T_{ij1},T_{ij2}) = \frac{\textrm{E}(T_{ij1}T_{ij2}) - \textrm{E}(T_{ij1})\textrm{E}(T_{ij2})}{\sqrt{\textrm{E}(T_{ij1}^2) - \textrm{E}(T_{ij1})^2}\sqrt{\textrm{E}(T_{ij2}^2) - \textrm{E}(T_{ij2})^2}}.$$
Similarly, we also calculate the spacial correlation for $T_{ijk}$ and $T_{i^{\ast}jk}$ for some $i$ and $i^{\ast}$. For a fixed $j$, to calculate the correlation between $T_{ijk}$ and $T_{i^{\ast}jk}$, we have
\begin{align*}
	\textrm{E}(T_{ijk},T_{i^{\ast}jk}) \approx  & \frac{\sum_{r = 1}^{n_r} t_{ijkr}t_{i^{\ast}jkr}\frac{1}{M}\sum_{l=1}^{n_l}f(t_{ijkr},t_{i^{\ast}jkr}|w_{ikl},w_{i^{\ast}kl})}{\sum_{r=1}^{n_r} \frac{1}{M}\sum_{l=1}^{n_l}f(t_{ijkr},t_{i^{\ast}jkr}|w_{ikl},w_{i^{\ast}kl})},\\
	\textrm{E}(T_{ijk}) \approx & \frac{\sum_{r = 1}^{n_r} t_{ijkr} \frac{1}{n_l}\sum_{l = 1}^M f(t_{ijkr}|w_{ikl})}{\sum_{r=1}^{n_r}\frac{1}{n_l}\sum_{l = 1}^{n_l} f(t_{ijkr}|w_{ikl})},\\
	\textrm{E}(T_{i^{\ast}jk}) \approx & \frac{\sum_{r = 1}^{n_r} t_{i^{\ast}jkr} \frac{1}{n_l}\sum_{l = 1}^{n_l} f(t_{i^{\ast}jkr}|w_{i^{\ast}kl})}{\sum_{r=1}^{n_r}\frac{1}{n_l}\sum_{l = 1}^{n_l} f(t_{i^{\ast}jkr}|w_{i^{\ast}kl})},\\
		\textrm{E}(T_{ijk}^2) \approx & \frac{\sum_{r = 1}^{n_r} t_{ijkr}^2 \frac{1}{n_l}\sum_{l = 1}^M f(t_{ijkr}|w_{ikl})}{\sum_{r=1}^{n_r}\frac{1}{n_l}\sum_{l = 1}^{n_l} f(t_{ijkr}|w_{ikl})},\\
	\textrm{E}(T_{i^{\ast}jk}^2) \approx & \frac{\sum_{r = 1}^{n_r} t_{i^{\ast}jkr}^2 \frac{1}{n_l}\sum_{l = 1}^{n_l} f(t_{i^{\ast}jkr}|w_{i^{\ast}kl})}{\sum_{r=1}^{n_r}\frac{1}{n_l}\sum_{l = 1}^{n_l} f(t_{i^{\ast}jkr}|w_{i^{\ast}kl})},\\
\textrm{Cor}(T_{ijk},T_{i^{\ast}jk}) = & \frac{\textrm{E}(T_{ijk}T_{i^{\ast}jk}) - \textrm{E}(T_{ijk})\textrm{E}(T_{i^{\ast}jk})}{\sqrt{\textrm{E}(T_{ijk}^2) - \textrm{E}(T_{ijk})^2}\sqrt{\textrm{E}(T_{i^{\ast}jk}^2) - \textrm{E}(T_{i^{\ast}jk})^2}}, \quad k = 1,2.
\end{align*}

%%%%%%%%%%%%%%%%%%%%%%%%%%%%%%%%%%%%%%%%%%%%%%%%%%%%%%%%%%%%%%%%%%%%%%%%%%%%%%%%%%%%%%%%%%%%%%%%%%%%%%%%%%%%%%%%%%%%%%%%%%%%%%%%%%%%%%%

%\bibliographystyle{chicago}
%\bibliography{ref}

\end{document}